\documentclass{aa}  

\usepackage{graphicx}
\usepackage{txfonts}
\usepackage{hyperref}
\begin{document}

   \title{Automated eccentricity measurement from raw eclipsing binary light curves with intrinsic variability\thanks{The summarising catalogue of our \textit{Kepler} analysis results is available at the CDS via anonymous ftp to \url{cdsarc.u-strasbg.fr} (130.79.128.5) or via \url{http://cdsweb.u-strasbg.fr/}}}

   \subtitle{}

   \author{L. W. IJspeert\inst{1}
          \and
          A. Tkachenko\inst{1}
          \and
          C. Johnston\inst{2, 3, 1}
          \and
          A. Prša\inst{4}
          \and
          M. A. Wells\inst{4}
          \and
          C. Aerts\inst{1, 2, 5}
          }

   \institute{Institute of Astronomy, KU Leuven,
            Celestijnenlaan 200D, 3001 Leuven, Belgium\\
            \email{luc.ijspeert@kuleuven.be}
        \and
            Department of Astrophysics, IMAPP, Radboud University Nijmegen, P. O. Box 9010, 6500 GL Nijmegen, the Netherlands
        \and
            Max-Planck-Institut f\"ur Astrophysik, Karl-Schwarzschild-Stra{\ss}e 1, 85741 Garching  bei M\"unchen, Germany
        \and
            Villanova University, Department of Astrophysics and Planetary Sciences, 800 East Lancaster Avenue, Villanova, PA 19085, USA
        \and
            Max Planck Institute for Astronomy, K\"onigstuhl 17, 69117 Heidelberg, Germany
             }

   \date{Received December 23, 2023; accepted February 02, 2023}

 
  \abstract
   {Eclipsing binary systems provide the opportunity to measure the fundamental parameters of their component stars in a stellar-model-independent way. This makes them ideal candidates for testing and calibrating theories of stellar structure and (tidal) evolution. Large photometric (space) surveys provide a wealth of data for both the discovery and the analysis of these systems. Even without spectroscopic follow-up there is often enough information in their photometric time series to warrant analysis, especially if there is an added value present in the form of intrinsic variability, such as pulsations.}
   {Our goal is to implement and validate a framework for the homogeneous analysis of large numbers of eclipsing binary light curves, such as the numerous high-duty-cycle observations from space missions like TESS. The aim of this framework is to be quick and simple to run and to limit the user's time investment when obtaining, amongst other parameters, orbital eccentricities.}
   {We developed a new and fully automated methodology for the analysis of eclipsing binary light curves with or without additional intrinsic variability. Our method includes a fast iterative pre-whitening procedure that results in a list of extracted sinusoids that is broadly applicable for purposes other than eclipses. After eclipses are identified and measured, orbital and stellar parameters are measured under the assumption of spherical stars of uniform brightness.}
   {We tested our methodology in two settings: a set of synthetic light curves with known input and the catalogue of \textit{Kepler} eclipsing binaries. The synthetic tests show that we can reliably recover the frequencies and amplitudes of the sinusoids included in the signal as well as the input binary parameters, albeit to varying degrees of accuracy. Recovery of the tangential component of eccentricity is the most accurate and precise. \textit{Kepler} results confirm a robust determination of orbital periods, with 80.5\% of periods matching the catalogued ones. We present the eccentricities for this analysis and show that they broadly follow the theoretically expected pattern as a function of the orbital period.}
   {Our analysis methodology is shown to be capable of analysing large numbers of eclipsing binary light curves with no user intervention, and in doing so provide a basis for a further in-depth analysis of systems of particular interest as well as for statistical analysis at the sample level. Furthermore, the computational performance of the frequency analysis, extracting hundreds of sinusoids from \textit{Kepler} light curves in a few hours, demonstrates its value as a tool for a field like asteroseismology.}


   \maketitle
%

\section{Introduction}

Eclipsing binary (EB) systems are an important source for determining the physical properties of stars \citep{torres2010}. By being locked in a gravitational dance they provide the opportunity to directly measure individual stellar masses, and by eclipsing each other they provide the opportunity to measure the stellar radii \citep[see][for a brief but clear introduction to the topic]{Southworth2012}. To achieve these measurements, the right observational data are needed: high-precision photometric time series of the brightness of the EB and enough data points of the radial velocities of both stars obtained through spectroscopy. Due to the nature of the observations, obtaining spectra is a lot more time-intensive than photometry, although both types of observations are seeing big boosts in the available amount of data: for example, from the Convection, Rotation and planetary Transits \citep[CoRoT;][]{corot2009}, \textit{Kepler} \citep{kepler2010}, and Transiting Exoplanet Survey Satellite \citep[TESS;][]{tess2015} missions for photometry and the \textit{Gaia}-ESO \citep{gaia_eso2012}, GALactic Archaeology with HERMES \citep[GALAH;][]{galah2015}, SDSS-IV Apache Point Observatory Galactic Evolution Experiment 2 \citep[APOGEE-2;][]{sdssiv2017}, and forthcoming SDSS-V Milky Way Mapper \citep[MWM;][]{sdssv2017} surveys for spectroscopy. 

Large photometric surveys are an ideal source for discovering variable stars such as EBs \citep[see e.g.][]{kepler_eb2011, tess_eb2022, armstrong2016, ijspeert2021, howard2022, rowan2022}. This means that there are large numbers of EBs with available photometric time series data that lack the multiple epochs of spectroscopic data needed to construct radial velocity curves. However, while less powerful, high-precision photometry alone can still provide valuable information about the physical properties of the system. To start, the orbital period determined from the eclipses can inform spectroscopic follow-up, and it is also an important ingredient in a further analysis of the system. Orbital eccentricities give the shape of the orbit and, by being tied to dissipation within the stars, provide a window into the stellar properties. The argument of periastron and inclination define two of the three angles of the orientation of the orbit\footnote{The third one being the longitude of the ascending node, which is generally not constrained.}. The sum of the radii scaled by the semi-major axis and the ratio of radii provide the sizes of the stars relative to the size of the orbit, which also ties back to dissipative processes. The surface brightness ratio is related to the effective temperature ratio and, when combined with the scaled radii, tells us whether the two stars are more like twins or distant cousins.

There are several more properties that influence the shape of the light curve, one example being limb darkening, but they are usually subtler effects. Furthermore, many stars, including those in binaries, are found to exhibit intrinsic variability, such as pulsations \citep[][and citations therein]{Lampens2021}. Systems that show both eclipses and pulsations offer the biggest opportunity to learn by virtue of the information encoded in their light variations \citep{Huber2015}. Various studies have taken advantage of this, as shown by the in-depth analyses by for example, \citet{Maceroni2009}, \citet{Schmid2015}, \citet{Johnston2019a, Johnston2021, Johnston2023}, \citet{Maxted2020}, \citet{Sekaran2020, Sekaran2021}, \citet{guo2022}, \citet{southworth2022a}, and \citet{southworth2022b}. 
However, these studies only treated between one and several dozen systems. The analysis of these systems is often complicated: the amplitudes of pulsations can be comparable to or even exceed the depth of eclipses in the light curve, rendering it impossible to treat a single source of light variation separately. Any generalised approach should therefore account for both the eclipses present and any intrinsic variability that one or both stars exhibit.

Many interesting eclipsing systems have been manually investigated in depth to great effect, providing physical insights on a case-by-case basis \citep[e.g.][to name just a few]{Schmid2016, Johnston2019b, Jennings2023, Pavlovski2023, Rosu2023}. It is through these kinds of studies that we can calibrate in detail our models of stars, for example with regards to the discrepancy between the dynamical and evolutionary masses of intermediate- to high-mass stars \citep[e.g.][]{Burkholder1997, Guinan2000, Weidner2010, Massey2012, Tkachenko2020, Johnston2021b}. This underlines the value of these systems for the calibration of stellar models. With the number of known EBs in the thousands and increasing, investigating every one of them by hand in a reasonable amount of time is infeasible. If a portion of the analysis could be done in an automated way, this would give the researcher more time to take a more detailed look at a select number of promising systems. Additionally, there is a lot of power in numbers, as certain aspects of stellar physics can only be constrained by being compared with the population as a whole \citep[see e.g.][]{vaneylen2016, Lurie2017, Wells2021}. Altogether, high levels of automation in data analysis are needed to achieve many of the goals in stellar astrophysical research today. 

Notable work towards automated pipelines for EB analysis includes Detached Eclipsing Binary Light curve fitter \citep[DEBiL;][]{Devor2005}, Eclipsing Binary Automated Solver \citep[EBAS;][]{Tamuz2006}, and Eclipsing Binaries via Artificial Intelligence \citep[EBAI;][]{Prsa2008}. The DEBiL analysis code is part of a larger EB detection pipeline and uses simulated annealing and the downhill simplex method \citep{NelderMead1965} to solve the inverse problem. EBAS has a strategy for making an initial guess of the parameters using grid searches, gradient descent, and light curve geometry and then employs simulated annealing as well. DEBiL uses its own light curve model implementation of limb-darkened spherical stars, whereas EBAS uses the more complex EBOP light curve generator. With EBAI, \citet{Prsa2008} took a different approach and used machine learning with neural networks trained on the rigorous Wilson \& Devinney \citep{WD1971} model to arrive at a fast estimator for EB parameters. All three pipelines were applied to the published database of Optical Gravitational Lensing Experiment \citep[OGLE; ][]{OGLE1992} EBs \citep{Wyrzykowski2003}, which at the time typically had of the order of a few hundred data points per target. However, while two of the pipelines include a measure of ellipsoidal variability, none account for the intrinsic variability in the form of pulsations that we are specifically pursuing in this work, given its high astrophysical value.

We present a new and fully automated method and its implementation, \texttt{STAR SHADOW}, for the analysis of EB light curves with and without intrinsic variability. Starting from an input light curve, the algorithm can provide the orbital period, several orbit- and eclipse-specific parameters such as eccentricity, and a set of sinusoids that are added to the model of eclipses to fully describe the light variations in the time series. If further investigation into a system's eclipses or the remaining variability is desired, the removal of the other signal, be it the eclipses or the pulsations, is greatly simplified. We validated our method on artificial data and applied it to a population of EBs from the \textit{Kepler} database \citep{kepler_eb2011, kepler_ebii2011, kepler_ebvii2016} to further validate the computed orbital periods. For the  \textit{Kepler} population, we determined the orbital configurations, in particular eccentricities, and the properties of the variability intrinsic to the binary components. We provide a detailed description of the method and its application to a number of realistic synthetic test cases in this paper.


\section{Light curve analysis}
\label{sec:method}

We developed an automated recipe for the analysis of EB light curves to handle large numbers of targets in a consistent, time-efficient way and without manual involvement. The implementation of the recipe, called \texttt{STAR SHADOW} (Satellite Time series Analysis Routine using Sinusoids and Harmonics Automatedly for Double stars with Occultations and Waves), is publicly available on \href{https://github.com/LucIJspeert/star_shadow}{GitHub}\footnote{\url{https://github.com/LucIJspeert/star_shadow}} (version 1.1.7a was used here) and comes with a short guide to get started. Since one of the main aims is the application to data from the TESS mission, the algorithm works best for high-duty-cycle satellite time series photometry and there are a few TESS-specific features like loading in TESS data products and recognition of TESS observing sectors. The algorithm is otherwise fully agnostic to the source of the input light curve, so in the following text if we mention `sectors', this can be taken to mean any subdivision of the data in the time domain. The method is meant to be applied to known EBs with space-based time series data of tens of ppm precision, which covers the TESS, \textit{Kepler,} and PLAnetary Transits and Oscillations of stars \citep[PLATO;][]{PLATO2022} missions, and it is assumed that care has been taken in detrending the raw light curves. Our own two main goals are the determination of the orbital eccentricity and a measure for the significance of variability in the light curve after subtraction of the eclipses. 

\begin{figure}[h!]
\centering
\includegraphics[width=\hsize]{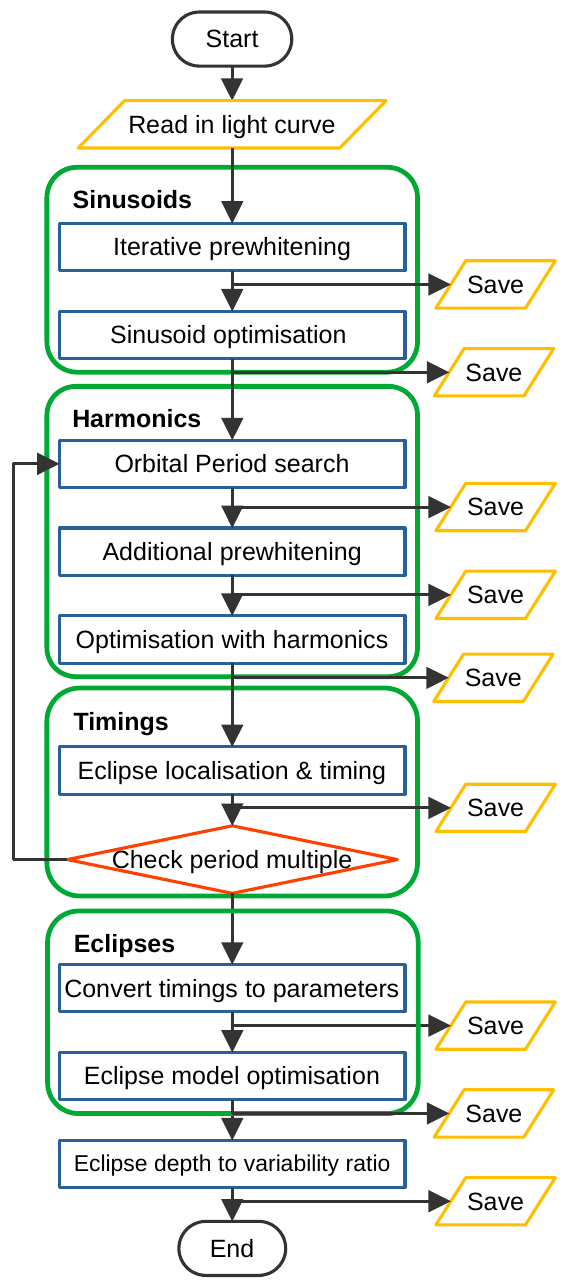}
    \caption{Flowchart of the \texttt{STAR SHADOW} analysis steps. Blue rectangles contain analysis steps that are themselves grouped into four more general functionalities. Read and write operations are in yellow parallelograms, and the orange diamond denotes a decision point to redo a number of steps concerning the orbital period.}
    \label{fig:flow}
\end{figure}

\subsection{Top level overview}

We provide a visual representation of the flow of the analysis steps in Figure \ref{fig:flow}, with additional example analysis output given in Appendix \ref{apx:overview}. The recipe starts with an iterative pre-whitening procedure meant to capture all the statistically relevant (with regards to the Bayesian information criterion) variability in the light curve in a mathematical model of sine waves. The parameters of the sinusoids are optimised and, if not given beforehand, the orbital period is determined. Frequencies in the previously extracted list that match a theoretical orbital harmonic within a tolerance are forced to the exact multiple of the orbital frequency, thus replacing all frequencies within the tolerance with a single harmonic. If frequencies are closely spaced around the harmonics, this approach will remove too many frequencies from the model: an additional round of iterative pre-whitening ensures that the model of the light curve is complete. The model is again optimised, now including the orbital period as a free parameter, although its value tends to only change by a fraction of the period error estimate at this point, if at all.

\begin{figure}
\centering
\includegraphics[width=\hsize]{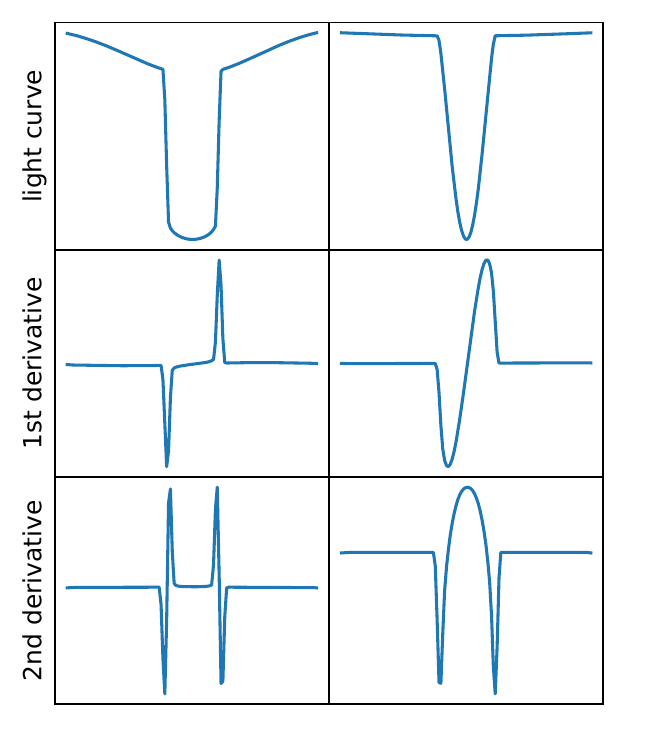}
    \caption{Typical eclipse signatures in the derivatives of the light curve. The left column shows a flat-bottomed eclipse resulting in two separated features in the derivatives, while the V-shaped eclipse on the right leads to peaks merged into one feature. Adapted with permission from \citet{ijspeert2021}}
    \label{fig:signature}
\end{figure}

The eclipses are located in the time series using their signatures in the first and second time derivatives, using only the model of harmonic sinusoids for the time derivatives to reduce the chance of mistaking intrinsic variability for eclipses. Examples of what is meant by such eclipse signatures can be seen in Figure \ref{fig:signature}. The eclipse timings and depths are measured by finding specific extrema and zero points in the two time derivatives. The eclipse detection is based on the methodology described in \citet[][Section 3]{ijspeert2021}.

The eclipse timings and depths are translated into some of the orbital and physical parameters of the system: eccentricity, argument of periastron, inclination, sum of the scaled radii, ratio of radii, and ratio of surface brightness. This is done using Kepler's second law and by iteratively solving a set of equations that relate specific points in the orbit to the system's properties. This is discussed further in Sect. \ref{sec:eclipse_analysis}.

The orbital and physical parameters are used as a starting point to fit a simple light curve model of eclipses to the data, to potentially increase the accuracy of some of the parameters. The model assumes spherical stars of uniform surface brightness and we refer to it later as just the `eclipse model' for brevity. The fit is done in three steps: fit only the eclipse model to the original light curve, subtract the eclipse model and iteratively pre-whiten the residuals, and then optimise the full model of eclipses and sinusoids together.

Finally, the ratios of eclipse depth to variability amplitude are computed to have a quick statistic of the balance between the two signals. At every step that adds, removes, or otherwise alters sinusoids, the extracted sinusoids are checked for significance against a signal-to-noise threshold, but not removed based on this check.

\subsection{Detailed description}

The model of the light curve constructed during the iterative pre-whitening procedure also includes a linear trend for each sector. For TESS data, the linear trend is split up according to the first and second half of each sector; each piece is described by an independent slope and a y-intercept. This captures the average level of the light curve as well as remaining linear instrumental trends in the TESS data products, which may include sudden jumps in the light curve of a sector due to momentum dumps of the gyroscopes. Where the linear trend is split up can be defined by the user. The slope and y-intercept are calculated through their maximum likelihood estimators (MLEs), and updated at each step of the subsequent iterations of the pre-whitening as well as during optimisation procedures. The formulae of the MLEs are provided in Appendix \ref{apx:mle}.

The time axis of the data is mean-subtracted throughout the analysis, which ensures a minimal correlation between parameters such as the slope and the y-intercept, and between the frequency and the phase. This means that sinusoid phases and eclipse times are measured with respect to this reference time. For the linear trends, the mean is computed for each separate sector, so their y-intercepts are with respect to those time points.

\subsubsection{Frequency analysis}
This subsection discusses the `sinusoids' block in the flow diagram (Figure \ref{fig:flow}) and the visualisation in Figure \ref{fig:schematic_1}. In the iterative pre-whitening procedure we build a model of the light curve in terms of sine waves: 
\begin{equation}
    y(t) = \sum_i A_i sin\left(2 \pi f_i (t - t_{mean}) + \phi_i\right),
    \label{eq:sines}
\end{equation}
where $t$ is time, $f$ is frequency, $A$ is amplitude, and $\phi$ is phase.
In this procedure, the next sinusoid to be subtracted is picked by finding the highest amplitude peak in the Lomb-Scargle periodogram \citep{1976Ap&SS..39..447L, 1982ApJ...263..835S}. The periodogram is over-sampled by a factor of 10 with respect to the Rayleigh criterion given by 1/T, where T stands for the total time base of observations.\footnote{So the periodogram frequency sampling is $1 / (10\cdot timespan)$.} The highest peak in the periodogram is over-sampled by an additional factor of 100 to obtain a precise position (i.e. frequency) and amplitude. The phase of the sinusoid is determined directly from the periodogram as well, using the formula described by \citet[][Equation 12]{Hocke1998}. 

A sinusoid is only accepted if the Bayesian information criterion (BIC) of the residuals is reduced by at least two, indicating at least positive evidence of its presence by rule of thumb \citep{Raftery1995, Kass1995}. The BIC decreases with more white-noise-like residuals (i.e. containing less information) but increases with additional free parameters, of which each sinusoid initially has three (see Appendix \ref{apx:bic} for a brief derivation). If the next sinusoid does not meet the requirement, the iteration is stopped. This was found to be a suitable procedure for CoRoT space photometry \citep{Degroote2009}. In the case red noise is present, and depending on its power spectral density, it is possible that frequency peaks of high enough signal-to-noise are missed if they have a frequency above the regime affected by the red noise but an amplitude below the amplitude of the red noise.

If at any point during the iterative pre-whitening a frequency is found within 1.5 times the Rayleigh criterion \citep{Loumos1978} of a previously extracted frequency, their peaks in the periodogram will have been blended prior to subtraction. This blending means that the top of the true peak in the periodogram got shifted and the extracted frequency is in a different location. A mechanism is introduced to counteract this shift, which checks each frequency that is being extracted for closely spaced frequencies in the previously extracted list. If a chain of closely spaced frequencies is found around the currently extracted frequency, all frequencies in the chain are iteratively updated by removing one of them at a time and re-extracting it from the updated periodogram around that frequency. Here `chain' is used to mean any consecutive set of frequencies that are closer than the frequency resolution (1.5/T) to their direct neighbour. This extends to frequencies that are further than the frequency resolution of the currently extracted frequency. The iteration of this sub-loop is terminated when the BIC of the residuals no longer decreases.

The sinusoid parameters may now have shifted with respect to their initial extraction, and the algorithm may have to remove some sinusoids again based on the BIC. This is done in two parts, starting with removing individual sinusoids one by one and checking whether it results in a decrease in the BIC of more than two. Secondly, it is checked whether chains of closely spaced frequencies (or any consecutive sub-chains) can be replaced by a single frequency and yield an improvement.

At the end of each step that involves extracting or updating sinusoid parameters, their significance is checked based on several criteria and stored alongside the parameters. This has no further effect on the analysis process. Sinusoids with an amplitude that is consistent with zero within three sigma are marked insignificant, and for frequencies that fall within three sigma of each other, the lowest amplitude sine wave is marked insignificant. The frequency-dependent periodogram noise level is computed by taking a running average over the Lomb-Scargle periodogram of the residuals of the light curve using a frequency window of one per day, as is common in the literature. The signal-to-noise for each sinusoid is the amplitude divided by this periodogram noise level and those not meeting the threshold are marked insignificant. We used the signal-to-noise threshold as determined by \citet[Equation 6]{Baran2021}, which is a function of the number of data points and computed for a false alarm probability of $0.1\%$. We selected candidate harmonics as well, marking only the sinusoids that are within three sigma of a theoretical harmonic frequency (this selection is only relevant later in the analysis). 
\\

At the end of the frequency extraction, it is necessary to do a multi-sinusoid non-linear optimisation to get closer still to the real frequencies present in the light curve \citep[Sect. 3]{Bowman2021}. We implemented the \texttt{minimize} function from the Scipy \citep{scipy} optimisation module and used the Limited-memory Broyden–Fletcher–Goldfarb–Shanno Bounds (L-BFGS-B) method with analytical gradients (see Appendix \ref{apx:details} for more insight into the choice of method and the mentioned gradient formulae). Doing such an optimisation whereby all sinusoids are simultaneously fitted is extremely time-consuming. Therefore, to limit the time taken the algorithm fits in groups of 20 to 25 sinusoids. This reduces the fit performance in terms of the final BIC value by about one-quarter of what it could have reached with a fully simultaneous fit in small-scale simulated tests, but the time saved is highly disproportional to this performance decrease (orders of magnitude). Groups are based on the amplitude of the sinusoids, bundling those with similar amplitudes together with the idea that those will have a similar level of influence on each other and on the likelihood function. The division between groups is made by sorting the amplitudes in descending order and making the cut where the largest jump in amplitude occurs within the given constraints of minimum and maximum group size. Group size is not fixed to one number in order to increase amplitude disparity between groups. Frequencies, amplitudes, and phases of the sinusoids in each group are fitted as free parameters in turn until all sinusoids have been optimised. Slopes and y-intercepts of all linear trends always remain free parameters throughout the fitting process.
\\

\subsubsection{Frequency analysis with harmonics}
This subsection discusses the `harmonics' block in the flow diagram (Figure \ref{fig:flow}) and the visualisation in Figure \ref{fig:schematic_2}. We now introduce the orbital period as a free parameter.\footnote{Up to this point, the pre-whitening steps are applicable to any time series and could be used as a convenient and quick method on its own.} If known, an initial period can be provided, which is then further optimised. Alternatively, a period-finding algorithm is used to obtain the unknown orbital period.

The period-finding algorithm makes use of the already obtained information of sinusoids by only searching at the extracted frequencies and fractions of those. A combined phase dispersion minimisation measure \citep{stellingwerf1978} and Lomb-Scargle amplitude statistic is computed as described in \citet{saha2017}. In addition, this is multiplied with two more statistics that use the known list of frequencies and are geared towards improving the odds of finding the correct EB orbital period. These two statistics are calculated as follows: (1) The number of harmonic sinusoids present in the extracted frequencies. At each tested orbital frequency, the theoretical orbital harmonic frequencies are computed and compared to the list of extracted frequencies. Those extracted frequencies that fall within the frequency resolution (1.5/T) of a theoretical harmonic are selected and count towards the length of the harmonic series for that orbital frequency. (2) The filling factor of the harmonic series. The second statistic is also based on the harmonic series and captures the filling factor calculated as the number of harmonics present in the extracted frequency list divided by the total number of theoretically possible harmonics below the Nyquist frequency. The Nyquist frequency is defined as the sampling rate divided by two; we used one over the minimum time step for the sampling rate. 

After multiplying these four statistics the highest scoring frequency is taken as the initial orbital frequency. The best orbital frequency is further refined locally by minimising the sum of squared distances between the series of extracted harmonic frequencies and the theoretical harmonic frequencies in a grid with a sampling of 0.001\% over a 1\% interval around the orbital frequency. 

As a final check, the two statistics on the series of harmonics as described above are calculated for several multiples of the orbital period: factors of 1/2, 2, 3, 4, and 5. There are two conditions for changing the period multiple, concerning both the filling factor itself and the multiplication of the two statistics of the filling factor and harmonic series length. If the multiplication of the two statistics results in more than 1.1 times the value in this statistic than for the original period and the filling factor is at least 0.85 times that of the original period, the final orbital period is set to the multiple achieving the best score. Alternatively, if both the mentioned values exceed 0.95 specifically for doubling the period, the period is doubled. The threshold values for this decision are empirically determined based on the tests described in Sections \ref{sec:testing} and \ref{sec:kepler}. 

The error on the period is estimated by taking the simple linear regression (SLR) uncertainty where each observed eclipse is represented by a single timestamp. This error is only a function of the number of observed eclipses and the uncertainty on the individual timestamps, which is taken to be half the integration time for the observations at this stage. The number of observed eclipses is approximated by rounding down the time base divided by the orbital period, and thus assuming no gaps.

Once the algorithm has a binary period, it selects the orbital harmonic frequencies. All frequencies falling within 1.5 times the Rayleigh criterion of the theoretical harmonics are removed and replaced by exact multiples of the orbital frequency. The new amplitudes and phases are obtained from the periodogram. From this point, the orbital harmonics will have a reduced number of free parameters (from three for unconstrained sinusoids to two, for example), and the orbital period is one additional free parameter. Despite the reduction in parameters and updated harmonic sinusoids, it is possible for this step to increase the overall BIC of the residuals, as too many sinusoids may have been removed.
\\

Since harmonics have one free parameter fewer than their base frequency, it is possible that additional harmonics can be extracted from the periodogram while decreasing the BIC of the residuals by more than two. This is attempted by calculating the amplitude and phase at harmonic frequencies not yet occupied in the frequency list, up to the Nyquist frequency, and checking the significance criterion before they are added to the sinusoidal model of the light curve. Additionally, due to the removal of candidate harmonics, it is likely that more unconstrained sinusoids can be extracted analogously to the initial pre-whitening. After adding sinusoids to the list, the algorithm checks again whether individual frequencies have to be removed, or whether closely spaced chains of frequencies can be replaced by a single frequency based on the BIC of the residuals. If a chain of close frequencies contains a coupled harmonic sinusoid, the algorithm attempts to replace the chain with the harmonic (and therefore does not remove the harmonic). The harmonic does not change in frequency, but its amplitude and phase can change.
\\

Finally, a second multi-sinusoid non-linear optimisation that is more or less analogous to the first is performed. The only differences are that a set of harmonic sinusoids is now coupled to the orbital period and the period is optimised along with the amplitudes and phases of the harmonics.

\subsubsection{Eclipse timing measurements}
\label{sec:eclipse_analysis}
\begin{figure}
\centering
\includegraphics[width=\hsize]{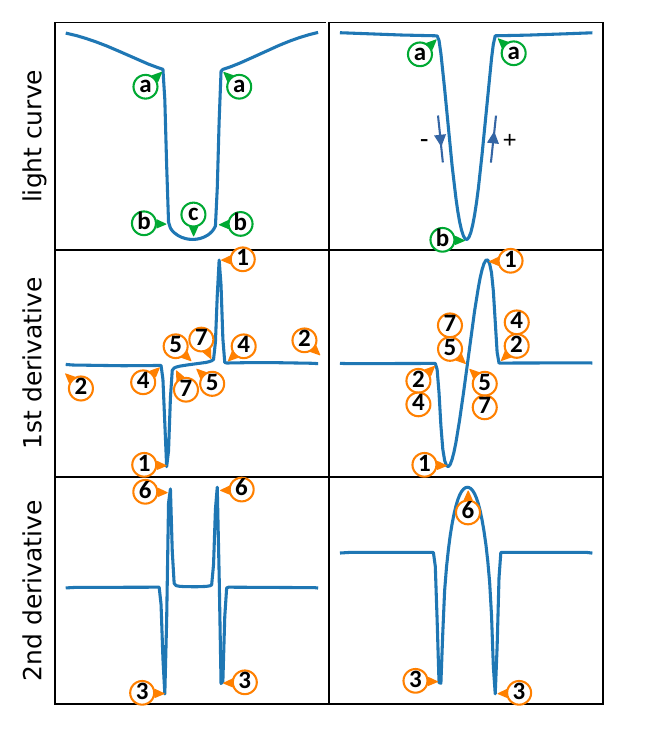}
    \caption{Typical eclipse signatures in the derivatives of the light curve, with markers pointing to important features for the automated eclipse detection and timing measurement. Note that points 7 in the left panel of first derivatives are not truly local minima but marked as such for illustrative purposes. Adapted with permission from \citet{ijspeert2021}}
    \label{fig:ecl_markers}
\end{figure}

This subsection discusses the `timings' block in the flow diagram (Figure \ref{fig:flow}) and the visualisation in Figure \ref{fig:schematic_3}. For the purpose of locating the eclipses, the eclipses are separated out from other variability in the light curve as much as possible by only using the model of orbital harmonic sinusoids (henceforth: harmonic model). Variability at the harmonic frequencies other than the eclipses cannot be separated yet, which can cause problems when that variability has high frequencies relative to the orbital frequency. For this reason, the algorithm initially determines the approximate eclipse positions using orbital harmonics up to and including the twentieth harmonic. Including fewer harmonics does not help as the eclipses lose too much detail. If this fails, it is attempted again with the first forty harmonics and finally with all the orbital harmonics. We provided a visualisation in Figure \ref{fig:ecl_markers} to aid in the explanation of the analysis steps below, matching the markers in the figure with numbers and letters between brackets in the text.

The first and second analytically computed time derivatives of the harmonic model inform the automated recognition of the eclipse signatures. The highest peaks in the absolute first derivative belong to the steepest slopes, which in most EB light curves will mean an eclipse ingress or egress. The sign of the first derivative at these locations allows the algorithm to distinguish the start of the eclipse from the end of the eclipse. The algorithm proceeds by determining the first and last contact points\footnote{Contact refers to the point in time when the observer sees the projected discs of the two stars touch while they are not (yet) covering each other. In other words the start and end of the eclipse. This always happens per definition of an eclipse.} of the eclipses by following several steps, as described below. Starting from the extrema in the first derivative (1) and moving away from the centre of the eclipse, we find its first zero point (2). Between the extrema and these zero points, there is a minimum in the second derivative (3) where the curvature of the eclipse is greatest (most negative). The minima in the second derivative are taken to be the inner limits for the position of the first and last contact points. The algorithm checks for the presence of local minima in the absolute first derivative (4) between the zero points in the first derivative and the minima in the second derivative. If present, these local minima replace the zero points in the first derivative as the outer limit for the position of the first and last contact points. The times of first and last contact (a) are taken to be midway between the described inner and outer limits, and the interval (3-4) is taken to be the three sigma error region. 

Towards the centre of the eclipse, the algorithm determines the first and last internal tangency points\footnote{Internal tangency refers to the point in time when the observer sees the projected discs of the two stars touch tangentially while one is in front of the other. This does not necessarily happen in every eclipse.}, analogously to the contact points albeit in the opposite direction. Starting again from the extrema in the first derivative (1) and moving towards the centre of the eclipse, we find its first inner zero point (5). Between the extrema and these zero points, there is a maximum in the second derivative (6) where the curvature of the eclipse is greatest (most positive). The maxima in the second derivative are taken to be the outer limits for the position of the first and last tangency points. The algorithm checks for the presence of local minima in the absolute first derivative (7) between the zero points in the first derivative and the maxima in the second derivative. If present, these local minima replace the inner zero points in the first derivative as the inner limit for the position of the first and last tangency points. If the algorithm finds two distinct maxima in the second derivative and the second derivative reaches zero in between them, the first and last internal tangency points (b) are defined as the midway between their respective limits and the interval (6-7) is taken to be the three sigma error region. The internal tangency points now delimit the flattened-off bottom of the eclipse. If instead, the algorithm finds that these two maxima coincide, or the curvature does not reach zero in between them, no such flat bottom is defined. In the case of a highly slanted (asymmetric) eclipse, the second derivative might not reach a strong enough maximum on one side of the eclipse for the algorithm to identify that edge.

The eclipse minima (c) are defined to be in the middle between the internal tangency points, and not as the actual minimum in the harmonic model. This is more robust in any cases with variability at harmonic frequencies that does not belong to the eclipses, and more closely corresponds to the time of conjunction. Possible asymmetry can still be accounted for because the internal tangency points may be anywhere between the contact points. 

The best consecutive primary and secondary eclipse combination is chosen by taking those with the largest combined depth. If no secondary is found, or no eclipses are found at all, the analysis is stopped. The method for determination of the orbital eccentricity does not hold for systems with no detected secondary eclipse. This method of locating the eclipses was already described in \citet[Section 3]{ijspeert2021}, but was adapted and simplified to analytic sums of sine waves instead of raw data. We note that in the left panel of first derivatives in Figure \ref{fig:ecl_markers}, points 7 are not truly local minima in the portrayed derivative of the theoretical eclipse model. Using finite sums of sine waves generally results in many more local minima and maxima than is the case for a theoretical model of eclipses, because of the inflection points around (sharp) bends, thus allowing for the simple identification of points 7 as local minima in the figure if a flat eclipse bottom is present. 

Once two eclipses have been detected their timing measurements are refined using all significant harmonics. Limiting to lower harmonics works to make detection more robust, but the downside is that eclipses look wider and shallower. An altered version of the eclipse detection algorithm is used that has the benefit of starting from the previously detected positions and is more robust against high-frequency interference. The characteristic points measured are the same, but the method of arriving at them is different. 

Starting off from previously determined inner zero points in the first derivative (5), the algorithm adjusts them to the different number of sinusoids included in the first derivative by moving to the nearest zero point. Subsequently, the algorithm determines the updated extrema in the first derivative (1) by searching between the inner zero points and the previous extrema in the first derivative. Moving away from the eclipse centre from the newly found extrema in the first derivative, we find the outer zero points (2). If secondary (local) extrema occur within the outer zero points, and these are sufficiently high (> 80\% of the height), the extrema in the first derivative (1) are shifted to those locations. The minima in the second derivative (3) are now found between the extrema and outer zero points in the first derivative. Local minima in the first derivative (4) are found by moving outwards from the minima in the second derivative. The inner zero points (5) are now adjusted again, by moving inwards to zero from the extrema in the first derivative. The maxima in the second derivative (6) are found between the extrema and inner zero points in the first derivative. Inner local minima in the first derivative (7) are obtained by moving inwards along the first derivative curve from the maxima in the second derivative. If the ingresses or egresses are found to be less than a quarter of the width of initial localisations (obtained with fewer harmonics), it may indicate that the algorithm did not capture the full extent of the eclipse due to variability not related to the eclipses. The algorithm adjusts the outer eclipse points (1, 2, 3, and 4) farther outwards by moving to the next extrema in the first derivative (1) and repeating the steps to find the other points. This adjustment is only accepted if it does not cross the original outer edges (4) as determined in the localisation step, and if it increases the depth by more than 20 percent. If the depth is not substantially increased, we are likely already outside of the eclipse.
\\

Improved error estimates for the individual timings and error estimates for the individual depths are determined using the noise level and the slope of the eclipse in- and egress. The noise level is calculated as the standard deviation of the residuals of the light curve. For errors in the individual eclipse times of ingress and egress, the algorithm determines an average slope of each ingress and egress by fitting a straight line to each side of the eclipse. Dividing the noise level by this slope gives the time it takes the flux level to cross the noise level, thus providing an estimate of the timing precision for contact points. Errors in the individual eclipse times are derived from this by taking half and adding in quadrature to the error estimates obtained during eclipse detection:
\begin{equation}
    \sigma_\textup{t, in/egress} = \sqrt{\frac{1}{4}\sigma_\textup{t, noise/slope}^2 + \sigma_\textup{t, detection}^2}.
    \label{eq:indiv_t}
\end{equation}
The factor 1/4 comes from halving the `noise-crossing time' to make an error estimate out of it. Analogous to the orbital period error the error on the final eclipse timings, over all observed eclipses, is determined from the SLR uncertainty, which is only a function of the number of observed eclipses and the uncertainty on the individual timings given by Equation \ref{eq:indiv_t}. The error for the orbital period is also recalculated, now using the averaged error on the primary and secondary eclipse times as input instead of the time stamp error.

The error in the flux level at first and last contact is estimated as being the noise level, and the error at the bottom of the eclipse is either the noise level (if there is no flat bottom) or the standard deviation of the flux measurements from first to last internal tangency (if there is a flat bottom). These flux level errors then translate into depth errors by adding them in quadrature:
\begin{equation}
    \sigma_\textup{depth} = \sqrt{\frac{1}{4}\sigma_\textup{first\ contact}^2 + \frac{1}{4}\sigma_\textup{last\ contact}^2 + \sigma_\textup{bottom}^2}.
    \label{eq:indiv_d}
\end{equation}
The factor 1/4 comes from averaging the flux levels for the first and last contact points in the calculation for the eclipse depth. To obtain the error on the final eclipse depths, over all observed eclipses, we divided Equation \ref{eq:indiv_d} by the square root of the number of eclipses.

Eclipses are rejected under certain conditions, at the end of the detection phase but also at points during the determination of the best candidates for primary and secondary eclipse. Most notably the algorithm applies a significance criterion to the depth and duration of the eclipses based on the estimated errors and a signal-to-noise threshold. The depth must exceed one sigma of the estimated error and one-half the standard deviation of the residuals, whereas the duration must exceed three sigma of the estimated error. Examples of other criteria, applied during the selection of candidate eclipses, look at differences between overlapping candidates and whether a candidate greatly changes in depth between harmonic models with all or only low-frequency harmonics (indicating the occurrence of unrelated signal). These selection criteria are meant to filter out as many false positive candidates as possible before choosing the final set of eclipses to use in the rest of the analysis. However, these criteria or the conditions applied at the end may result in only a primary eclipse (or none), in which case this is logged and the analysis ends.
\\

\subsubsection{Eclipse analysis}
This subsection discusses the `eclipses' block in the flow diagram (Figure \ref{fig:flow}) and the visualisation in Figure \ref{fig:schematic_4}. Determination of orbital and physical properties -- including the eccentricity, argument of periastron, orbital inclination, sum of the scaled radii, ratio of the radii, and surface brightness ratio -- is initially done through the use of formulae coupling these properties to the timings and depths of the primary and secondary eclipse. These formulae can be found in, for example, \citet{Kopal1959}, but some will be repeated here in altered form for completeness. As the main link between time and orbital elements, we need Kepler's second law in integral form:
\begin{equation}
    \frac{2\pi(t_b-t_a)}{P} = \left(1-e^2\right)^{\frac{3}{2}} \int_{\nu_a}^{\nu_b}\frac{d\nu}{(1 + e\,cos(\nu))^2},
    \label{eq:kepler}
\end{equation}
where $t_a$ and $t_b$ represent time points in the orbit, $P$ is the orbital period, $e$ the eccentricity and $\nu_a$ and $\nu_b$ the true anomaly angles corresponding to the time points. This equation has an analytic solution, which we show in Appendix \ref{eq:kepler3}. Since we do not know the true anomaly ($\nu$) angles beforehand, we have broken it down into argument of periastron ($\omega$) and phase angle ($\theta$): 
\begin{equation}
    \nu = \frac{\pi}{2} - \omega + \theta.
    \label{eq:true_anom}
\end{equation}
The argument of periastron is a free parameter defining part of the orbital orientation in the sky, the phase angle gives the position along the orbit and is zero near primary minimum and $\pi$ near secondary minimum. The algorithm now only needs to determine the phase angles at several strategic points in the orbit to calculate useful values with the integral (\ref{eq:kepler}). These points constitute the times of minima, times of contact and for certain cases times of internal tangency as well. Times of minima are reached when the projected orbital separation reaches a minimum, hence when its derivative becomes zero:
\begin{equation}
    0 = e\ cos(\theta - \omega) + sin^2(i)\ cos(\theta)\left(sin(\theta) - e\ cos(\omega)\right),
    \label{eq:ecl_minima}
\end{equation}
with $i$ the orbital inclination. From this formula, we can see that in circular orbits the minima are at phase angles of exactly zero and $\pi$. Times of contact are reached when the projected distance between the two stars is equal to the sum of both radii (scaled by the semi-major axis), leading to the equation\begin{equation}
    \sqrt{1 - sin^2(i)\ cos^2(\phi)} = \frac{r_1 + r_2}{a (1 - e^2)} (1 \pm e\ sin(\omega \pm \phi)),
    \label{eq:ecl_edges}
\end{equation}
where we introduce the angles $\phi$, which are the phase angles of the contact points as measured from zero and from $\pi$ for the primary and secondary eclipse, respectively. In their exact form, Equations \ref{eq:ecl_minima} and \ref{eq:ecl_edges} do not have analytical solutions. Approximate analytical solutions do exist, but these are only valid for low eccentricities. 

To be able to obtain accurate results for all eccentricities, we used an iterative form of Kepler's second law and let the algorithm solve it numerically for $\psi$:
\begin{equation}
    \frac{2 \pi (t_2 - t_1)}{P} = \psi - sin(\psi),
    \label{eq:psi}
\end{equation}
where $\psi$ relates to the component $ecos(\omega)$ of the eccentricity. We write\begin{equation}
    e cos(\omega) = \frac{tan\left(\frac{\psi - \pi}{2}\right) \sqrt{1 - (esin(\omega))^2}}{\sqrt{1 + tan^2\left(\frac{\psi - \pi}{2}\right)}},
    \label{eq:ecosw}
\end{equation}
which depends on an estimate of $e sin(\omega)$. For $e sin(\omega)$ we write a formula with explicit dependence on inclination, as well as auxiliary angle $\phi_0$, which further simplifies with the assumption of small $\phi_0$ and $i = 90^\circ$:
\begin{align}
    e sin(\omega) &=  \frac{\pi (\tau_{1,1}+\tau_{1,2}-\tau_{2,1}-\tau_{2,2})}{2 P sin(\phi_0)} \frac{sin^2(i) sin^2(\phi_0)}{1 - sin^2(i) (1 + sin^2(\phi_0))}, \label{eq:esinw} \\
     &\approx \frac{d_2 - d_1}{d_2 + d_1},
\end{align}
introducing the notation $\tau$ as the time duration from conjunction to either contact point and the subscript $\tau_{n,m}$ is used to indicate primary ($n=1$) or secondary ($n=2$) eclipse and first ($m=1$) or last ($m=2$) contact. Eclipse durations are denoted with $d_n$ here. We used formula \ref{eq:esinw} with $i = 90^\circ$ and the approximate formula for the auxiliary angle, $\phi_0$, as in \citet{Kopal1959}:
\begin{equation}
    \phi_0 \approx \frac{\pi}{2 P} \left(\tau_{1,1} + \tau_{1,2} + \tau_{2,1} + \tau_{2,2}\right).
    \label{eq:phi0_approx}
\end{equation}
Angle $\phi_0$ relates to the sum of the scaled radii in the following manner:
\begin{equation}
    \sqrt{1 - sin^2(i)\ cos^2(\phi_0)} = \frac{r_1 + r_2}{a \left(1-e^2\right)},
    \label{eq:phi0_radius}
\end{equation}
with $a$ the semi-major axis. After obtaining the components of the eccentricity, the algorithm uses an iterative procedure to improve the value of $\phi_0$ by computing the sum of eclipse durations using Kepler's second law (Equation \ref{eq:kepler3}) and Equation \ref{eq:ecl_edges}. A fast global minimisation method is used to find inclination, ratio of radii and ratio of surface brightness by comparing measured eclipse times and depths against the values computed using Equations \ref{eq:kepler3}, \ref{eq:ecl_minima}, \ref{eq:ecl_edges}, and \ref{eq:simple_model}. 

All of the above assumes spherical stars of uniform brightness. Error estimates are made using a combination of analytical error formulae and the Monte Carlo approach of importance sampling. We provide more details for this step and the error formulae in Appendix \ref{apx:translation}.
\\

\begin{figure*}
\resizebox{\hsize}{!}
    {\includegraphics[width=\hsize,clip]{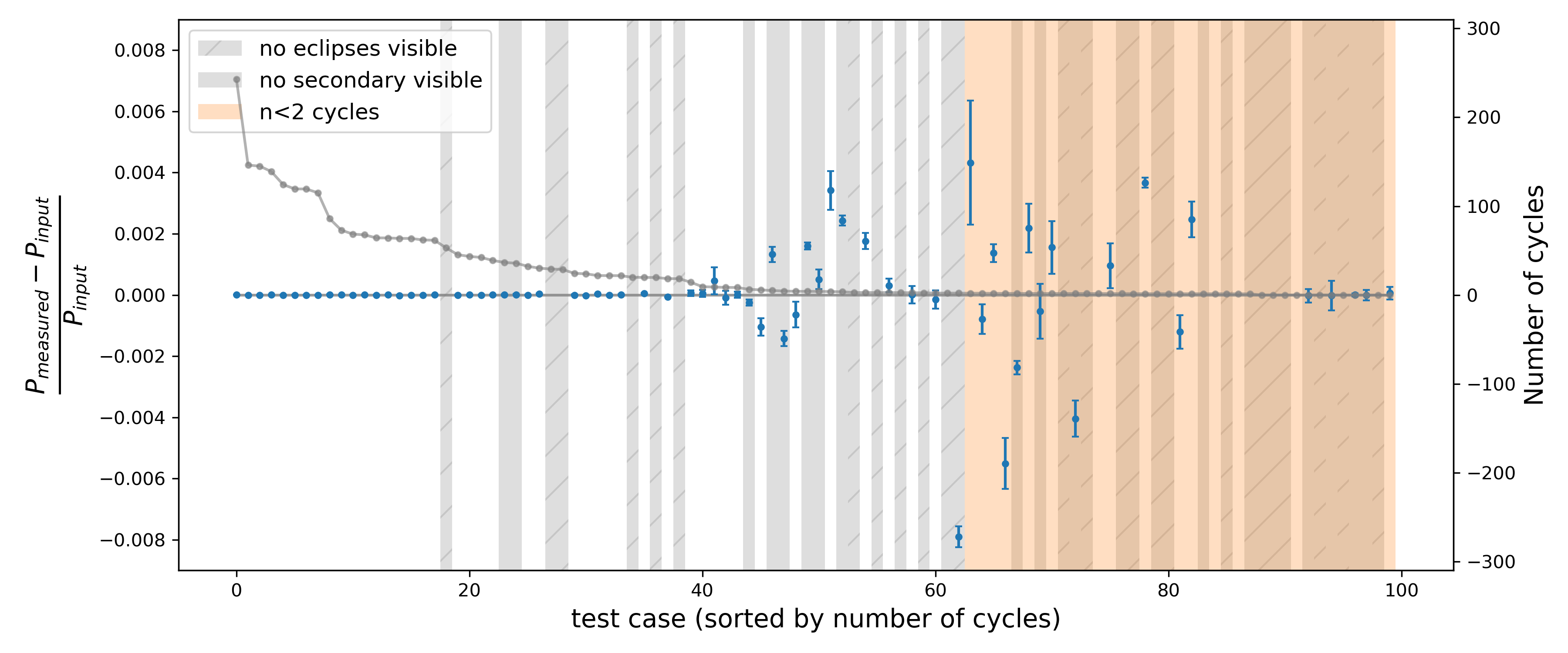}}
    \caption{Performance of the period search algorithm as a function of the number of orbital cycles. The blue points with error bars represent the fractional difference between the measured and the input orbital period with SLR error estimates, sorted by descending number of cycles. The grey line with points shows the number of cycles, calculated as the period divided by the  time base, plotted on the right axis. The period error scales strongly with the number of cycles. Note that the number of cycles for cases $\sim$60 to 100 is close to one, and cases that are shaded grey and hatched do not show eclipses.}
    \label{fig:period_cycles}
\end{figure*}

Fitting eclipse models to a light curve with no prior information usually does not work since the parameter space is degenerate and the optimisation is prone to getting stuck in local minima. Having obtained the parameters from the eclipse timings and depths in the previous step, our pipeline avoids this problem by being able to start close to the optimal solution. 

We constructed a simple physical eclipse model by assuming spherical stars of uniform surface brightness and calculating the fraction of light lost due to one star covering the other, which we refer to as the eclipse model. We have not included light from a third body in this model. Subtracting the fractional light loss from one gives us the leftover normalised flux, $l$:
\begin{equation}
    l = 1 - \frac{A_{covered}}{\pi\ r_1^2 + \pi\ r_2^2\ \frac{sb_2}{sb_1}},
    \label{eq:simple_model}
\end{equation}
where $sb_1$ and $sb_2$ stand for the surface brightness of stars 1 and 2, respectively. This formula gives the normalised flux for the primary eclipse: we have multiplied the right-hand fraction by $\frac{sb_2}{sb_1}$ to obtain the flux for the secondary eclipse. The covered area $A_{covered}$ is the result of two overlapping circles as a function of the distance between their centres $s$ and is computed as follows:
\begin{equation}
\begin{split}
    A &= r_1^2\ arccos\left(\frac{s^2 + r_1^2 - r_2^2}{2 s r_1}\right) \\
    &\quad + r_2^2\ arccos\left(\frac{s^2 + r_2^2 - r_1^2}{2 s r_2}\right) \\
    &\quad - r_1 r_2 \sqrt{1 - \left(\frac{r_1^2 + r_2^2 - s^2}{2 r_1 r_2}\right)^2}.
    \label{eq:area}
\end{split}
\end{equation}
The separation between the two centres is expressed in terms of the orbital parameters and as a function of the phase angle \citep{Kopal1959}:
\begin{equation}
    s = \sqrt{\frac{\left(1 - e^2\right)^2 \left(1 - sin^2(i)\ cos^2(\theta)\right)}{\left(1 - e\ sin(\theta - \omega)\right)^2}}.
    \label{eq:distance}
\end{equation}
In the three equations above the stellar radii $r_i$ and the separation $s$ are in units of the semi-major axis $a$, but this is left implicit for the simplicity of the formulae. The model is calculated at a regular interval of 0.001 rad in $\theta$, which is converted into times using Kepler's law (Equation \ref{eq:kepler3}). Linear interpolation is used to get the values at the times of the observations. 

This eclipse model is first fit to the data, starting from the previously obtained parameters, after which it is subtracted from the light curve and the residuals are pre-whitened analogously to the earlier steps. The obtained model of sinusoids, the linear model and the eclipse model are then optimised simultaneously. We implemented two approaches here: Markov chain Monte Carlo (MCMC) sampling with PyMC3 \citep{pymc3} and Scipy \texttt{minimize} with method L-BFGS-B as before (default option)\footnote{For the fitting method, the sinusoids are again subdivided into groups.}. Since MCMC can deal with high dimensionality, it gives the full correlation structure as an output and thus additionally provides statistical error estimates. However, these estimates are found to be small compared to our other error estimates (and are thus underestimated; see Section \ref{sec:testing}). This procedure is slower than the fitting method. We note that the orbital period and time of primary minimum are fixed and no longer free parameters at this point. As time-dependent parameters are not included in our model of eclipses at this moment, we recommend to perform the analysis of the light curve in parts. This can reveal possible time dependence of the parameters due to a changing orbit.

\subsection{Remaining variability}
The level of intrinsic variability compared to the eclipse depths is computed to get an insight into the strength of the presence of potential pulsations or the harmonic sinusoids that are not captured by the eclipse model. Standard deviations are computed for the residuals of four different regression models: an eclipse model with no extra sinusoids, a model with only the first and second harmonic sinusoids, a model with only non-harmonic sinusoids and the full model including all sinusoids. All four models include the linear trend and the eclipse model. Then the algorithm takes the ratio of primary and secondary eclipse depth to each of these four standard deviations. While we have not further explored the variability in this work, additional steps of analysis on the remaining variability could be easily integrated into the framework.

\begin{figure}
\centering
\includegraphics[width=\hsize]{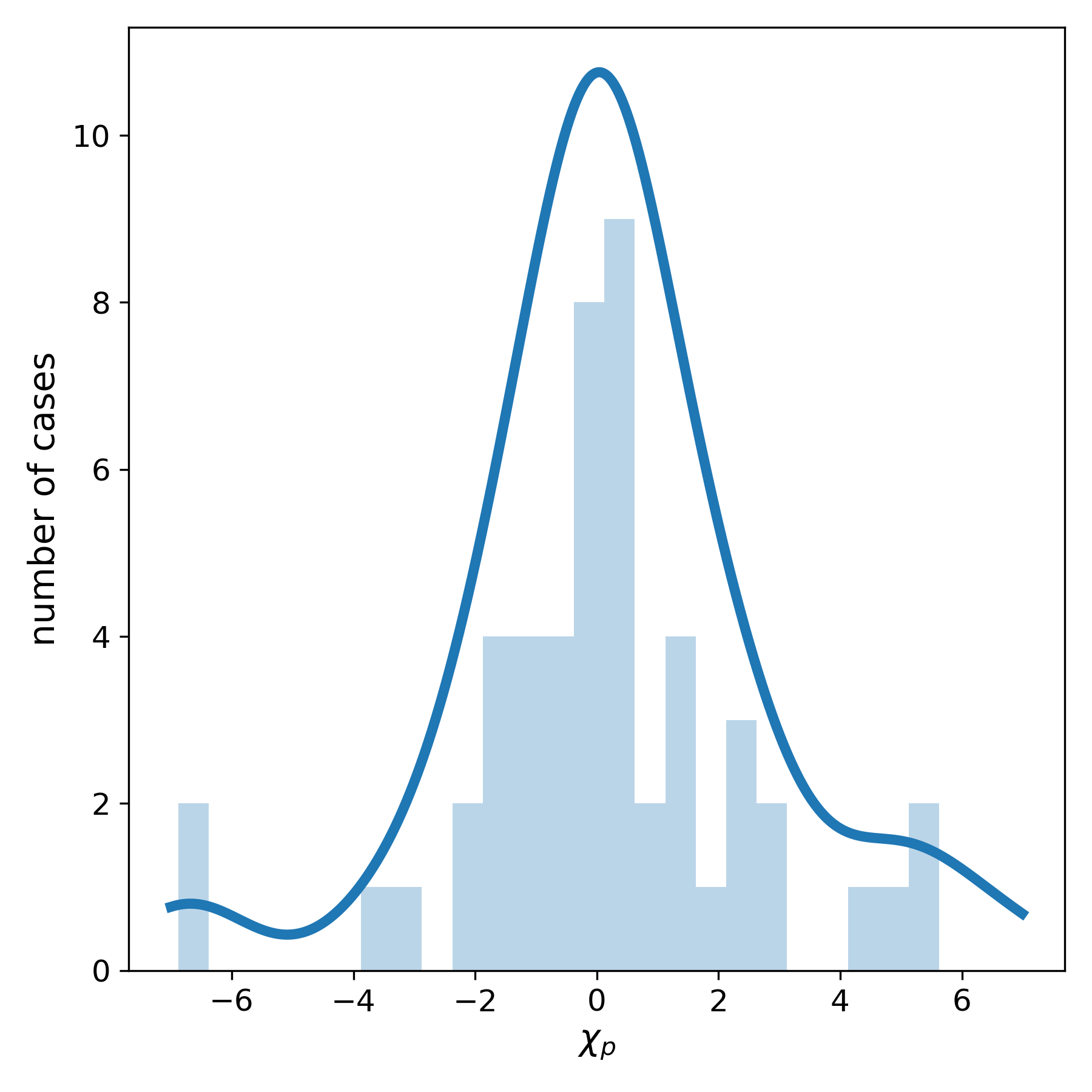}
  \caption{Histogram and KDE of the orbital period $\chi$ values ($(P_{measured} - P_{input})/\sigma_P$), using the SLR uncertainties. The KDE is scaled to the height of the histogram.}
     \label{fig:period_dev}
\end{figure}

\section{Application to synthetic data}
\label{sec:testing}
We produced a set of 100 synthetic test light curves with randomly sampled parameters, including orbital period, eccentricity, argument of periastron, inclination, scaled radii, surface brightness ratio, third light and a number of pulsations represented by sinusoids. Light curves of a single TESS sector (27.4 days) and a full year (13 sectors) at a 30-minute cadence were constructed to mimic some of the TESS time series. The EB light curve models were produced with \texttt{ellc} \citep{ellc} and sine waves were added to represent the intrinsic variability of various frequencies and amplitudes, finally adding Gaussian noise to imitate real data. Table \ref{tab:synth_par} summarises the ranges of input parameters considered. Appendix \ref{apx:plots_examples} includes examples of some of the generated light curves; all light curves and their parameters are made available through the \href{https://github.com/LucIJspeert/star_shadow}{GitHub} page of \texttt{STAR SHADOW}.

\begin{table}
        \centering
        \caption{Ranges of synthetic light curve parameters.}
        \label{tab:synth_par}
        \begin{tabular}{c l l l l l l}
        \hline
        Parameter & Min & Max & Description\\
        \hline
        P & 1 & 20 & period (days)\\
    e & 0.0 & 0.6 & eccentricity \\
    $\omega$ & $-\pi$ & $\pi$ & argument of periastron \\
    i & 75$^\circ$ & 90$^\circ$ & inclination \\
    r$_1$ + r$_2$ & 0.05 & 0.75 & sum of scaled radii \\
    r$_2$ / r$_1$ & 0.15 & 1.75 & ratio of radii \\
    sb$_2$ / sb$_1$ & 0.25 & 1.1 & surface brightness ratio \\
    q & 0.1 & 0.9 & mass ratio \\
    sma & 5 & 95 & semi-major axis (solar radii) \\
    l3 & 0 & 0.2 & third light \\
    n$_\textup{puls}$ & 1 & 100 & number of pulsations \\
    f min & 0.05 & 5 & minimum frequencies \\
    f max & 4 & 12 & maximum frequencies \\
        a min & 0.001 & 0.002 & minimum amplitudes \\
    a max & 0.004 & 0.006 & maximum amplitudes \\
    $\phi$ & -$\pi$ & $\pi$ & phases \\
    \hline
        \end{tabular}
    \tablefoot{Frequency and amplitude ranges differ on a case-by-case basis, so we state the ranges of minima and maxima over all cases.}
\end{table}

Due to the random nature of the generated parameters, some of the synthetic light curves contain no eclipses or have such shallow eclipses that they are impossible to spot by eye. These systems are marked in Figure \ref{fig:period_cycles} and later removed for a cleaner overview of the results. From the viewpoint of the analysis, there are several indicators for such difficult cases, most importantly the completion of all stages of the analysis. If the analysis was completed, the following parameters provide a handle on possible erroneous results: the number of eclipse cycles (period over time base) and depth of the secondary eclipse. We go into some more detail on both of these in the following.

The analysis algorithm was run with no prior knowledge of the orbital period, which is the most important parameter for the rest of the analysis to produce accurate results. We show the performance of the global period search in Figure \ref{fig:period_cycles}. In all relevant cases ($n_{cycles} \geq 2$, visible eclipses), the algorithm was able to pick the correct multiple of the input period to better than 0.34\% precision. More specifically, using the same selection criteria, the mean absolute deviation from the input period is 0.031\%, with a median of 0.0009\%. Further restricting our selection to cases that have more than ten eclipse cycles reduces those numbers to a mean of 0.0011\% and median of 0.00048\%.

\begin{figure}
\centering
\includegraphics[width=\hsize]{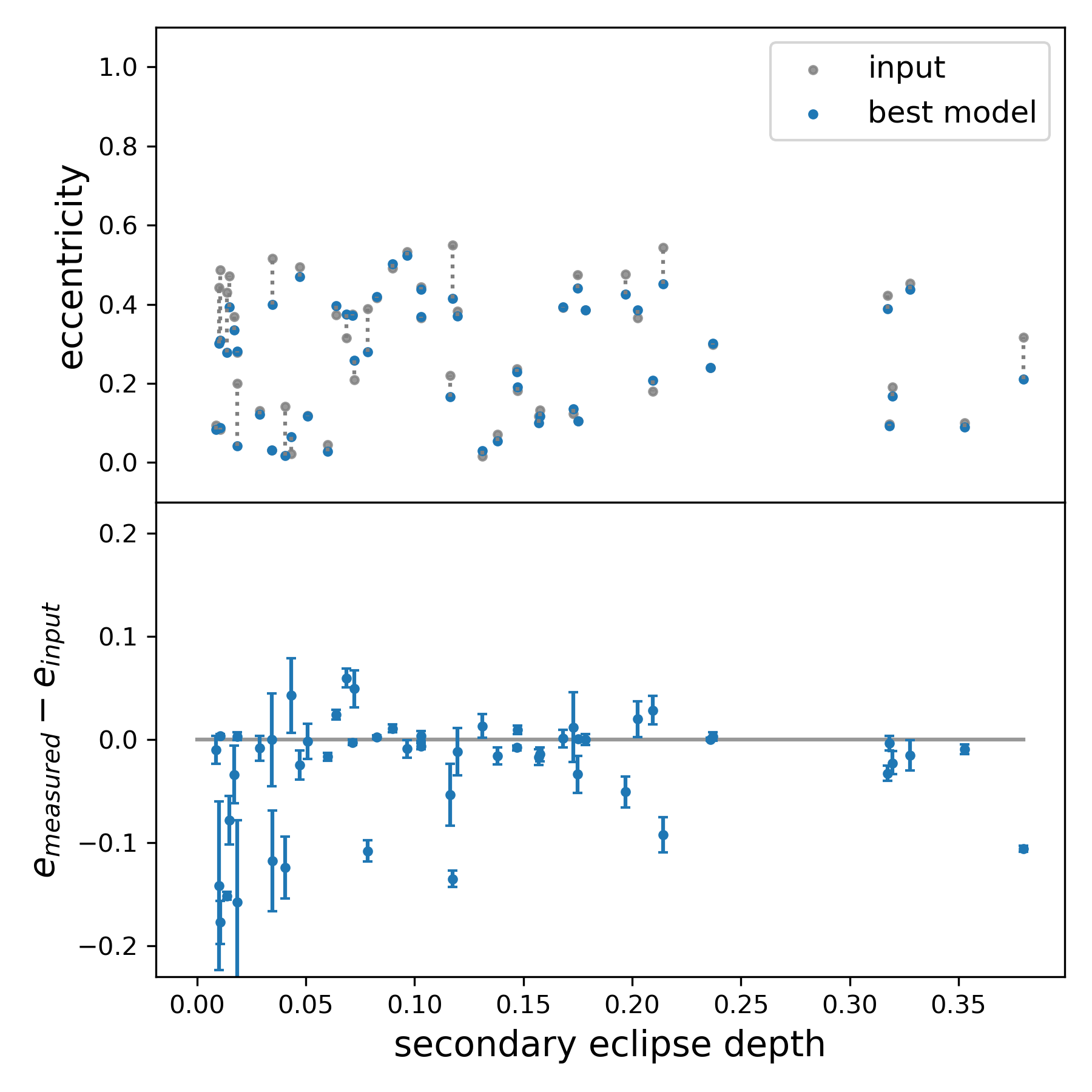}
    \caption{Eccentricity results for the synthetic test cases. Top panel: Input (grey) and measured (coloured) eccentricities.\ Bottom panel: Deviation from the input eccentricity against the secondary eclipse depth. The scatter in the bottom panel is seen to increase to the left, with decreasing eclipse depth, as the eccentricity depends critically on the identification and measurement of the secondary eclipse.}
    \label{fig:ecc_depth}
\end{figure}

\begin{figure}
\centering
\includegraphics[width=\hsize]{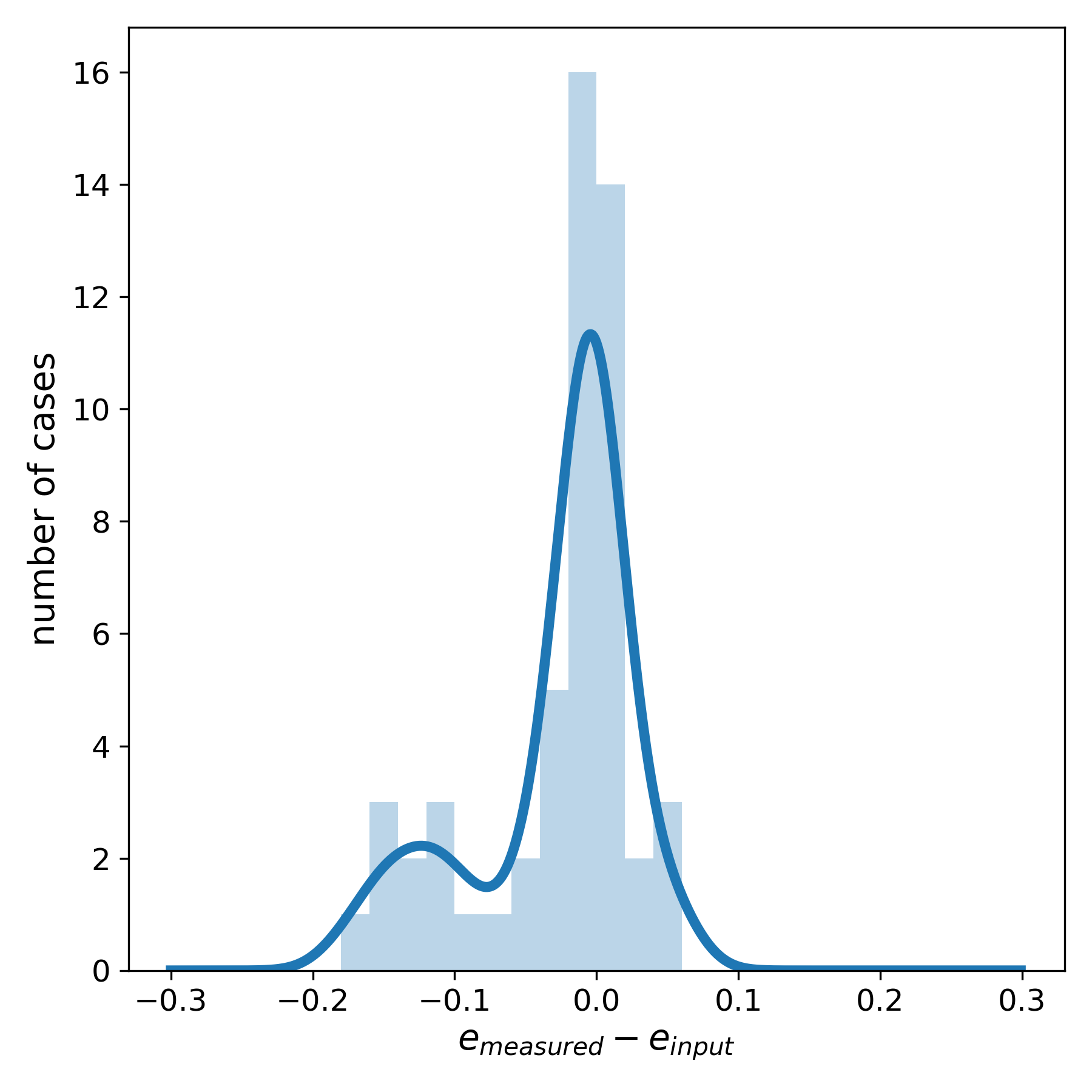}
    \caption{Histogram and KDE of the differences between measured and input eccentricity. Next to the sharp peak at zero, and within the 0.1 deviation, there is a relatively small second maximum where eccentricities are underestimated by between -0.2 and -0.1. The KDE is scaled to the height of the histogram.}
    \label{fig:ecc_dev_abs}
\end{figure}

\begin{figure}
\centering
\includegraphics[width=\hsize]{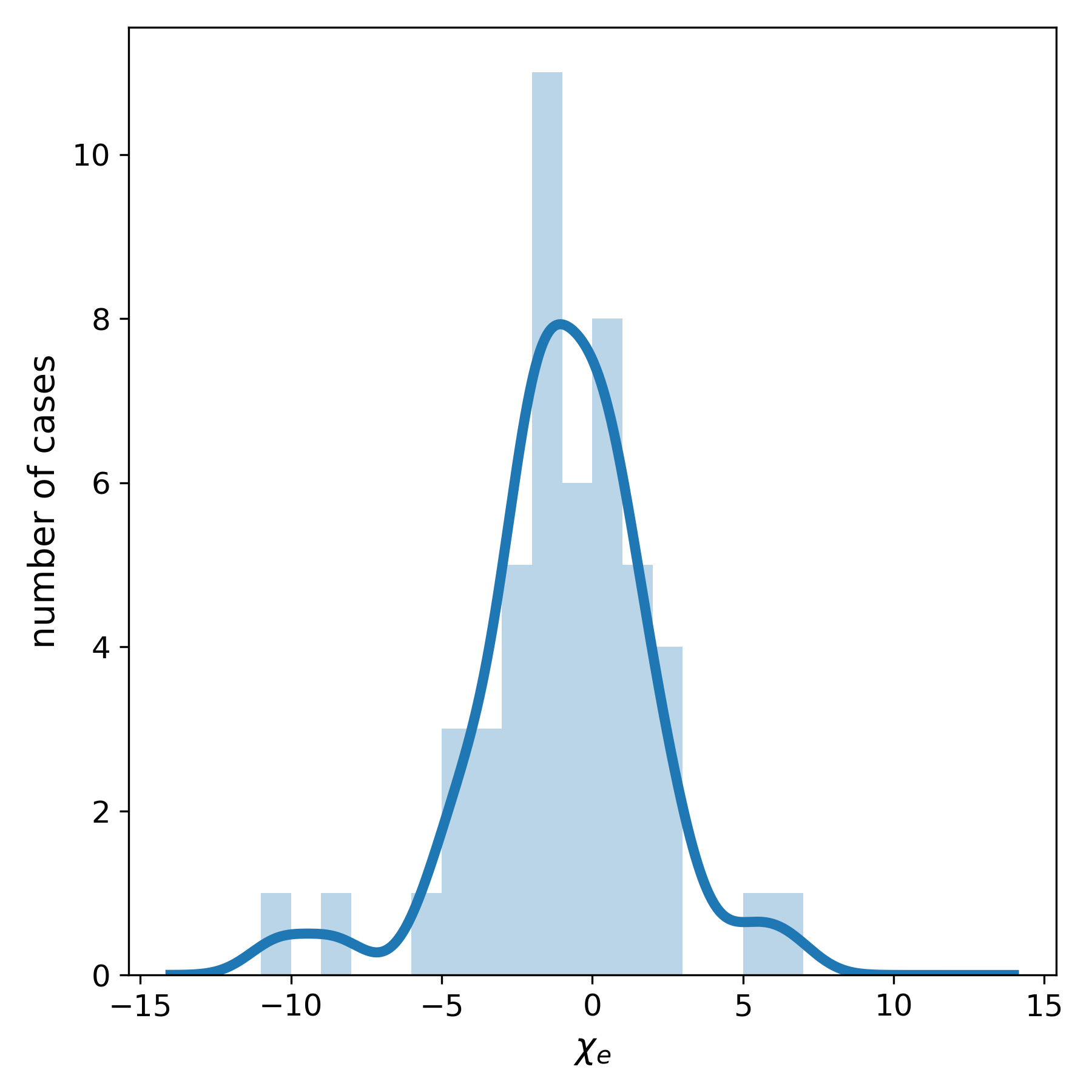}
    \caption{Histogram and KDE of the eccentricity $\chi$ values ($(e_{measured} - e_{input})/\sigma_e$), using the error formula uncertainties. The bi-modality seen in the absolute differences has largely disappeared. The KDE is scaled to the height of the histogram.}
    \label{fig:ecc_dev}
\end{figure}

As Figure \ref{fig:period_cycles} is sorted by the number of cycles, we can see the importance of this parameter as an indicator of the accuracy in the orbital period. This dependence on the number of cycles is taken into account in determining error values, and in the majority of cases, the estimated SLR errors include the input period value within one or two sigma. Figure \ref{fig:period_dev} shows the difference between the measured and input period divided by the SLR errors ($\chi_p$) in a histogram and kernel density estimation (KDE); the KDE is scaled such that its height is comparable to the histogram. The distribution has a broad peak at zero (width > 1), indicating accurate period measurements with estimates of their precision that are overall underestimating the true errors made. 
\\

\subsection*{Eccentricities}

Further analysis depends on the secondary eclipse and thus its correct identification and the measurements of time of minimum, points of in- and egress and depth. Figure \ref{fig:ecc_depth} shows the eccentricity measurements and inputs versus the depth of the secondary eclipse in the top panel and the deviation from the input in the bottom panel. The identification of the secondary becomes more difficult with decreasing secondary eclipse depth, as evidenced by the increasing scatter to the left in Figure \ref{fig:ecc_depth}. What actually causes the measurements to be more difficult for lower secondary depths is both the decrease in signal-to-noise and the ratio between the secondary eclipse depth and the amplitude of any intrinsic variability with high-frequency compared to the orbital frequency. Both of these affect the precision with which we can localise the eclipse. Intrinsic variability can additionally impact the accuracy by either shifting the eclipse position or outright causing the method to miss the second eclipse entirely in the most extreme of cases. We refer to Figure \ref{fig:ecosw_depth} for the tangential components of the eccentricity. Specifically looking into the few cases with large (negative) eccentricity offsets at eclipse depths above 0.05 we see that the tangential components only constitute a fraction of the difference, up to about 0.03 in $e cos(\omega)$.

Deviations in eccentricity generally stay within 0.1. Although error estimates are representative in most cases, we also see cases with underestimated errors that are not explained by their secondary eclipse depth. We see a second maximum in the distribution of absolute differences between measured and input eccentricities (see Figure \ref{fig:ecc_dev_abs}), between values of -0.2 and -0.1, underestimating the input eccentricities. Inspecting these cases, we find no overarching indicator for their deviation, as their eclipses are correctly identified. We do find their deviations to stem nearly completely from the radial component, $e sin(\omega)$. This is caused by an inaccurate determination of eclipse start and end times. They are in some cases easily seen in their analysis output, but not in others (see Figures \ref{fig:sim7} and \ref{fig:sim99} for examples). Figure \ref{fig:ecc_dev} shows the distribution of the true deviations relative to the estimated errors, $\chi_e = (e_{measured} - e_{input})/\sigma_e$, which has a large width (> 1), meaning errors are underestimated. The bi-modality of the former distribution has largely disappeared, which indicates that the errors are sufficiently large for the second maximum to be incorporated into the main distribution of points. The eccentricities obtained from translating the timings (not shown here; see Appendix \ref{apx:plots_synth}) and those obtained from the eclipse model (shown here) have a similar spread. The mean for both distributions lies at zero, suggesting no significant bias towards lower or higher values.

\begin{figure}
\centering
\includegraphics[width=\hsize]{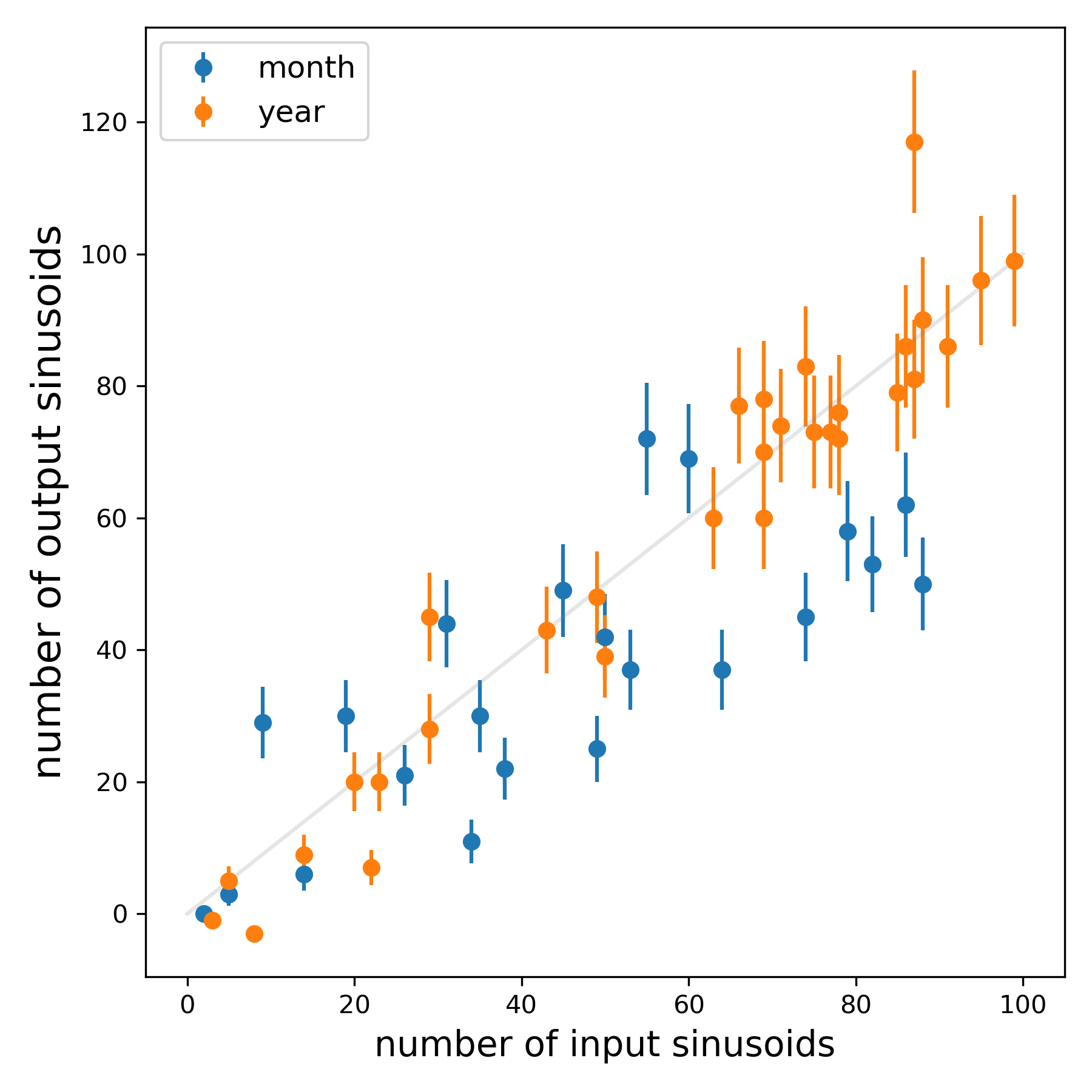}
    \caption{Number of extracted independent sinusoids ($n - n_h$) versus the input number of sinusoids put into the light curve ($n_{input}$). Harmonic sinusoids are filtered out as they capture residuals related to the eclipse model that are unrelated to the number of input sinusoids. Points are divided into month-long time series (blue) and year-long time series (orange), and the error bars are calculated as the square root of the number $n - n_h$.}
    \label{fig:n_sin}
\end{figure}

\begin{figure}
\centering
\includegraphics[width=\hsize]{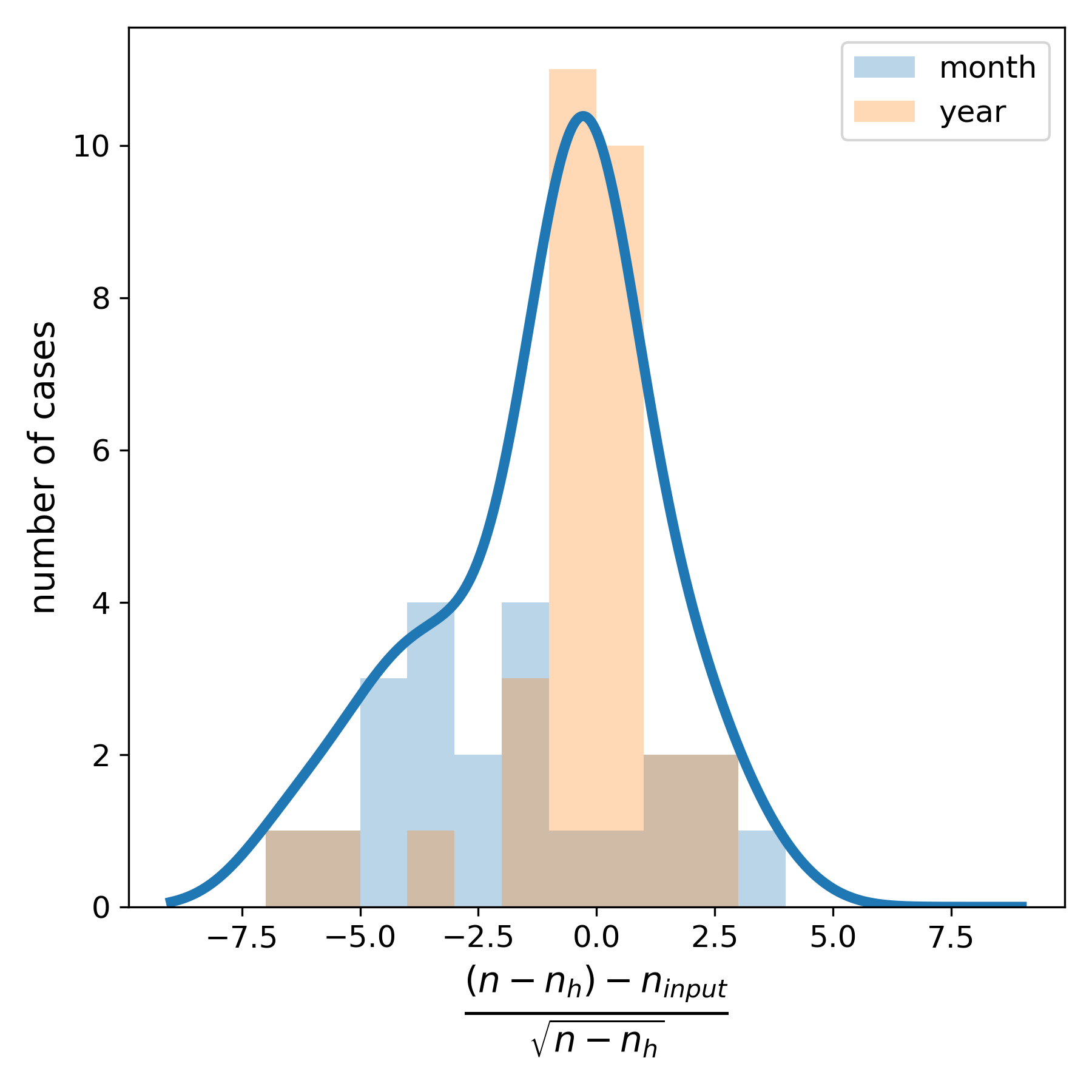}
    \caption{Histogram and KDE of the number of sinusoids $\chi$ values. Errors are taken to be the square root of the amount $(n - n_h)$. The histogram is split into light curves of a month in length and those of a year in length: the shorter time base tends to lead to an underestimation of the number of sinusoids present. The KDE is scaled to the height of the histogram.}
    \label{fig:n_sin_hist}
\end{figure}

\begin{figure}
\centering
\includegraphics[width=\hsize]{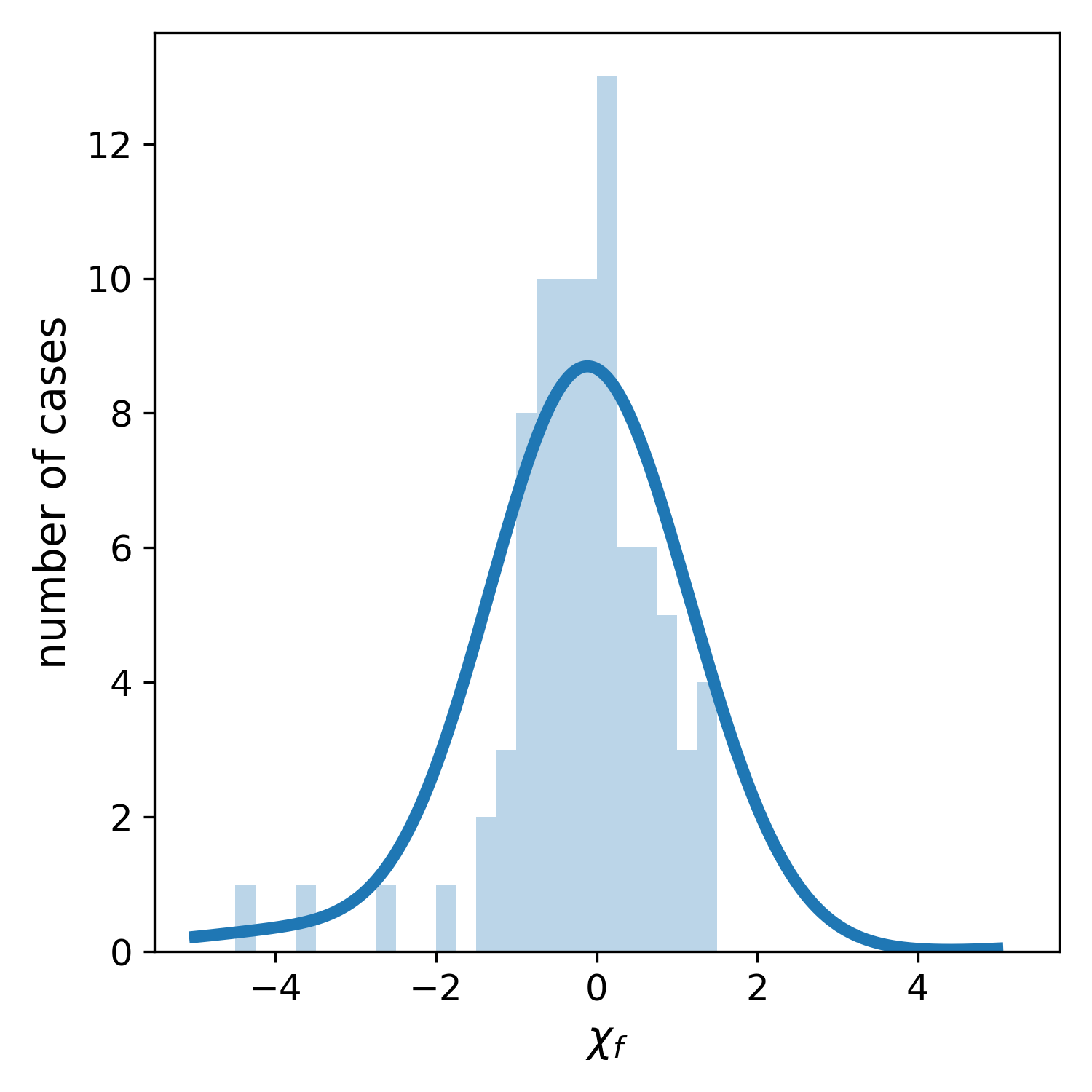}
    \caption{Histogram and KDE of the frequency $\chi$ values ($(f_{measured} - f_{input})/\sigma_f$) for one case of the synthetic light curves. Errors are calculated with the standard formulae. The KDE is scaled to the height of the histogram.}
    \label{fig:case_26}
\end{figure}

The MCMC method also provides error values (not shown here). These are generally smaller than the previous estimates and more often than not missing the input values. This shows that while the MCMC sampling provides an overview of the full correlation structure, it does not capture the full extent of the error; it only captures the statistical error made when fitting a model with no uncertainties to noisy data. Results obtained using the L-BFGS-B fitting method are qualitatively the same as those obtained through the MCMC sampling method.

Distributions for the deviations from the input of the parameters argument of periastron, inclination, sum of the scaled radii, ratio of radii, and ratio of surface brightness are given in Appendix \ref{apx:plots_synth}. We note a few key takeaway points from those figures. There is a good agreement between the overall distributions of the parameter values from translating the timings and those obtained after the eclipse model optimisation step. Two exceptions seem to be the radial component of eccentricity, where the distribution for the eclipse model is shifted slightly towards overestimating this parameter by about two percent, and the auxiliary angle $\phi_0$, which is slightly underestimated by the eclipse model by less than a percent. The tangential component of the eccentricity is accurate and precise with a distribution width of about 0.01, but errors in a minority of cases are underestimated, resulting in long tails for its distribution of true deviations relative to the estimated errors. The sum of the scaled radii shows good agreement with the inputs: its distribution is less than 0.05 in width, be it with a heavy tail on the high end. The orbital inclination shows a slightly wider spread, and no correlation is seen between inclination deviations and third light. The ratio of radii measurement performs worst overall: we attribute this to the fact that for grazing eclipses, and ignoring limb darkening, there is not enough information in the light curve to strongly constrain the individual stellar radii. We do note that the errors the for ratio of radii appear overestimated for the majority of cases, narrowing down the central peak in its distribution of deviations relative to the error estimates. The addition of limb darkening to our eclipse model may improve the predictive power in this area, but for this specific set of synthetic light curves, other variability would dominate over this effect in most cases. The surface brightness ratio performs better than the radii ratio on an absolute scale, but its error estimates are more underestimated and this gives it a poor result on the relative scale.

We also note that we see no correlation between $\phi_0$ (or similarly sum of radii) and deviation in eccentricity across our parameter space, with maximum $\phi_0=0.63$. The limitation of the spherical eclipse model is not a simple function of one parameter, but rather depends on the shape of the eclipses as well as on the shape of the light curve at out-of-eclipse phases.
\\

The number of sinusoids used in generating the synthetic light curves is randomly generated as well as their parameters. It is not evident a priori that we would recover this number: iterative pre-whitening procedures tend to vary in the number of sinusoids found based on stopping criteria and other aspects \citep[e.g.][]{VanBeeck2021}. Our approach is to rely on a statistical measure for the information held in the data (the BIC) during pre-whitening, and only select based on significance criteria at the very end. Our final light curve model constitutes an eclipse model, a linear trend and a sum of sinusoids. We now look only at those sinusoids in the final model that pass the significance criteria, which includes the signal-to-noise threshold. Most of the harmonics found in this filtered list contain the leftovers of the eclipse signal not captured by the eclipse model, like ellipsoidal variability and effects of limb darkening. Therefore, it makes sense that we do not see a correlation with the number of harmonics in this list and the number of input sinusoids. We subtract the number of harmonics ($n_h$) from the total number of filtered sinusoids ($n$) to arrive at the number of independent frequencies ($n - n_h$), which is shown in Figure \ref{fig:n_sin}.  We find that this number strongly correlates with the input number of frequencies and that the results for longer time series tend to lie closer to the diagonal over all.

We take the square root of $n - n_h$ as the error estimate and compute the histogram and KDE for the deviations from the input value divided by the error to arrive at the distribution shown in Figure \ref{fig:n_sin_hist}. When separated into cases with month-long and year-long light curves, the former have consistently slightly underestimated numbers of sinusoids while the latter have unbiased estimates centred at zero. 

One case with a large number of added sinusoids is picked out to look further into the accuracy of the parameters of the extracted sinusoids. We compare the input frequency values with the measured ones by pairing the closest matches together. Figure \ref{fig:case_26} shows the accuracy and precision of measured frequencies, showing that the error formulae agree with the deviations from the input. The same is true for the amplitudes and, to a lesser degree, for the phase measurements; the distribution of the latter is wider  by about 50. Figures \ref{fig:sim26} and \ref{fig:sim26b} depict the light curve model and the periodogram for this synthetic test case.

\section{\textit{Kepler} EB Catalog}
\label{sec:kepler}

For the purpose of further testing the period-finding capability of \texttt{STAR SHADOW} against a vetted sample of real stars, we apply our methods to the 30-minute cadence light curves for the \textit{Kepler} EB Catalog targets collected by \citet{kepler_eb2011}, \citet{kepler_ebii2011}, and \citet{kepler_ebvii2016}.\footnote{found at \href{http://keplerebs.villanova.edu/}{keplerebs.villanova.edu}} In addition we present the results for the eccentricities found for this dataset. Of the 2920 detrended light curves in the catalogue, all completed analysis at least up to and including the period measurement and for 2001 of those we obtained an eccentricity measurement with our analysis (for all morphologies in the catalogue). 

\begin{figure}
\centering
\includegraphics[width=\hsize]{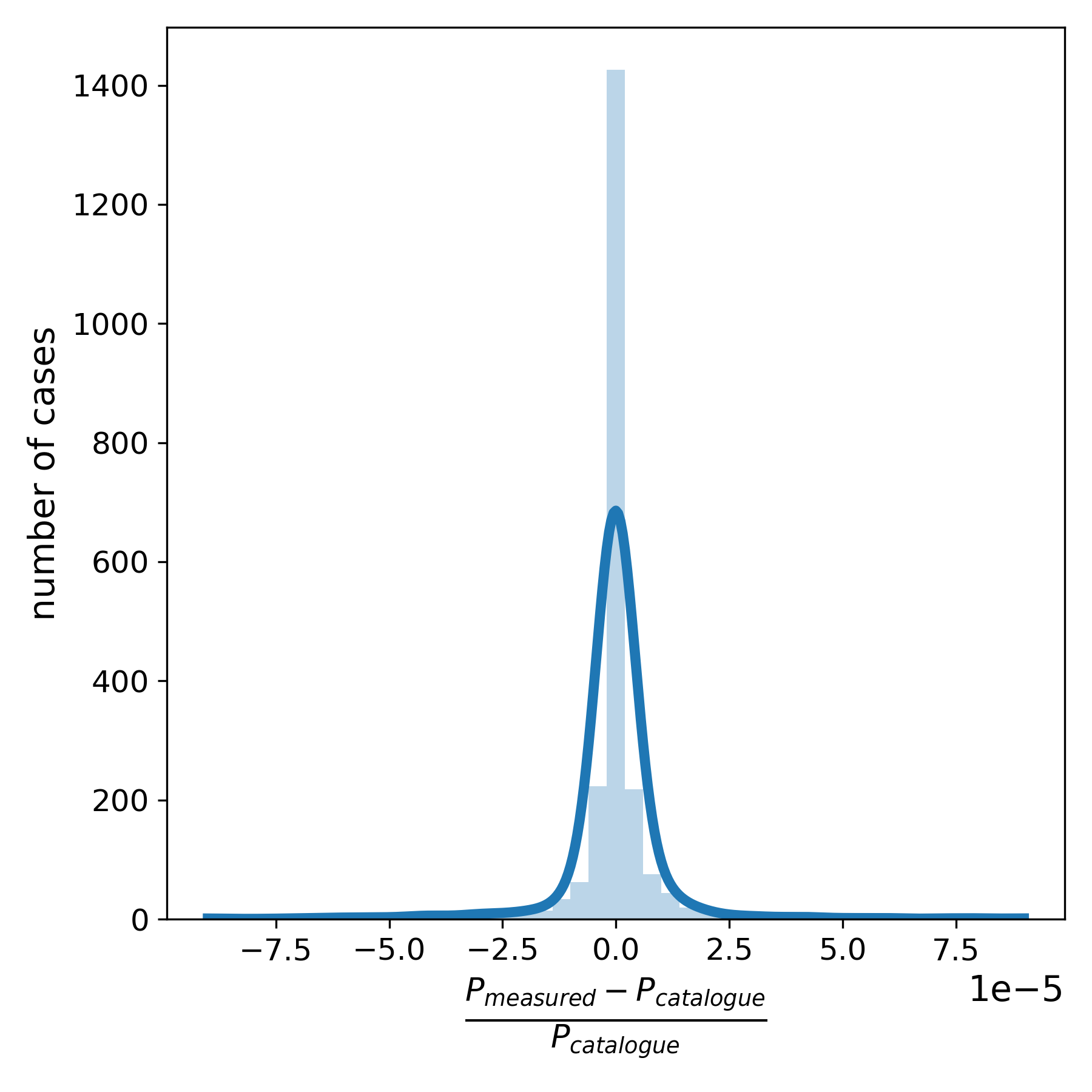}
    \caption{Histogram and KDE of the fractional difference between the \textit{Kepler} catalogue orbital periods and those measured in this work. Here we select the periods within 1\% of the catalogue values. The KDE is scaled to the height of the histogram.}
    \label{fig:kepler_period_dev_fractional}
\end{figure}

\begin{figure}
\centering
\includegraphics[width=\hsize]{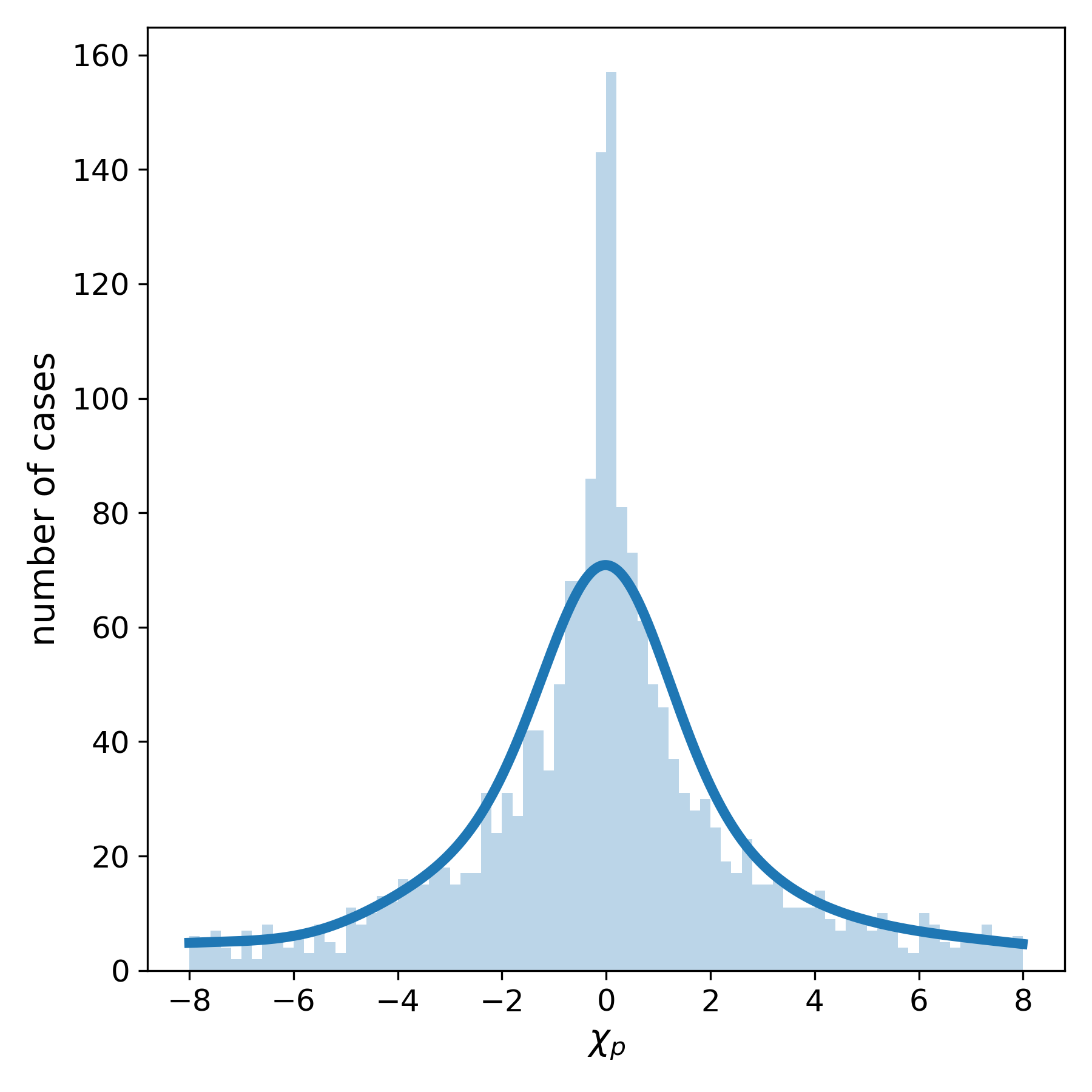}
    \caption{Histogram and KDE of the orbital period $\chi$ values ($(P_{measured} - P_{catalogue})/\sigma_P$) comparing the \textit{Kepler} catalogue values to those measured in this work. We select the periods within 1\% of the catalogue values. The relatively pronounced wings of the distribution indicate underestimated period errors for a portion of systems. The KDE is scaled to the height of the histogram.}
    \label{fig:kepler_period_dev}
\end{figure}

Generally, we see a high level of agreement between the periods found by our method and the periods reported in the catalogue. The proportion of targets where the period is within one percent of the reported period is 80.5\%. The amount of targets where the period is within a percent of half the reported period is 7.9\%, while other close multiplicative factors together constitute 1.3\%, adding to a total agreement level of 89.7\%. Cases where half the period is designated as the orbital period consist of systems with a primary and secondary eclipse that look very similar, and thus create only weak evidence in Fourier space and in phase dispersion for the correct multiple of the period. Figure \ref{fig:kepler_period_dev_fractional} shows the fractional differences between measured and catalogue periods, around the zero point: we see a very sharp central peak and wide but low tails extending to around 7e-5 that are imperceptible on the scale of the plot. In the subset with the correct period multiple, we see a mean absolute fractional difference of 0.0027\% and a median of 0.00010\%. We divide the differences from the catalogue period by our estimated errors to get the distribution shown in Figure \ref{fig:kepler_period_dev}. The strong central peak seen in the fractional differences remains but the wings of the distribution become more pronounced, indicating an underestimation of a subset of period errors. 

The remaining 10.3\% of cases that did not have a multiple of the catalogued period consist of three groups of light curves: (1) very low signal-to-noise cases (62\%), (2) those where the binary signal is close to a pure sinusoid, which does not result in a strong series of harmonics to pick up on (36\%), and (3) those from multiple systems that have more than one period reported in the catalogue; for these systems, the strongest periodic signal is found (correctly), but we then miss the other periods as we do not continue to search for signals caused by higher-order multiples (4\%). 

The residuals of the 2001 fully analysed light curves show a range in noise level (standard deviation) from 24 ppm to 8.8 ppt, with a median of 526 ppm and mean of 731 ppm. The signal-to-noise, measured as the smallest eclipse depth divided by the noise level, has a minimum of 0.4, a median of 68, and a mean of 151.

\subsection*{Eccentricities}

The \textit{Kepler} EB Catalog does not provide eccentricity measurements to directly compare ours against. However, we can present our results in an eccentricity-versus-period plot that gives an idea of whether our measured eccentricity values are globally as we expect them to be. Generally speaking the orbital circularisation timescale is strongly dependent on the orbital period \citep{Tassoul1990}, which means we would expect all short-period binaries to have zero or otherwise small eccentricity. Conversely we expect a wide distribution of eccentricities at the high period end as these systems are still at their birth configurations. 

\begin{figure}
\centering
\includegraphics[width=\hsize]{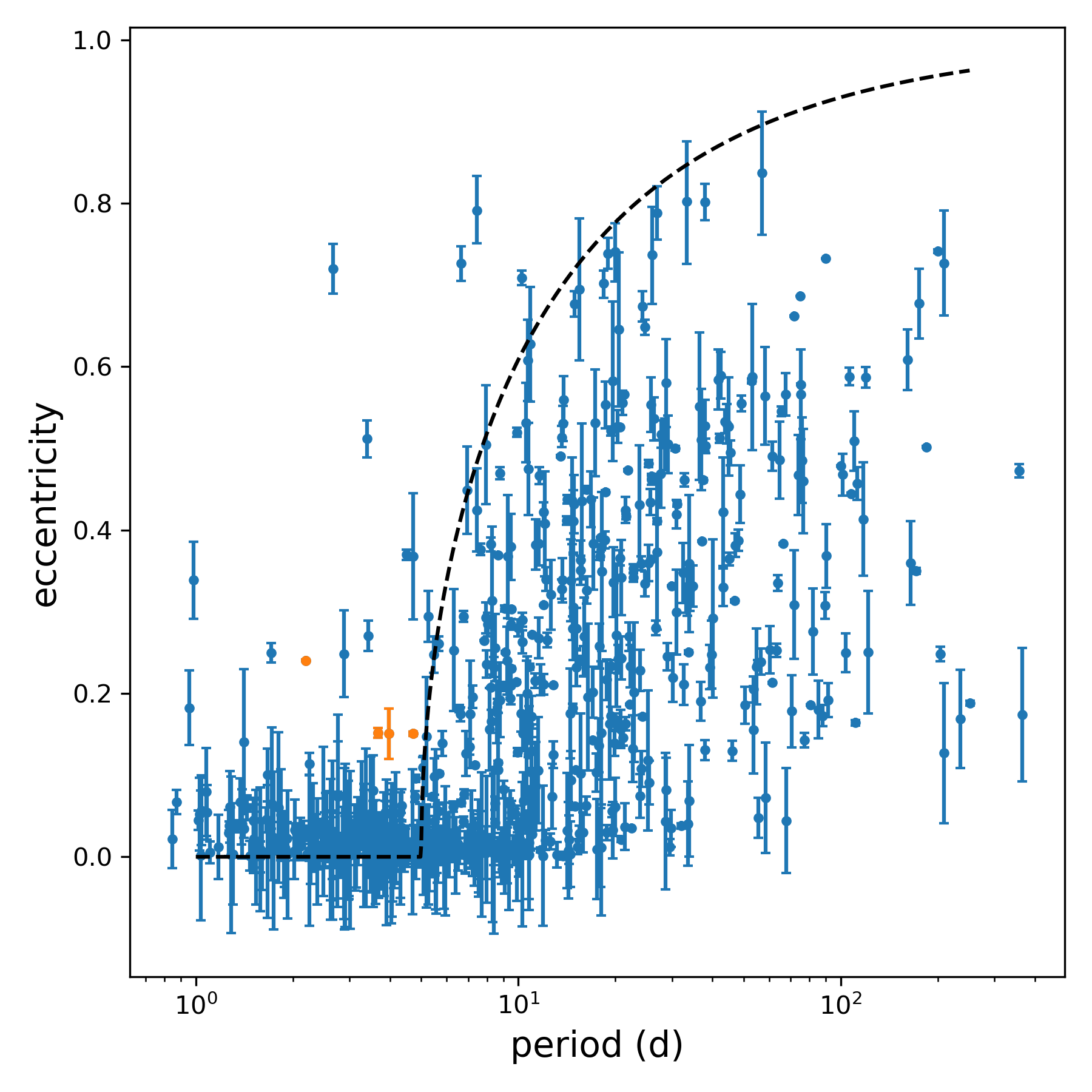}
    \caption{Eccentricity measurements for a subset of targets from the \textit{Kepler} EB Catalog. Selections were made in eccentricity error (<0.1) and morphology parameter (<0.5), the latter being the more stringent criterion. We use the maximum eccentricity as a function of orbital period in \citet[][Equation 3]{Halbwachs2005} with the cut-off period set to 5 days to plot the dashed line. Orange points indicate four systems of interest, with well-fitting models.}
    \label{fig:kepler_ecc}
\end{figure}

\begin{figure}
\centering
\includegraphics[width=\hsize]{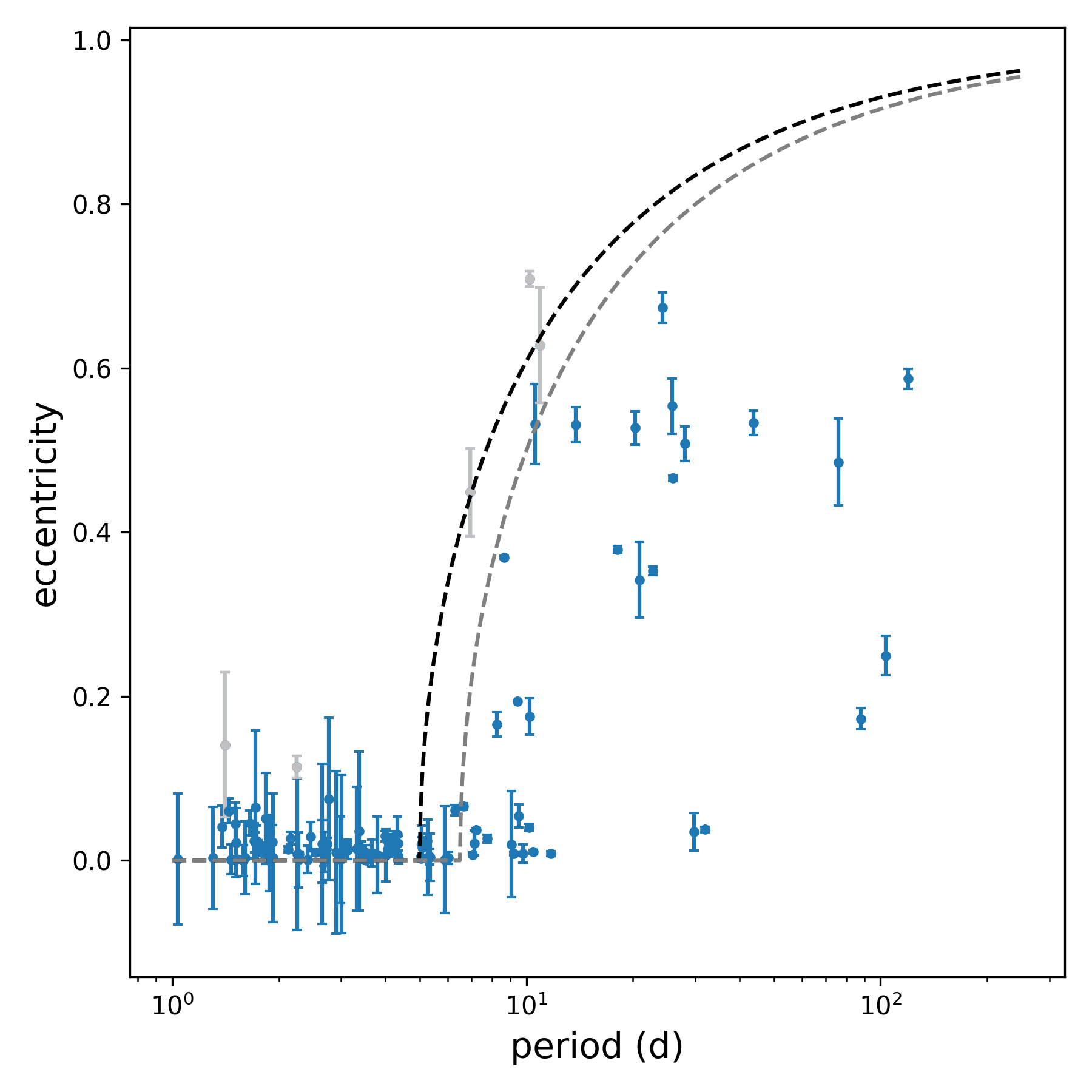}
    \caption{Eccentricity measurements for a subset of targets from the \textit{Kepler} EB Catalog. Selections were made in eccentricity error (<0.1) and morphology parameter (<0.5), and measure of signal-to-noise of the intrinsic variability at non-harmonic frequencies (>6). We use the maximum eccentricity as a function of orbital period in \citet[][Equation 3]{Halbwachs2005} with the cut-off period set to 5 and 6.5 days to plot the dashed black and grey lines. Grey points indicate systems that do not belong in this selection or have otherwise faulty measurements.}
    \label{fig:kepler_ecc_var}
\end{figure}

Figure \ref{fig:kepler_ecc} shows the eccentricity results for a selection of \textit{Kepler} targets. The selection criteria are an eccentricity measurement error below 0.1 and a catalogue morphology parameter below 0.5, leaving 715 targets of the 2001 systems with an eccentricity measurement. The catalogue morphology parameter is an indicator of the level of detachment between the stars in the system, running from zero to one, with lower values being more detached. We show results that include all morphologies in Figure \ref{fig:kepler_ecc_morph}. As an integral part of our methodology, we analyse the sinusoidal content in the light curve that is present in addition to the eclipses. To separate targets that are likely pulsating at independent frequencies, we compute the standard deviation of the residuals after subtracting the eclipse model and all harmonic sinusoids from the light curve and dividing this by the standard deviation of the residuals of the full light curve model. This is interpreted as a measure of signal-to-noise of the intrinsic variability at non-harmonic frequencies, and we select targets above a value of 6. These are shown in Figure \ref{fig:kepler_ecc_var}. Grey markers indicate systems that were found to either be wrongly assigned to this selection of variables due to an ill-fitting eclipse model or have faulty eccentricity measurement due to the absence of one or both eclipses by manual inspection.

\begin{figure}
\centering
\includegraphics[width=\hsize]{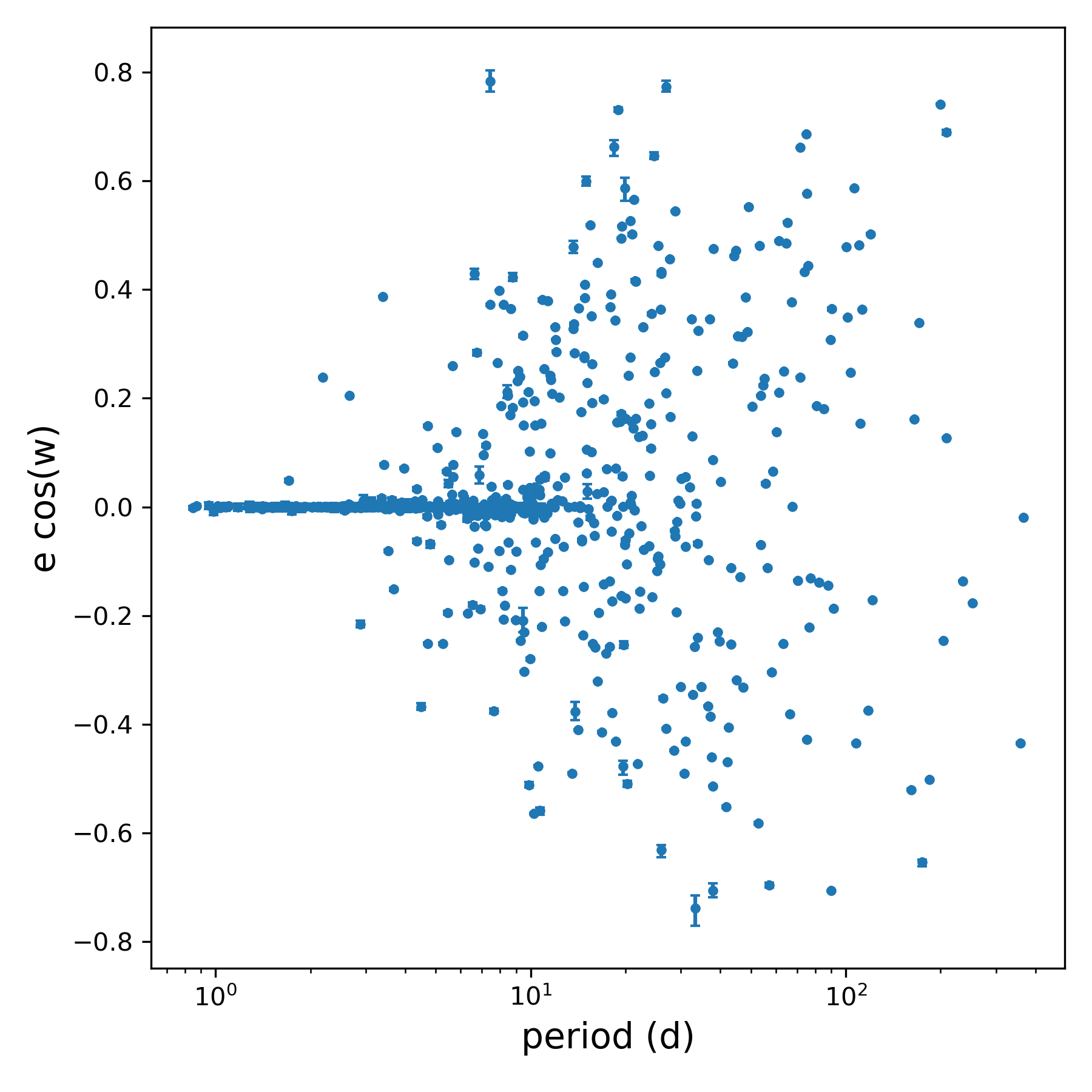}
    \caption{Measurements of the tangential component of eccentricity for a subset of targets from the \textit{Kepler} EB Catalog. The subset results from cut-offs in eccentricity error (<0.1) and morphology parameter (<0.5), the latter being the more stringent criterion.}
    \label{fig:kepler_ecosw}
\end{figure}

The overall picture is consistent with expectation: we see a buildup of systems at zero eccentricity below a period of some ten days, and a large range above that period. There are a few clear outliers among the lowest periods that have significant eccentricity measurements, the majority of which are due to an inaccurate identification or measurement of the secondary eclipse. As mentioned in \citet{vaneylen2016}, the measurement of the radial component of eccentricity, $e sin(\omega)$, is less straightforward than that of the tangential component, $e cos(\omega)$. This is also apparent in our methodology, which becomes visible when we take a look at the measurements of the tangential component in Figure \ref{fig:kepler_ecosw}. Apart from the much smaller error estimates, the scatter around zero tangential eccentricity of the stars below ten days orbital period is tight. 

The theory of close binary tides predicts the circularisation timescales of binary orbits (see e.g. \citealt{Zahn1975, Zahn1977, Zahn1989, ZahnBouchet1989, Goldreich1977, Savonije1983, Savonije1984}). Circularisation is the result of dissipation mechanisms (tidal friction) in the stellar interior, which in turn decrease the orbital energy. Mechanisms include, but are not limited to, damping of the tidal deformation through turbulent flow in a convective region, and the damping of tidally forced oscillations through radiative diffusion. The state of circularisation for a single binary depends on its age, which is generally not straightforward to constrain. Instead, the timescales on which binaries of certain types are expected to circularise can be used to predict the distribution of eccentricities in a population of binary systems. As explained by \citet{vaneylen2016}, age is involved as a factor for high-temperature stars. Due to their much shorter lifespans, comparable to or shorter than the circularisation timescale, they have not had time to circularise even at short periods where tidal friction is strongest. Additionally, hot stars with radiative envelopes have a different mechanism for circularisation than cool stars with convective envelopes. Therefore, we expect hot stars to exhibit larger eccentricities at shorter orbital periods than cool stars. 

The \textit{Kepler} sample studied by \citet{vaneylen2016}, and revisited here, has a majority of cool stars. If we use the same differentiating temperature of 6250K only one-sixth of the selected eccentricity measurements remain in the high-temperature group. In that publication, the authors used the tangential component of eccentricity, $e cos(\omega)$ as a proxy for eccentricity, whereas we determine both components in our analysis. The two temperature groups look broadly consistent with each other in our measurements, due in part to outliers that may or may not be at correct measurement values. We looked in more detail at some of the clear outliers at low periods with eccentricity measurements at or above 0.1. Most of these measurements arise from misidentifications of noise as a secondary eclipse, or broad ellipsoidal variability as an eclipse (the \textit{Kepler} EB Catalog includes singly eclipsing and even non-eclipsing systems with ellipsoidal variability). Nevertheless, there is     a small number of correctly identified systems in this regime, around an eccentricity of 0.15 with well-fitting light curve solutions\footnote{And one with a poorer light curve fit that has a high enough tangential component of eccentricity to be sure of its classification in this high eccentricity group.}: KIC 4544587, 7943535, 8196180, and 11867071 (orange points in Figure \ref{fig:kepler_ecc}). Using the temperatures provided by the \textit{Kepler} EB Catalog, all four of these systems fall in the high-temperature category, two of them having temperatures above 7000K. This seems to support the conclusion by \citet{vaneylen2016}, who reported an observed difference between the tangential eccentricity distributions of hot-hot binaries as compared to cool-hot and cool-cool binaries, where hot-hot systems are more eccentric at shorter periods. This is in broad lines consistent with theory.

We use a simple formula from \citet[][Equation 3]{Halbwachs2005}, based on the strength of tides varying with distance as $1/r^6$ \citep{Lecar1976}, to visualise the envelope of maximum eccentricity in Figure \ref{fig:kepler_ecc} more clearly. The cut-off period, or circularisation period, was set to the value reported by \citet{vaneylen2016}, calculated for cool stars. Apart from the outliers, both those misidentified and the hot stars discussed above, this envelope qualitatively fits the sample. We see the same extension of the low eccentricity cluster towards periods longer than 5 days as in the \citet{vaneylen2016} sample, which may arise from dependences on age and/or temperature of the circularisation timescale. Our selection of systems with intrinsic variability at non-harmonic frequencies, Figure \ref{fig:kepler_ecc_var}, contains 108 systems. Even though only a few systems constrain the maximum eccentricity envelope in this selection, we may visually fit a line at a slightly longer cut-off period of 6.5 days. This could be an effect of the low numbers, but it may be a physical effect if pulsations aid the redistribution of orbital angular momentum, increasing the effectiveness of tides. We will investigate this effect further in a future study based on a TESS sample with a factor of about 4 more systems.

The summarising catalogue of our \textit{Kepler} analysis results is available at the CDS\footnote{Via anonymous ftp to \url{cdsarc.u-strasbg.fr} (130.79.128.5) or via \url{http://cdsweb.u-strasbg.fr/}}. The results are also made available through the \textit{Kepler} EB Catalog main web page\footnote{\href{http://keplerebs.villanova.edu/}{keplerebs.villanova.edu}}.

\section{Conclusions}

We have presented a novel photometric time series analysis methodology and its implementation, \texttt{STAR SHADOW}. Its aim is, in the first place, to analyse large numbers of EBs homogeneously, but it is also broadly applicable for computationally efficient iterative pre-whitening. The method has been tested against a set of synthetic light curves with known input parameters and high levels of intrinsic variability from sources other than eclipses. We find good levels of agreement between the input and extracted sinusoidal frequencies and amplitudes, as well as between the input and computed binary parameters.  The accuracy and precision of the binary parameters vary, with the tangential component of eccentricity being the most accurate and precise and the radial component approximately an order of magnitude less well constrained. This is unsurprising as most of the information about the radial component is in radial velocity curves. The number of extracted frequencies after the subtraction of orbital harmonics and after filtering with significance criteria strongly correlates with the number of input sinusoids for the one-year-long simulated light curves. The number is underestimated for shorter light curves of one month durations. 

We further tested the period-finding capabilities of our implementation against the \textit{Kepler} EB Catalog and examined the eccentricities obtained for this sample. The method finds the correct orbital period multiple, or half the orbital period, in 90\% of cases, showing its capability to be used with minimal predetermined knowledge of the data. Cases where it finds half the period are those where the primary and secondary eclipses are nearly identical. Period error estimates are found to be appropriate for the vast majority of targets, with heavy tails on the distribution of the deviation over the error revealing an underestimation for a fraction of targets. 

\begin{figure}
\centering
\includegraphics[width=\hsize]{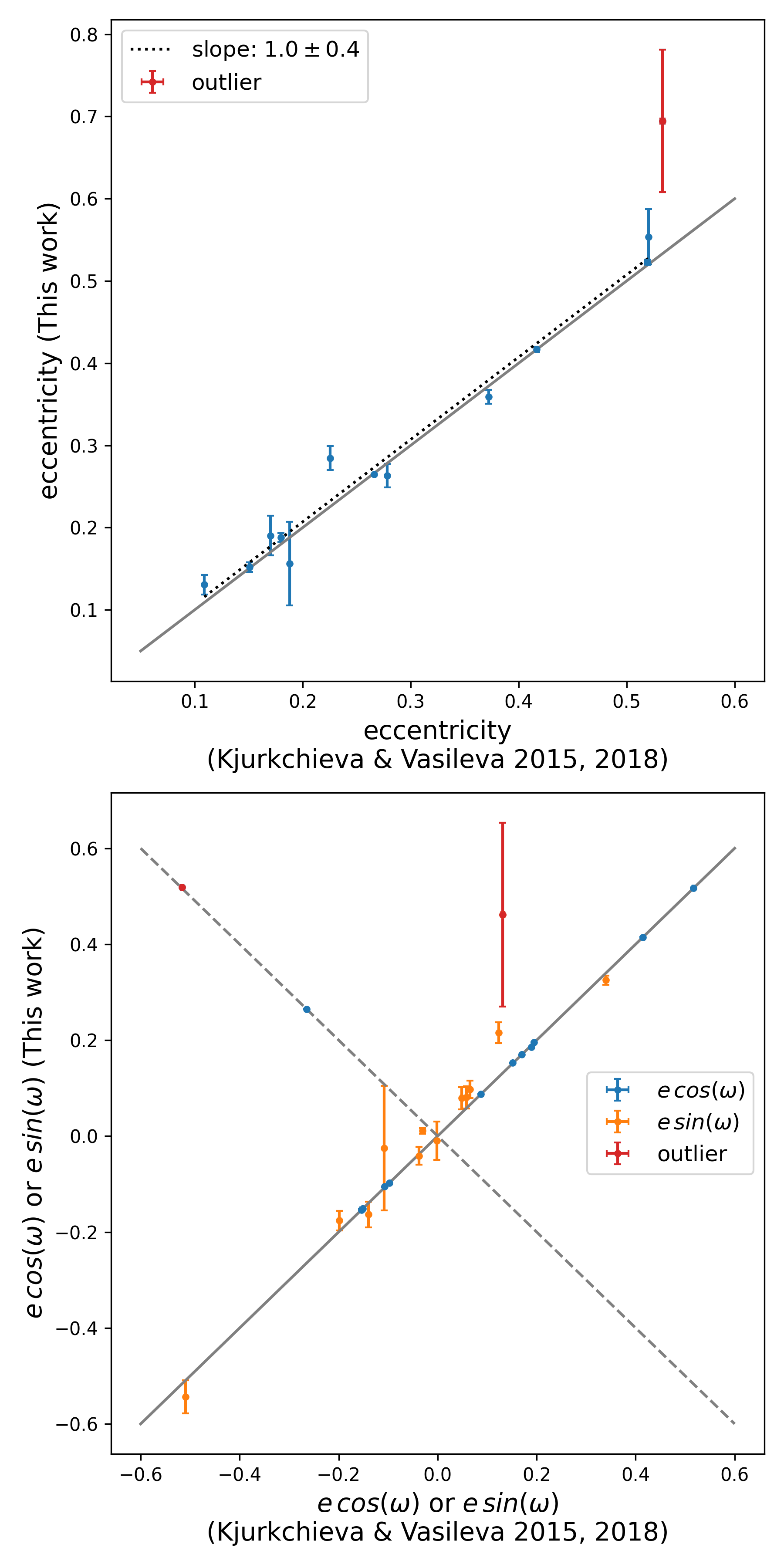}
    \caption{Comparison of eccentricity measurements for 13 \textit{Kepler} targets. Top panel: Eccentricity measurements from this work versus those obtained in \citet{Kjurkchieva2015, Kjurkchieva2018}, along the diagonal (grey line). The linear regression model is plotted as a black dotted line. Bottom panel: Comparison between the tangential and radial components of eccentricity. The dashed line indicates where points occur if the definition of primary and secondary eclipse is swapped.}
    \label{fig:benchmark}
\end{figure}

A few of our \textit{Kepler} eccentricity determinations were compared to results obtained during an in-depth analysis by \citet{Kjurkchieva2015, Kjurkchieva2018} as a benchmark. Out of 13 targets, all but two eccentricity values agree within two sigma of our error estimates. The mean absolute deviation in eccentricity is 0.015, giving us confidence in the precision of the method for this parameter. We plot the eccentricities and the two components of eccentricity of our result against those found in the literature in separate panels of Figure \ref{fig:benchmark}. Inspecting the output of our methodology, we find one clear outlier (KIC 8111622) for which the secondary eclipse is misidentified; this measurement is indicated in red. The reason for this misidentification is a very narrow eclipse that is also shallow; because of this, determining its positioning based on the time derivatives of the time series (limited to low frequencies) is difficult. The linear regression coefficient of the eccentricities, excluding this outlier, is $1.00 \pm 0.41$. In the bottom panel of Figure \ref{fig:benchmark}, the diagonal with a negative slope indicates the position where points are expected if the definition of the primary and secondary eclipse is swapped. The one system where this is the case (KIC 6949550) happens to have equally deep eclipses, so assigning the tag primary or secondary to its eclipses is an ambiguous choice.

The eccentricities obtained for the \textit{Kepler} targets generally follow the expected pattern as a function of the orbital period. Below around ten days, eccentricities are broadly consistent with zero, and above ten days there is a wide distribution of higher eccentricities. We do see a few outliers that are caused by a misidentification of the secondary eclipse, due to a combination of its low eclipse depth and the presence of other variability in the out-of-eclipse part of the light curve. Our methodology was developed with the TESS mission in mind, since our aim is to apply it to the sample of EBs from \citet{ijspeert2021}.

We qualitatively compared our \textit{Kepler} results to the work by \citet{vaneylen2016}, who analysed their own selection of \textit{Kepler} EBs to obtain the tangential part of the eccentricities. Our results support their conclusion that, at low orbital periods, hot stars (>6250K) have higher eccentricities than cool stars, which is broadly consistent with theory. Our eccentricity distribution (for cool stars) looks qualitatively consistent with the cut-off period of 5 days computed by \citet{vaneylen2016} for cool stars with convective envelopes.

The average run time of the analysis was ten minutes for one-year time base simulated light curves, with an average of two minutes of that spent on the frequency analysis. For the \textit{Kepler} light curves, the recorded average time to completion is just over two hours, including one hour for frequency analysis. The scaling with the time series length (four years for \textit{Kepler}) is not linear since the number of extracted sinusoids also increases with the time base. This rapid performance is achieved through various optimisations in the algorithm of iterative pre-whitening and multi-sinusoid non-linear optimisation; the main optimisations are further described in Appendix \ref{apx:details}. We encourage others to implement a gradient-based approach for sinusoid optimisation for the free performance boost it gives, in order to save on both time and energy.\\

\begin{acknowledgements}
The research leading to these results has received funding from the KU\,Leuven Research Council (grant C16/18/005: PARADISE), from the Research Foundation Flanders (FWO) under grant agreements 1124321N (Aspirant Fellowship to LIJ), G089422N (AT), as well as from the BELgian federal Science Policy Office (BELSPO) through PRODEX grant PLATO. CJ gratefully acknowledges support from the Netherlands Research School of Astronomy (NOVA) and from the Research Foundation Flanders (FWO) under grant agreement G0A2917N (BlackGEM). AP acknowledges support of the NSF grant \#2306996 and NASA 23-ADAP23-0068.
CA acknowledges funding by the European Research Council under grant ERC SyG 101071505. Funded by the European Union. Views and opinions expressed are however those of the author(s) only and do not necessarily reflect those of the European Union or the European Research Council. Neither the European Union nor the granting authority can be held responsible for them. 
This paper includes data collected by the Kepler mission, which are publicly available from the Mikulski Archive for Space Telescopes (MAST) at the Space Telescope Science Institute (STScI). Funding for the Kepler mission is provided by the NASA Science Mission Directorate. STScI is operated by the Association of Universities for Research in Astronomy, Inc., under NASA contract NAS 5–26555.
Some of the computing resources and services used in this work were provided by the VSC (Flemish Supercomputer Center), funded by the Research Foundation - Flanders (FWO) and the Flemish Government.
\\
\\

\textit{Software.}
This work and the presented code (\texttt{STAR SHADOW}) make use of Python (Python Software Foundation. Python Language Reference, version 3.7. Available at \href{http://www.python.org}{www.python.org}) and the Python packages Numpy \citep{numpy}, Numba \citep{numba}, Scipy \citep{scipy}, Astropy \citep{astropy:2013, astropy:2018, astropy:2022}, PyMC3 \citep{pymc3,pymc3_code}, Theano \citep{theano}, Arviz \citep{arviz}, h5py \citep{h5py_2014}, Matplotlib \citep{matplotlib} and corner \citep{corner}. The authors thank A. Kemp and D. Fritzewski for their valuable input as beta testers of parts of the code presented in this work. 
 
\end{acknowledgements}

%
%

\bibliographystyle{aa}
\bibliography{references}

\begin{thebibliography}{92}
\expandafter\ifx\csname natexlab\endcsname\relax\def\natexlab#1{#1}\fi

\bibitem[{{Armstrong} {et~al.}(2016){Armstrong}, {Kirk}, {Lam}, {McCormac},
  {Osborn}, {Spake}, {Walker}, {Brown}, {Kristiansen}, {Pollacco}, {West}, \&
  {Wheatley}}]{armstrong2016}
{Armstrong}, D.~J., {Kirk}, J., {Lam}, K.~W.~F., {et~al.} 2016, \mnras, 456,
  2260

\bibitem[{{Astropy Collaboration} {et~al.}(2022){Astropy Collaboration},
  {Price-Whelan}, {Lim}, {Earl}, {Starkman}, {Bradley}, {Shupe}, {Patil},
  {Corrales}, {Brasseur}, {N{"o}the}, {Donath}, {Tollerud}, {Morris},
  {Ginsburg}, {Vaher}, {Weaver}, {Tocknell}, {Jamieson}, {van Kerkwijk},
  {Robitaille}, {Merry}, {Bachetti}, {G{"u}nther}, {Aldcroft},
  {Alvarado-Montes}, {Archibald}, {B{'o}di}, {Bapat}, {Barentsen}, {Baz{'a}n},
  {Biswas}, {Boquien}, {Burke}, {Cara}, {Cara}, {Conroy}, {Conseil}, {Craig},
  {Cross}, {Cruz}, {D'Eugenio}, {Dencheva}, {Devillepoix}, {Dietrich},
  {Eigenbrot}, {Erben}, {Ferreira}, {Foreman-Mackey}, {Fox}, {Freij}, {Garg},
  {Geda}, {Glattly}, {Gondhalekar}, {Gordon}, {Grant}, {Greenfield}, {Groener},
  {Guest}, {Gurovich}, {Handberg}, {Hart}, {Hatfield-Dodds}, {Homeier},
  {Hosseinzadeh}, {Jenness}, {Jones}, {Joseph}, {Kalmbach}, {Karamehmetoglu},
  {Ka{l}uszy{'n}ski}, {Kelley}, {Kern}, {Kerzendorf}, {Koch}, {Kulumani},
  {Lee}, {Ly}, {Ma}, {MacBride}, {Maljaars}, {Muna}, {Murphy}, {Norman},
  {O'Steen}, {Oman}, {Pacifici}, {Pascual}, {Pascual-Granado}, {Patil},
  {Perren}, {Pickering}, {Rastogi}, {Roulston}, {Ryan}, {Rykoff}, {Sabater},
  {Sakurikar}, {Salgado}, {Sanghi}, {Saunders}, {Savchenko}, {Schwardt},
  {Seifert-Eckert}, {Shih}, {Jain}, {Shukla}, {Sick}, {Simpson},
  {Singanamalla}, {Singer}, {Singhal}, {Sinha}, {Sip{H{o}}cz}, {Spitler},
  {Stansby}, {Streicher}, {{{S}}umak}, {Swinbank}, {Taranu}, {Tewary},
  {Tremblay}, {Val-Borro}, {Van Kooten}, {Vasovi{'c}}, {Verma}, {de Miranda
  Cardoso}, {Williams}, {Wilson}, {Winkel}, {Wood-Vasey}, {Xue}, {Yoachim},
  {Zhang}, {Zonca}, \& {Astropy Project Contributors}}]{astropy:2022}
{Astropy Collaboration}, {Price-Whelan}, A.~M., {Lim}, P.~L., {et~al.} 2022,
  apj, 935, 167

\bibitem[{{Astropy Collaboration} {et~al.}(2018){Astropy Collaboration},
  {Price-Whelan}, {Sip{\H{o}}cz}, {G{\"u}nther}, {Lim}, {Crawford}, {Conseil},
  {Shupe}, {Craig}, {Dencheva}, {Ginsburg}, {Vand erPlas}, {Bradley},
  {P{\'e}rez-Su{\'a}rez}, {de Val-Borro}, {Aldcroft}, {Cruz}, {Robitaille},
  {Tollerud}, {Ardelean}, {Babej}, {Bach}, {Bachetti}, {Bakanov}, {Bamford},
  {Barentsen}, {Barmby}, {Baumbach}, {Berry}, {Biscani}, {Boquien}, {Bostroem},
  {Bouma}, {Brammer}, {Bray}, {Breytenbach}, {Buddelmeijer}, {Burke},
  {Calderone}, {Cano Rodr{\'\i}guez}, {Cara}, {Cardoso}, {Cheedella}, {Copin},
  {Corrales}, {Crichton}, {D'Avella}, {Deil}, {Depagne}, {Dietrich}, {Donath},
  {Droettboom}, {Earl}, {Erben}, {Fabbro}, {Ferreira}, {Finethy}, {Fox},
  {Garrison}, {Gibbons}, {Goldstein}, {Gommers}, {Greco}, {Greenfield},
  {Groener}, {Grollier}, {Hagen}, {Hirst}, {Homeier}, {Horton}, {Hosseinzadeh},
  {Hu}, {Hunkeler}, {Ivezi{\'c}}, {Jain}, {Jenness}, {Kanarek}, {Kendrew},
  {Kern}, {Kerzendorf}, {Khvalko}, {King}, {Kirkby}, {Kulkarni}, {Kumar},
  {Lee}, {Lenz}, {Littlefair}, {Ma}, {Macleod}, {Mastropietro}, {McCully},
  {Montagnac}, {Morris}, {Mueller}, {Mumford}, {Muna}, {Murphy}, {Nelson},
  {Nguyen}, {Ninan}, {N{\"o}the}, {Ogaz}, {Oh}, {Parejko}, {Parley}, {Pascual},
  {Patil}, {Patil}, {Plunkett}, {Prochaska}, {Rastogi}, {Reddy Janga},
  {Sabater}, {Sakurikar}, {Seifert}, {Sherbert}, {Sherwood-Taylor}, {Shih},
  {Sick}, {Silbiger}, {Singanamalla}, {Singer}, {Sladen}, {Sooley},
  {Sornarajah}, {Streicher}, {Teuben}, {Thomas}, {Tremblay}, {Turner},
  {Terr{\'o}n}, {van Kerkwijk}, {de la Vega}, {Watkins}, {Weaver}, {Whitmore},
  {Woillez}, {Zabalza}, \& {Astropy Contributors}}]{astropy:2018}
{Astropy Collaboration}, {Price-Whelan}, A.~M., {Sip{\H{o}}cz}, B.~M., {et~al.}
  2018, \aj, 156, 123

\bibitem[{{Astropy Collaboration} {et~al.}(2013){Astropy Collaboration},
  {Robitaille}, {Tollerud}, {Greenfield}, {Droettboom}, {Bray}, {Aldcroft},
  {Davis}, {Ginsburg}, {Price-Whelan}, {Kerzendorf}, {Conley}, {Crighton},
  {Barbary}, {Muna}, {Ferguson}, {Grollier}, {Parikh}, {Nair}, {Unther},
  {Deil}, {Woillez}, {Conseil}, {Kramer}, {Turner}, {Singer}, {Fox}, {Weaver},
  {Zabalza}, {Edwards}, {Azalee Bostroem}, {Burke}, {Casey}, {Crawford},
  {Dencheva}, {Ely}, {Jenness}, {Labrie}, {Lim}, {Pierfederici}, {Pontzen},
  {Ptak}, {Refsdal}, {Servillat}, \& {Streicher}}]{astropy:2013}
{Astropy Collaboration}, {Robitaille}, T.~P., {Tollerud}, E.~J., {et~al.} 2013,
  \aap, 558, A33

\bibitem[{{Auvergne} {et~al.}(2009){Auvergne}, {Bodin}, {Boisnard}, {Buey},
  {Chaintreuil}, {Epstein}, {Jouret}, {Lam-Trong}, {Levacher}, {Magnan},
  {Perez}, {Plasson}, {Plesseria}, {Peter}, {Steller}, {Tiph{\`e}ne}, {Baglin},
  {Agogu{\'e}}, {Appourchaux}, {Barbet}, {Beaufort}, {Bellenger}, {Berlin},
  {Bernardi}, {Blouin}, {Boumier}, {Bonneau}, {Briet}, {Butler}, {Cautain},
  {Chiavassa}, {Costes}, {Cuvilho}, {Cunha-Parro}, {de Oliveira Fialho},
  {Decaudin}, {Defise}, {Djalal}, {Docclo}, {Drummond}, {Dupuis}, {Exil},
  {Faur{\'e}}, {Gaboriaud}, {Gamet}, {Gavalda}, {Grolleau}, {Gueguen},
  {Guivarc'h}, {Guterman}, {Hasiba}, {Huntzinger}, {Hustaix}, {Imbert},
  {Jeanville}, {Johlander}, {Jorda}, {Journoud}, {Karioty}, {Kerjean},
  {Lafond}, {Lapeyrere}, {Landiech}, {Larqu{\'e}}, {Laudet}, {Le Merrer},
  {Leporati}, {Leruyet}, {Levieuge}, {Llebaria}, {Martin}, {Mazy}, {Mesnager},
  {Michel}, {Moalic}, {Monjoin}, {Naudet}, {Neukirchner}, {Nguyen-Kim},
  {Ollivier}, {Orcesi}, {Ottacher}, {Oulali}, {Parisot}, {Perruchot},
  {Piacentino}, {Pinheiro da Silva}, {Platzer}, {Pontet}, {Pradines},
  {Quentin}, {Rohbeck}, {Rolland}, {Rollenhagen}, {Romagnan}, {Russ}, {Samadi},
  {Schmidt}, {Schwartz}, {Sebbag}, {Smit}, {Sunter}, {Tello}, {Toulouse},
  {Ulmer}, {Vandermarcq}, {Vergnault}, {Wallner}, {Waultier}, \&
  {Zanatta}}]{corot2009}
{Auvergne}, M., {Bodin}, P., {Boisnard}, L., {et~al.} 2009, \aap, 506, 411

\bibitem[{{Baran} \& {Koen}(2021)}]{Baran2021}
{Baran}, A.~S. \& {Koen}, C. 2021, \actaa, 71, 113

\bibitem[{{Blanton} {et~al.}(2017){Blanton}, {Bershady}, {Abolfathi},
  {Albareti}, {Allende Prieto}, {Almeida}, {Alonso-Garc{\'\i}a}, {Anders},
  {Anderson}, {Andrews}, \& et~al.}]{sdssiv2017}
{Blanton}, M.~R., {Bershady}, M.~A., {Abolfathi}, B., {et~al.} 2017, \aj, 154,
  28

\bibitem[{{Bowman} \& {Michielsen}(2021)}]{Bowman2021}
{Bowman}, D.~M. \& {Michielsen}, M. 2021, \aap, 656, A158

\bibitem[{{Burkholder} {et~al.}(1997){Burkholder}, {Massey}, \&
  {Morrell}}]{Burkholder1997}
{Burkholder}, V., {Massey}, P., \& {Morrell}, N. 1997, \apj, 490, 328

\bibitem[{Collette(2013)}]{h5py_2014}
Collette, A. 2013, Python and HDF5 (O'Reilly)

\bibitem[{{De Silva} {et~al.}(2015){De Silva}, {Freeman}, {Bland-Hawthorn},
  {Martell}, {de Boer}, {Asplund}, {Keller}, {Sharma}, {Zucker}, {Zwitter},
  {Anguiano}, {Bacigalupo}, {Bayliss}, {Beavis}, {Bergemann}, {Campbell},
  {Cannon}, {Carollo}, {Casagrande}, {Casey}, {Da Costa}, {D'Orazi}, {Dotter},
  {Duong}, {Heger}, {Ireland}, {Kafle}, {Kos}, {Lattanzio}, {Lewis}, {Lin},
  {Lind}, {Munari}, {Nataf}, {O'Toole}, {Parker}, {Reid}, {Schlesinger},
  {Sheinis}, {Simpson}, {Stello}, {Ting}, {Traven}, {Watson}, {Wittenmyer},
  {Yong}, \& {{\v{Z}}erjal}}]{galah2015}
{De Silva}, G.~M., {Freeman}, K.~C., {Bland-Hawthorn}, J., {et~al.} 2015,
  \mnras, 449, 2604

\bibitem[{{Degroote} {et~al.}(2009){Degroote}, {Briquet}, {Catala},
  {Uytterhoeven}, {Lefever}, {Morel}, {Aerts}, {Carrier}, {Auvergne}, {Baglin},
  \& {Michel}}]{Degroote2009}
{Degroote}, P., {Briquet}, M., {Catala}, C., {et~al.} 2009, \aap, 506, 111

\bibitem[{{Devor}(2005)}]{Devor2005}
{Devor}, J. 2005, \apj, 628, 411

\bibitem[{Foreman-Mackey(2016)}]{corner}
Foreman-Mackey, D. 2016, The Journal of Open Source Software, 1, 24

\bibitem[{{Gilmore} {et~al.}(2012){Gilmore}, {Randich}, {Asplund}, {Binney},
  {Bonifacio}, {Drew}, {Feltzing}, {Ferguson}, {Jeffries}, {Micela},
  {Negueruela}, {Prusti}, {Rix}, {Vallenari}, {Alfaro}, {Allende-Prieto},
  {Babusiaux}, {Bensby}, {Blomme}, {Bragaglia}, {Flaccomio}, {Fran{\c{c}}ois},
  {Irwin}, {Koposov}, {Korn}, {Lanzafame}, {Pancino}, {Paunzen},
  {Recio-Blanco}, {Sacco}, {Smiljanic}, {Van Eck}, {Walton}, {Aden}, {Aerts},
  {Affer}, {Alcala}, {Altavilla}, {Alves}, {Antoja}, {Arenou}, {Argiroffi},
  {Asensio Ramos}, {Bailer-Jones}, {Balaguer-Nunez}, {Bayo}, {Barbuy},
  {Barisevicius}, {Barrado y Navascues}, {Battistini}, {Bellas Velidis},
  {Bellazzini}, {Belokurov}, {Bergemann}, {Bertelli}, {Biazzo}, {Bienayme},
  {Bland-Hawthorn}, {Boeche}, {Bonito}, {Boudreault}, {Bouvier}, {Brandao},
  {Brown}, {de Bruijne}, {Burleigh}, {Caballero}, {Caffau}, {Calura},
  {Capuzzo-Dolcetta}, {Caramazza}, {Carraro}, {Casagrande}, {Casewell},
  {Chapman}, {Chiappini}, {Chorniy}, {Christlieb}, {Cignoni}, {Cocozza},
  {Colless}, {Collet}, {Collins}, {Correnti}, {Covino}, {Crnojevic}, {Cropper},
  {Cunha}, {Damiani}, {David}, {Delgado}, {Duffau}, {Edvardsson}, {Eldridge},
  {Enke}, {Eriksson}, {Evans}, {Eyer}, {Famaey}, {Fellhauer}, {Ferreras},
  {Figueras}, {Fiorentino}, {Flynn}, {Folha}, {Franciosini}, {Frasca},
  {Freeman}, {Fremat}, {Friel}, {Gaensicke}, {Gameiro}, {Garzon}, {Geier},
  {Geisler}, {Gerhard}, {Gibson}, {Gomboc}, {Gomez}, {Gonzalez-Fernandez},
  {Gonzalez Hernandez}, {Gosset}, {Grebel}, {Greimel}, {Groenewegen},
  {Grundahl}, {Guarcello}, {Gustafsson}, {Hadrava}, {Hatzidimitriou}, {Hambly},
  {Hammersley}, {Hansen}, {Haywood}, {Heber}, {Heiter}, {Held}, {Helmi},
  {Hensler}, {Herrero}, {Hill}, {Hodgkin}, {Huelamo}, {Huxor}, {Ibata},
  {Jackson}, {de Jong}, {Jonker}, {Jordan}, {Jordi}, {Jorissen}, {Katz},
  {Kawata}, {Keller}, {Kharchenko}, {Klement}, {Klutsch}, {Knude}, {Koch},
  {Kochukhov}, {Kontizas}, {Koubsky}, {Lallement}, {de Laverny}, {van Leeuwen},
  {Lemasle}, {Lewis}, {Lind}, {Lindstrom}, {Lobel}, {Lopez Santiago}, {Lucas},
  {Ludwig}, {Lueftinger}, {Magrini}, {Maiz Apellaniz}, {Maldonado}, {Marconi},
  {Marino}, {Martayan}, {Martinez-Valpuesta}, {Matijevic}, {McMahon},
  {Messina}, {Meyer}, {Miglio}, {Mikolaitis}, {Minchev}, {Minniti}, {Moitinho},
  {Momany}, {Monaco}, {Montalto}, {Monteiro}, {Monier}, {Montes}, {Mora},
  {Moraux}, {Morel}, {Mowlavi}, {Mucciarelli}, {Munari}, {Napiwotzki},
  {Nardetto}, {Naylor}, {Naze}, {Nelemans}, {Okamoto}, {Ortolani}, {Pace},
  {Palla}, {Palous}, {Parker}, {Penarrubia}, {Pillitteri}, {Piotto}, {Posbic},
  {Prisinzano}, {Puzeras}, {Quirrenbach}, {Ragaini}, {Read}, {Read}, {Reyle},
  {De Ridder}, {Robichon}, {Robin}, {Roeser}, {Romano}, {Royer}, {Ruchti},
  {Ruzicka}, {Ryan}, {Ryde}, {Santos}, {Sanz Forcada}, {Sarro Baro},
  {Sbordone}, {Schilbach}, {Schmeja}, {Schnurr}, {Schoenrich}, {Scholz},
  {Seabroke}, {Sharma}, {De Silva}, {Smith}, {Solano}, {Sordo}, {Soubiran},
  {Sousa}, {Spagna}, {Steffen}, {Steinmetz}, {Stelzer}, {Stempels},
  {Tabernero}, {Tautvaisiene}, {Thevenin}, {Torra}, {Tosi}, {Tolstoy}, {Turon},
  {Walker}, {Wambsganss}, {Worley}, {Venn}, {Vink}, {Wyse}, {Zaggia},
  {Zeilinger}, {Zoccali}, {Zorec}, {Zucker}, {Zwitter}, \& {Gaia-ESO Survey
  Team}}]{gaia_eso2012}
{Gilmore}, G., {Randich}, S., {Asplund}, M., {et~al.} 2012, The Messenger, 147,
  25

\bibitem[{{Goldreich} \& {Nicholson}(1977)}]{Goldreich1977}
{Goldreich}, P. \& {Nicholson}, P.~D. 1977, \icarus, 30, 301

\bibitem[{{Guinan} {et~al.}(2000){Guinan}, {Ribas}, {Fitzpatrick},
  {Gim{\'e}nez}, {Jordi}, {McCook}, \& {Popper}}]{Guinan2000}
{Guinan}, E.~F., {Ribas}, I., {Fitzpatrick}, E.~L., {et~al.} 2000, \apj, 544,
  409

\bibitem[{{Guo} {et~al.}(2022){Guo}, {Ogilvie}, {Li}, {Townsend}, \&
  {Sun}}]{guo2022}
{Guo}, Z., {Ogilvie}, G.~I., {Li}, G., {Townsend}, R. H.~D., \& {Sun}, M. 2022,
  \mnras, 517, 437

\bibitem[{{Halbwachs} {et~al.}(2005){Halbwachs}, {Mayor}, \&
  {Udry}}]{Halbwachs2005}
{Halbwachs}, J.~L., {Mayor}, M., \& {Udry}, S. 2005, \aap, 431, 1129

\bibitem[{Harris {et~al.}(2020)Harris, Millman, van~der Walt, Gommers,
  Virtanen, Cournapeau, Wieser, Taylor, Berg, Smith, Kern, Picus, Hoyer, van
  Kerkwijk, Brett, Haldane, del R{'{\i}}o, Wiebe, Peterson,
  G{'{e}}rard-Marchant, Sheppard, Reddy, Weckesser, Abbasi, Gohlke, \&
  Oliphant}]{numpy}
Harris, C.~R., Millman, K.~J., van~der Walt, S.~J., {et~al.} 2020, Nature, 585,
  357

\bibitem[{{Hocke}(1998)}]{Hocke1998}
{Hocke}, K. 1998, Annales Geophysicae, 16, 356

\bibitem[{{Howard} {et~al.}(2022){Howard}, {Davenport}, \&
  {Covey}}]{howard2022}
{Howard}, E.~L., {Davenport}, J. R.~A., \& {Covey}, K.~R. 2022, Research Notes
  of the American Astronomical Society, 6, 96

\bibitem[{{Huber}(2015)}]{Huber2015}
{Huber}, D. 2015, in Astrophysics and Space Science Library, Vol. 408, Giants
  of Eclipse: The \&zeta; Aurigae Stars and Other Binary Systems, ed. T.~B.
  {Ake} \& E.~{Griffin}, 169

\bibitem[{Hunter(2007)}]{matplotlib}
Hunter, J.~D. 2007, Computing in Science \& Engineering, 9, 90

\bibitem[{{IJspeert} {et~al.}(2021){IJspeert}, {Tkachenko}, {Johnston},
  {Garcia}, {De Ridder}, {Van Reeth}, \& {Aerts}}]{ijspeert2021}
{IJspeert}, L.~W., {Tkachenko}, A., {Johnston}, C., {et~al.} 2021, \aap, 652,
  A120

\bibitem[{{Jennings} {et~al.}(2023){Jennings}, {Southworth}, {Maxted}, \&
  {Mancini}}]{Jennings2023}
{Jennings}, Z., {Southworth}, J., {Maxted}, P.~F.~L., \& {Mancini}, L. 2023,
  \mnras, 521, 3405

\bibitem[{{Johnston}(2021)}]{Johnston2021b}
{Johnston}, C. 2021, \aap, 655, A29

\bibitem[{{Johnston} {et~al.}(2021){Johnston}, {Aimar}, {Abdul-Masih},
  {Bowman}, {White}, {Hawcroft}, {Sana}, {Sekaran}, {Dsilva}, {Tkachenko}, \&
  {Aerts}}]{Johnston2021}
{Johnston}, C., {Aimar}, N., {Abdul-Masih}, M., {et~al.} 2021, \mnras, 503,
  1124

\bibitem[{{Johnston} {et~al.}(2019{\natexlab{a}}){Johnston}, {Pavlovski}, \&
  {Tkachenko}}]{Johnston2019b}
{Johnston}, C., {Pavlovski}, K., \& {Tkachenko}, A. 2019{\natexlab{a}}, \aap,
  628, A25

\bibitem[{{Johnston} {et~al.}(2019{\natexlab{b}}){Johnston}, {Tkachenko},
  {Aerts}, {Molenberghs}, {Bowman}, {Pedersen}, {Buysschaert}, \&
  {P{\'a}pics}}]{Johnston2019a}
{Johnston}, C., {Tkachenko}, A., {Aerts}, C., {et~al.} 2019{\natexlab{b}},
  \mnras, 482, 1231

\bibitem[{{Johnston} {et~al.}(2023){Johnston}, {Tkachenko}, {Van Reeth},
  {Bowman}, {Pavlovski}, {Sana}, \& {Sekaran}}]{Johnston2023}
{Johnston}, C., {Tkachenko}, A., {Van Reeth}, T., {et~al.} 2023, \aap, 670,
  A167

\bibitem[{Kass \& Raftery(1995)}]{Kass1995}
Kass, R.~E. \& Raftery, A.~E. 1995, Journal of the American Statistical
  Association, 90, 773

\bibitem[{{Kirk} {et~al.}(2016){Kirk}, {Conroy}, {Pr{\v{s}}a}, {Abdul-Masih},
  {Kochoska}, {Matijevi{\v{c}}}, {Hambleton}, {Barclay}, {Bloemen}, {Boyajian},
  {Doyle}, {Fulton}, {Hoekstra}, {Jek}, {Kane}, {Kostov}, {Latham}, {Mazeh},
  {Orosz}, {Pepper}, {Quarles}, {Ragozzine}, {Shporer}, {Southworth},
  {Stassun}, {Thompson}, {Welsh}, {Agol}, {Derekas}, {Devor}, {Fischer},
  {Green}, {Gropp}, {Jacobs}, {Johnston}, {LaCourse}, {Saetre}, {Schwengeler},
  {Toczyski}, {Werner}, {Garrett}, {Gore}, {Martinez}, {Spitzer}, {Stevick},
  {Thomadis}, {Vrijmoet}, {Yenawine}, {Batalha}, \&
  {Borucki}}]{kepler_ebvii2016}
{Kirk}, B., {Conroy}, K., {Pr{\v{s}}a}, A., {et~al.} 2016, \aj, 151, 68

\bibitem[{{Kjurkchieva} \& {Vasileva}(2015)}]{Kjurkchieva2015}
{Kjurkchieva}, D. \& {Vasileva}, D. 2015, \pasa, 32, e023

\bibitem[{{Kjurkchieva} \& {Vasileva}(2018)}]{Kjurkchieva2018}
{Kjurkchieva}, D.~P. \& {Vasileva}, D.~L. 2018, \apss, 363, 19

\bibitem[{{Koch} {et~al.}(2010){Koch}, {Borucki}, {Basri}, {Batalha}, {Brown},
  {Caldwell}, {Christensen-Dalsgaard}, {Cochran}, {DeVore}, {Dunham},
  {Gautier}, {Geary}, {Gilliland}, {Gould}, {Jenkins}, {Kondo}, {Latham},
  {Lissauer}, {Marcy}, {Monet}, {Sasselov}, {Boss}, {Brownlee}, {Caldwell},
  {Dupree}, {Howell}, {Kjeldsen}, {Meibom}, {Morrison}, {Owen}, {Reitsema},
  {Tarter}, {Bryson}, {Dotson}, {Gazis}, {Haas}, {Kolodziejczak}, {Rowe}, {Van
  Cleve}, {Allen}, {Chandrasekaran}, {Clarke}, {Li}, {Quintana}, {Tenenbaum},
  {Twicken}, \& {Wu}}]{kepler2010}
{Koch}, D.~G., {Borucki}, W.~J., {Basri}, G., {et~al.} 2010, \apjl, 713, L79

\bibitem[{{Kollmeier} {et~al.}(2017){Kollmeier}, {Zasowski}, {Rix}, {Johns},
  {Anderson}, {Drory}, {Johnson}, {Pogge}, {Bird}, {Blanc}, {Brownstein},
  {Crane}, {De Lee}, {Klaene}, {Kreckel}, {MacDonald}, {Merloni}, {Ness},
  {O'Brien}, {Sanchez-Gallego}, {Sayres}, {Shen}, {Thakar}, {Tkachenko},
  {Aerts}, {Blanton}, {Eisenstein}, {Holtzman}, {Maoz}, {Nandra}, {Rockosi},
  {Weinberg}, {Bovy}, {Casey}, {Chaname}, {Clerc}, {Conroy}, {Eracleous},
  {G{\"a}nsicke}, {Hekker}, {Horne}, {Kauffmann}, {McQuinn}, {Pellegrini},
  {Schinnerer}, {Schlafly}, {Schwope}, {Seibert}, {Teske}, \& {van
  Saders}}]{sdssv2017}
{Kollmeier}, J.~A., {Zasowski}, G., {Rix}, H.-W., {et~al.} 2017, arXiv
  e-prints, arXiv:1711.03234

\bibitem[{{Kopal}(1959)}]{Kopal1959}
{Kopal}, Z. 1959, {Close binary systems} (John Wiley \& Sons)

\bibitem[{{Kumar} {et~al.}(2019){Kumar}, {Carroll}, {Hartikainen}, \&
  {Martin}}]{arviz}
{Kumar}, R., {Carroll}, C., {Hartikainen}, A., \& {Martin}, O. 2019, The
  Journal of Open Source Software, 4, 1143

\bibitem[{Lam {et~al.}(2015)Lam, Pitrou, \& Seibert}]{numba}
Lam, S.~K., Pitrou, A., \& Seibert, S. 2015, in Proceedings of the Second
  Workshop on the LLVM Compiler Infrastructure in HPC, LLVM '15 (New York, NY,
  USA: Association for Computing Machinery)

\bibitem[{{Lampens}(2021)}]{Lampens2021}
{Lampens}, P. 2021, Galaxies, 9, 28

\bibitem[{{Lecar} {et~al.}(1976){Lecar}, {Wheeler}, \& {McKee}}]{Lecar1976}
{Lecar}, M., {Wheeler}, J.~C., \& {McKee}, C.~F. 1976, \apj, 205, 556

\bibitem[{{Lomb}(1976)}]{1976Ap&SS..39..447L}
{Lomb}, N.~R. 1976, \apss, 39, 447

\bibitem[{{Loumos} \& {Deeming}(1978)}]{Loumos1978}
{Loumos}, G.~L. \& {Deeming}, T.~J. 1978, \apss, 56, 285

\bibitem[{{Lurie} {et~al.}(2017){Lurie}, {Vyhmeister}, {Hawley}, {Adilia},
  {Chen}, {Davenport}, {Juri{\'c}}, {Puig-Holzman}, \&
  {Weisenburger}}]{Lurie2017}
{Lurie}, J.~C., {Vyhmeister}, K., {Hawley}, S.~L., {et~al.} 2017, \aj, 154, 250

\bibitem[{{Maceroni} {et~al.}(2009){Maceroni}, {Montalb{\'a}n}, {Michel},
  {Harmanec}, {Prsa}, {Briquet}, {Niemczura}, {Morel}, {Ladjal}, {Auvergne},
  {Baglin}, {Baudin}, {Catala}, {Samadi}, \& {Aerts}}]{Maceroni2009}
{Maceroni}, C., {Montalb{\'a}n}, J., {Michel}, E., {et~al.} 2009, \aap, 508,
  1375

\bibitem[{{Massey} {et~al.}(2012){Massey}, {Morrell}, {Neugent}, {Penny},
  {DeGioia-Eastwood}, \& {Gies}}]{Massey2012}
{Massey}, P., {Morrell}, N.~I., {Neugent}, K.~F., {et~al.} 2012, \apj, 748, 96

\bibitem[{{Maxted}(2016)}]{ellc}
{Maxted}, P.~F.~L. 2016, \aap, 591, A111

\bibitem[{{Maxted} {et~al.}(2020){Maxted}, {Gaulme}, {Graczyk}, {He{\l}miniak},
  {Johnston}, {Orosz}, {Pr{\v{s}}a}, {Southworth}, {Torres}, {Davies}, {Ball},
  \& {Chaplin}}]{Maxted2020}
{Maxted}, P.~F.~L., {Gaulme}, P., {Graczyk}, D., {et~al.} 2020, \mnras, 498,
  332

\bibitem[{Nelder \& Mead(1965)}]{NelderMead1965}
Nelder, J.~A. \& Mead, R. 1965, The Computer Journal, 7, 308

\bibitem[{{Pavlovski} {et~al.}(2023){Pavlovski}, {Southworth}, {Tkachenko},
  {Van Reeth}, \& {Tamajo}}]{Pavlovski2023}
{Pavlovski}, K., {Southworth}, J., {Tkachenko}, A., {Van Reeth}, T., \&
  {Tamajo}, E. 2023, \aap, 671, A139

\bibitem[{{Press} \& {Rybicki}(1989)}]{astropy_ls_fast}
{Press}, W.~H. \& {Rybicki}, G.~B. 1989, \apj, 338, 277

\bibitem[{{Pr{\v{s}}a} {et~al.}(2011){Pr{\v{s}}a}, {Batalha}, {Slawson},
  {Doyle}, {Welsh}, {Orosz}, {Seager}, {Rucker}, {Mjaseth}, {Engle}, {Conroy},
  {Jenkins}, {Caldwell}, {Koch}, \& {Borucki}}]{kepler_eb2011}
{Pr{\v{s}}a}, A., {Batalha}, N., {Slawson}, R.~W., {et~al.} 2011, \aj, 141, 83

\bibitem[{{Pr{\v{s}}a} {et~al.}(2008){Pr{\v{s}}a}, {Guinan}, {Devinney},
  {DeGeorge}, {Bradstreet}, {Giammarco}, {Alcock}, \& {Engle}}]{Prsa2008}
{Pr{\v{s}}a}, A., {Guinan}, E.~F., {Devinney}, E.~J., {et~al.} 2008, \apj, 687,
  542

\bibitem[{{Pr{\v{s}}a} {et~al.}(2022){Pr{\v{s}}a}, {Kochoska}, {Conroy},
  {Eisner}, {Hey}, {IJspeert}, {Kruse}, {Fleming}, {Johnston}, {Kristiansen},
  {LaCourse}, {Mortensen}, {Pepper}, {Stassun}, {Torres}, {Abdul-Masih},
  {Chakraborty}, {Gagliano}, {Guo}, {Hambleton}, {Hong}, {Jacobs}, {Jones},
  {Kostov}, {Lee}, {Omohundro}, {Orosz}, {Page}, {Powell}, {Rappaport}, {Reed},
  {Schnittman}, {Schwengeler}, {Shporer}, {Terentev}, {Vanderburg}, {Welsh},
  {Caldwell}, {Doty}, {Jenkins}, {Latham}, {Ricker}, {Seager}, {Schlieder},
  {Shiao}, {Vanderspek}, \& {Winn}}]{tess_eb2022}
{Pr{\v{s}}a}, A., {Kochoska}, A., {Conroy}, K.~E., {et~al.} 2022, \apjs, 258,
  16

\bibitem[{Raftery(1995)}]{Raftery1995}
Raftery, A.~E. 1995, Sociological Methodology, 25, 111

\bibitem[{{Rauer} {et~al.}(2022){Rauer}, {Aerts}, {Deleuil}, {Gizon}, {Goupil},
  {Heras}, {Mas-Hesse}, {Pagano}, {Piotto}, {Pollacco}, {Ragazzoni}, {Ramsay},
  \& {Udry}}]{PLATO2022}
{Rauer}, H., {Aerts}, C., {Deleuil}, M., {et~al.} 2022, in European Planetary
  Science Congress, EPSC2022--453

\bibitem[{{Ricker} {et~al.}(2015){Ricker}, {Winn}, {Vanderspek}, {Latham},
  {Bakos}, {Bean}, {Berta-Thompson}, {Brown}, {Buchhave}, {Butler}, {Butler},
  {Chaplin}, {Charbonneau}, {Christensen-Dalsgaard}, {Clampin}, {Deming},
  {Doty}, {De Lee}, {Dressing}, {Dunham}, {Endl}, {Fressin}, {Ge}, {Henning},
  {Holman}, {Howard}, {Ida}, {Jenkins}, {Jernigan}, {Johnson}, {Kaltenegger},
  {Kawai}, {Kjeldsen}, {Laughlin}, {Levine}, {Lin}, {Lissauer}, {MacQueen},
  {Marcy}, {McCullough}, {Morton}, {Narita}, {Paegert}, {Palle}, {Pepe},
  {Pepper}, {Quirrenbach}, {Rinehart}, {Sasselov}, {Sato}, {Seager},
  {Sozzetti}, {Stassun}, {Sullivan}, {Szentgyorgyi}, {Torres}, {Udry}, \&
  {Villasenor}}]{tess2015}
{Ricker}, G.~R., {Winn}, J.~N., {Vanderspek}, R., {et~al.} 2015, Journal of
  Astronomical Telescopes, Instruments, and Systems, 1, 014003

\bibitem[{{Rosu} {et~al.}(2023){Rosu}, {Quintero}, {Rauw}, \&
  {Eenens}}]{Rosu2023}
{Rosu}, S., {Quintero}, E.~A., {Rauw}, G., \& {Eenens}, P. 2023, \mnras, 521,
  2988

\bibitem[{{Rowan} {et~al.}(2022){Rowan}, {Jayasinghe}, {Stanek}, {Kochanek},
  {Thompson}, {Shappee}, {Holoien}, {Prieto}, \& {Giles}}]{rowan2022}
{Rowan}, D.~M., {Jayasinghe}, T., {Stanek}, K.~Z., {et~al.} 2022, \mnras, 517,
  2190

\bibitem[{{Saha} \& {Vivas}(2017)}]{saha2017}
{Saha}, A. \& {Vivas}, A.~K. 2017, \aj, 154, 231

\bibitem[{Salvatier {et~al.}(2016)Salvatier, Wiecki, \& Fonnesbeck}]{pymc3}
Salvatier, J., Wiecki, T.~V., \& Fonnesbeck, C. 2016, PeerJ Computer Science,
  2, e55

\bibitem[{{Salvatier} {et~al.}(2016){Salvatier}, {Wiecki{\^a}}, \&
  {Fonnesbeck}}]{pymc3_code}
{Salvatier}, J., {Wiecki{\^a}}, T.~V., \& {Fonnesbeck}, C. 2016, ascl:1610.016

\bibitem[{{Savonije} \& {Papaloizou}(1983)}]{Savonije1983}
{Savonije}, G.~J. \& {Papaloizou}, J.~C.~B. 1983, \mnras, 203, 581

\bibitem[{{Savonije} \& {Papaloizou}(1984)}]{Savonije1984}
{Savonije}, G.~J. \& {Papaloizou}, J.~C.~B. 1984, \mnras, 207, 685

\bibitem[{{Scargle}(1982)}]{1982ApJ...263..835S}
{Scargle}, J.~D. 1982, \apj, 263, 835

\bibitem[{{Schmid} \& {Aerts}(2016)}]{Schmid2016}
{Schmid}, V.~S. \& {Aerts}, C. 2016, \aap, 592, A116

\bibitem[{{Schmid} {et~al.}(2015){Schmid}, {Tkachenko}, {Aerts}, {Degroote},
  {Bloemen}, {Murphy}, {Van Reeth}, {P{\'a}pics}, {Bedding}, {Keen},
  {Pr{\v{s}}a}, {Menu}, {Debosscher}, {Hrudkov{\'a}}, {De Smedt}, {Lombaert},
  \& {N{\'e}meth}}]{Schmid2015}
{Schmid}, V.~S., {Tkachenko}, A., {Aerts}, C., {et~al.} 2015, \aap, 584, A35

\bibitem[{{Sekaran} {et~al.}(2020){Sekaran}, {Tkachenko}, {Abdul-Masih},
  {Pr{\v{s}}a}, {Johnston}, {Huber}, {Murphy}, {Banyard}, {Howard}, {Isaacson},
  {Bowman}, \& {Aerts}}]{Sekaran2020}
{Sekaran}, S., {Tkachenko}, A., {Abdul-Masih}, M., {et~al.} 2020, \aap, 643,
  A162

\bibitem[{{Sekaran} {et~al.}(2021){Sekaran}, {Tkachenko}, {Johnston}, \&
  {Aerts}}]{Sekaran2021}
{Sekaran}, S., {Tkachenko}, A., {Johnston}, C., \& {Aerts}, C. 2021, \aap, 648,
  A91

\bibitem[{{Slawson} {et~al.}(2011){Slawson}, {Pr{\v{s}}a}, {Welsh}, {Orosz},
  {Rucker}, {Batalha}, {Doyle}, {Engle}, {Conroy}, {Coughlin}, {Gregg},
  {Fetherolf}, {Short}, {Windmiller}, {Fabrycky}, {Howell}, {Jenkins}, {Uddin},
  {Mullally}, {Seader}, {Thompson}, {Sand erfer}, {Borucki}, \&
  {Koch}}]{kepler_ebii2011}
{Slawson}, R.~W., {Pr{\v{s}}a}, A., {Welsh}, W.~F., {et~al.} 2011, \aj, 142,
  160

\bibitem[{{Southworth}(2012)}]{Southworth2012}
{Southworth}, J. 2012, in Orbital Couples: Pas de Deux in the Solar System and
  the Milky Way, ed. F.~{Arenou} \& D.~{Hestroffer}, 51--58

\bibitem[{{Southworth} \& {Bowman}(2022)}]{southworth2022a}
{Southworth}, J. \& {Bowman}, D.~M. 2022, \mnras, 513, 3191

\bibitem[{{Southworth} \& {Van Reeth}(2022)}]{southworth2022b}
{Southworth}, J. \& {Van Reeth}, T. 2022, \mnras, 515, 2755

\bibitem[{{Stellingwerf}(1978)}]{stellingwerf1978}
{Stellingwerf}, R.~F. 1978, \apj, 224, 953

\bibitem[{{Tamuz} {et~al.}(2006){Tamuz}, {Mazeh}, \& {North}}]{Tamuz2006}
{Tamuz}, O., {Mazeh}, T., \& {North}, P. 2006, \mnras, 367, 1521

\bibitem[{{Tassoul} \& {Tassoul}(1990)}]{Tassoul1990}
{Tassoul}, J.-L. \& {Tassoul}, M. 1990, \apj, 359, 155

\bibitem[{{The Theano Development Team} {et~al.}(2016){The Theano Development
  Team}, {Al-Rfou}, {Alain}, {Almahairi}, {Angermueller}, {Bahdanau}, {Ballas},
  {Bastien}, {Bayer}, {Belikov}, {Belopolsky}, {Bengio}, {Bergeron},
  {Bergstra}, {Bisson}, {Bleecher Snyder}, {Bouchard}, {Boulanger-Lewandowski},
  {Bouthillier}, {de Br{\'e}bisson}, {Breuleux}, {Carrier}, {Cho}, {Chorowski},
  {Christiano}, {Cooijmans}, {C{\^o}t{\'e}}, {C{\^o}t{\'e}}, {Courville},
  {Dauphin}, {Delalleau}, {Demouth}, {Desjardins}, {Dieleman}, {Dinh},
  {Ducoffe}, {Dumoulin}, {Ebrahimi Kahou}, {Erhan}, {Fan}, {Firat}, {Germain},
  {Glorot}, {Goodfellow}, {Graham}, {Gulcehre}, {Hamel}, {Harlouchet}, {Heng},
  {Hidasi}, {Honari}, {Jain}, {Jean}, {Jia}, {Korobov}, {Kulkarni}, {Lamb},
  {Lamblin}, {Larsen}, {Laurent}, {Lee}, {Lefrancois}, {Lemieux},
  {L{\'e}onard}, {Lin}, {Livezey}, {Lorenz}, {Lowin}, {Ma}, {Manzagol},
  {Mastropietro}, {McGibbon}, {Memisevic}, {van Merri{\"e}nboer}, {Michalski},
  {Mirza}, {Orlandi}, {Pal}, {Pascanu}, {Pezeshki}, {Raffel}, {Renshaw},
  {Rocklin}, {Romero}, {Roth}, {Sadowski}, {Salvatier}, {Savard},
  {Schl{\"u}ter}, {Schulman}, {Schwartz}, {Vlad Serban}, {Serdyuk},
  {Shabanian}, {Simon}, {Spieckermann}, {Ramana Subramanyam}, {Sygnowski},
  {Tanguay}, {van Tulder}, {Turian}, {Urban}, {Vincent}, {Visin}, {de Vries},
  {Warde-Farley}, {Webb}, {Willson}, {Xu}, {Xue}, {Yao}, {Zhang}, \&
  {Zhang}}]{theano}
{The Theano Development Team}, {Al-Rfou}, R., {Alain}, G., {et~al.} 2016, arXiv
  e-prints, arXiv:1605.02688

\bibitem[{{Tkachenko} {et~al.}(2020){Tkachenko}, {Pavlovski}, {Johnston},
  {Pedersen}, {Michielsen}, {Bowman}, {Southworth}, {Tsymbal}, \&
  {Aerts}}]{Tkachenko2020}
{Tkachenko}, A., {Pavlovski}, K., {Johnston}, C., {et~al.} 2020, \aap, 637, A60

\bibitem[{{Torres} {et~al.}(2010){Torres}, {Andersen}, \&
  {Gim{\'e}nez}}]{torres2010}
{Torres}, G., {Andersen}, J., \& {Gim{\'e}nez}, A. 2010, \aapr, 18, 67

\bibitem[{{Udalski} {et~al.}(1992){Udalski}, {Szymanski}, {Kaluzny}, {Kubiak},
  \& {Mateo}}]{OGLE1992}
{Udalski}, A., {Szymanski}, M., {Kaluzny}, J., {Kubiak}, M., \& {Mateo}, M.
  1992, \actaa, 42, 253

\bibitem[{{Van Beeck} {et~al.}(2021){Van Beeck}, {Bowman}, {Pedersen}, {Van
  Reeth}, {Van Hoolst}, \& {Aerts}}]{VanBeeck2021}
{Van Beeck}, J., {Bowman}, D.~M., {Pedersen}, M.~G., {et~al.} 2021, \aap, 655,
  A59

\bibitem[{{Van Eylen} {et~al.}(2016){Van Eylen}, {Winn}, \&
  {Albrecht}}]{vaneylen2016}
{Van Eylen}, V., {Winn}, J.~N., \& {Albrecht}, S. 2016, \apj, 824, 15

\bibitem[{Virtanen {et~al.}(2020)Virtanen, Gommers, Oliphant, Haberland, Reddy,
  Cournapeau, Burovski, Peterson, Weckesser, Bright, {van der Walt}, Brett,
  Wilson, Millman, Mayorov, Nelson, Jones, Kern, Larson, Carey, Polat, Feng,
  Moore, {VanderPlas}, Laxalde, Perktold, Cimrman, Henriksen, Quintero, Harris,
  Archibald, Ribeiro, Pedregosa, {van Mulbregt}, \& {SciPy 1.0
  Contributors}}]{scipy}
Virtanen, P., Gommers, R., Oliphant, T.~E., {et~al.} 2020, Nature Methods, 17,
  261

\bibitem[{{Weidner} \& {Vink}(2010)}]{Weidner2010}
{Weidner}, C. \& {Vink}, J.~S. 2010, \aap, 524, A98

\bibitem[{{Wells} \& {Pr{\v{s}}a}(2021)}]{Wells2021}
{Wells}, M.~A. \& {Pr{\v{s}}a}, A. 2021, \apjs, 253, 32

\bibitem[{{Wilson} \& {Devinney}(1971)}]{WD1971}
{Wilson}, R.~E. \& {Devinney}, E.~J. 1971, \apj, 166, 605

\bibitem[{{Wyrzykowski} {et~al.}(2003){Wyrzykowski}, {Udalski}, {Kubiak},
  {Szymanski}, {Zebrun}, {Soszynski}, {Wozniak}, {Pietrzynski}, \&
  {Szewczyk}}]{Wyrzykowski2003}
{Wyrzykowski}, L., {Udalski}, A., {Kubiak}, M., {et~al.} 2003, \actaa, 53, 1

\bibitem[{{Zahn}(1975)}]{Zahn1975}
{Zahn}, J.~P. 1975, \aap, 41, 329

\bibitem[{{Zahn}(1977)}]{Zahn1977}
{Zahn}, J.~P. 1977, \aap, 57, 383

\bibitem[{{Zahn}(1989)}]{Zahn1989}
{Zahn}, J.~P. 1989, \aap, 220, 112

\bibitem[{{Zahn} \& {Bouchet}(1989)}]{ZahnBouchet1989}
{Zahn}, J.~P. \& {Bouchet}, L. 1989, \aap, 223, 112

\end{thebibliography}

\begin{appendix}
\section{Additional formulae}
\label{apx:formulae}

\subsection*{Maximum likelihood estimators}
\label{apx:mle}
Model parameters can sometimes be determined via direct calculation using MLEs under the assumption that our observations are independent and identically distributed according to a normal distribution. We could do this for the two parameters needed to capture linear trends, as well as for higher-order polynomials; we have gone up to third order here. 

We can describe each observation as being drawn from a normal distribution (our first assumption):
\begin{equation}
    f(x|\mu,\sigma) = \frac{1}{\sqrt{2\pi\sigma^2}} exp\left(\frac{-(x - \mu)^2}{2\sigma^2}\right),
    \label{eq:normal}
\end{equation}
where $x$ are the observations, $\mu$ the mean, and $\sigma$ the standard deviation. The likelihood function is the joint probability density of our observations, which under the assumption that our observations are independent, can be written as the multiplication of the individual distributions:
\begin{equation}
    L(\theta) = \Pi_i f(x_i|\theta) = \left(\frac{1}{\sqrt{2\pi\sigma^2}}\right)^n exp\left(\frac{-\sum_i(x_i - \mu)^2}{2\sigma^2}\right),
    \label{eq:likelihood}
\end{equation}
with $\theta$ representing the model parameters and $n$ is the number of data points. An MLE is obtained by maximising the likelihood, or taking the derivative of the likelihood with respect to each model parameter $\theta$ and setting this to zero. We take the (natural) logarithm of the likelihood because summation is easier to work with and the location of the maximum is conserved. Another advantage in the case of an iterative use of the likelihood is that it is computationally more stable to use the logarithm:
\begin{equation}
    l(\theta) \equiv ln(L(\theta)) = -\frac{n}{2}ln\left(2\pi\sigma^2\right) - \frac{\sum_i(x_i - \mu)^2}{2\sigma^2}.
    \label{eq:log_likelihood}
\end{equation}
We can now confirm that indeed the sample mean ($\frac{1}{n}\sum_i x_i = \bar{x}$) is the MLE of the mean ($\mu$) and the square root of the sample variance ($\sqrt{}\frac{1}{n}\sum_i(x_i - \bar{x})^2$) is the MLE of the standard deviation ($\sigma$) by taking the derivative of $l$ to $\mu$ and $\sigma$, respectively, and setting this to zero. Substituting these parameters for their estimators, we obtain
\begin{equation}
    l(\theta) = -\frac{n}{2}ln\left(\frac{2\pi}{n}\sum_i(x_i - \bar{x})^2\right) - \frac{n}{2}.
    \label{eq:log_likelihood_2}
\end{equation}
The light curve model that we use, whether that is a straight line, sinusoid or more complex, can be seen as the mean of the normal distributions that each observation follows, in the sense that subtracting the model from the data results in a zero mean for all data points. With $y$ our observations and $\hat{y}$ the model, we can then write\begin{equation}
    l(\theta) = -\frac{n}{2} \left(ln\left(2\pi\sigma_r^2\right) + 1\right), 
    \label{eq:ln_l}
\end{equation}
and $\sigma_r^2$ is the variance of the residuals, estimated as\begin{equation*}
    \sigma_r^2 = \frac{1}{n}\sum_i (y_i - \hat{y}_i)^2.
\end{equation*}
We now need an analytical model to work out the estimators for the model parameters. A linear trend is used extensively in this work, but we also include the quadratic and cubic case here for completeness:
\begin{align*}
    \hat{y} &= \theta_1 x + \theta_0, \\
    \hat{y_q} &= \theta_2 x^2 + \theta_1 x + \theta_0, \\
    \hat{y_c} &= \theta_3 x^3 + \theta_2 x^2 + \theta_1 x + \theta_0.
\end{align*}

To shorten the following formulae, we define several sums using subscripts denoting the power of the variable within each term as follows:
\begin{equation}
    S_{\hat{y}} = \sum_i (y_i - \hat{y}_i)^2 = n\sigma_r^2,
    \label{eq:s_y}
\end{equation}
\begin{equation}
    S_{xx} = \sum_i (x_i - \bar{x})^2,
    \label{eq:s_xx}
\end{equation}
\begin{equation}
    S_{x2x} = \sum_i (x_i^2 - \bar{x^2})(x_i - \bar{x}),
    \label{eq:s_x2x}
\end{equation}
\begin{equation}
    S_{x2x2} = \sum_i (x_i^2 - \bar{x^2})^2,
    \label{eq:s_x2x2}
\end{equation}
\begin{equation}
    S_{x3x} = \sum_i (x_i^3 - \bar{x^3})(x_i - \bar{x}),
    \label{eq:s_x3x}
\end{equation}
\begin{equation}
    S_{x3x2} = \sum_i (x_i^3 - \bar{x^3})(x_i^2 - \bar{x^2}),
    \label{eq:s_x3x2}
\end{equation}
\begin{equation}
    S_{x3x3} = \sum_i (x_i^3 - \bar{x^3})^2,
    \label{eq:s_x3x3}
\end{equation}
\begin{equation}
    S_{xy} = \sum_i (x_i - \bar{x})(y_i - \bar{y}),
    \label{eq:s_xy}
\end{equation}
\begin{equation}
    S_{x2y} = \sum_i (x_i^2 - \bar{x^2})(y_i - \bar{y}),
    \label{eq:s_x2y}
\end{equation}
\begin{equation}
    S_{x3y} = \sum_i (x_i^3 - \bar{x^3})(y_i - \bar{y}),
    \label{eq:s_x3y}
\end{equation}
where $\bar{x^2}$ is the mean of squares and $\bar{x^3}$ is the mean of cubes. 

Setting ${\partial l}/{\partial\theta_0}$ equal to zero for the linear case we find\begin{equation}
    \theta_0 = \bar{y} - \theta_1 \bar{x}.
    \label{eq:intercept}
\end{equation}
Setting ${\partial l}/{\partial\theta_1}$ equal to zero and substituting \ref{eq:intercept} we find\begin{equation}
    \theta_1 = \frac{S_{xy}}{S_{xx}}.
    \label{eq:slope}
\end{equation}
Maximum likelihood estimators for the standard deviations of the above estimators are
\begin{equation}
    S_{\theta_0}^2 = S_{y,x}^2 \left(\frac{1}{n_{data}}  + \frac{\bar{x}^2}{S_{xx}}\right), 
    \label{eq:std_intercept}
\end{equation}
\begin{equation}
    S_{\theta_1}^2 = \frac{S_{y,x}^2}{S_{xx}},
    \label{eq:std_slope}
\end{equation}
where the (corrected) standard deviation of $y(x)$ is
\begin{equation}
    S_{y,x}^2 = \frac{S_{\hat{y}}}{n_{dof}}.
    \label{eq:std_y}
\end{equation}


The quadratic case yields
\begin{align}
    \theta_2 &= \frac{S_{x2y} S_{xx} - S_{xy} S_{x2x}}{S_{x2x2} S_{xx} - S_{x2x}^2}, \\
    \theta_1 &= \frac{S_{xy} - \theta_2 S_{x2x}}{S_{xx}}, \\
    \theta_0 &= \bar{y} - \theta_2 \bar{x^2} - \theta_1 \bar{x}.
    \label{eq:quadratic_params}
\end{align}

For the cubic case, we broke up $\theta_3$ into its numerator and denominator:
\begin{align*}
    \theta_{3,a} &= S_{x3y} \left(S_{x2x2} S_{xx} - S_{x2x}^2\right) \\ &\quad - S_{x2y} \left(S_{x3x2} S_{xx} - S_{x3x} S_{x2x}\right) \\ &\quad + S_{xy} \left(S_{x3x2} S_{x2x} - S_{x3x} S_{x2x2}\right), \\
    \theta_{3,b} &= S_{x3x3} \left(S_{x2x2} S_{xx} - S_{x2x}^2\right) \\ &\quad - S_{x3x2} \left(S_{x3x2} S_{xx} - 2 S_{x3x} S_{x2x}\right) \\ &\quad - S_{x3x}^2 S_{x2x2} \\
\end{align*}
and obtain the following results:
\begin{align}
    \theta_3 &= \frac{\theta_{3,a}}{\theta_{3,b}}, \\
    \theta_2 &= \frac{S_{x2y} S_{xx} - S_{xy} S_{x2x} - \theta_3 \left(S_{x3x2} S_{xx} - S_{x3x} S_{x2x}\right)}{S_{x2x2} S_{xx} - S_{x2x}^2}, \\
    \theta_1 &= \frac{S_{xy} - \theta_3 S_{x3x} - \theta_2 S_{x2x}}{S_{xx}}, \\
    \theta_0 &= \bar{y} - \theta_3 \bar{x^3} - \theta_2 \bar{x^2} - \theta_1 \bar{x}.
    \label{eq:cubic_params}
\end{align}

\subsection*{Bayesian information criterion}
\label{apx:bic}
The BIC of the residuals decreases with more white-noise-like residuals (i.e. containing less information), but increases with additional free parameters of which each sinusoid initially has three:
\begin{equation}
    BIC = -2\cdot ln(L(\theta)) + k\cdot ln(n), 
    \label{eq:bic}
\end{equation}
where $L$ is the likelihood as a function of the parameters $\theta$, $n$ the number of data points and $k$ the number of free parameters. Under the simplifying assumption that the observations are independent and identically distributed according to a normal distribution, the likelihood as in \ref{eq:ln_l} can be plugged in to get\begin{equation}
    BIC = n\cdot ln\left(2\pi\sigma_r^2\right) + n + k\cdot ln(n).
\end{equation}

\subsection*{Kepler's second law}
The law of equal swept-out areas, or Kepler's second law, couples the orbital properties to time intervals. This means it enables the computation of said properties if we can measure enough `special' time points along the orbit. Special here means that we know something about the geometric configuration of the binary at that specific time. For example, at the time of eclipse minimum, we know that the star in front is covering a maximal portion of the surface area of the star in the back and thus the projected (in the plane of the sky) distance between their centres is minimal. 

For an eccentric orbit, the instantaneous distance between the centres of the two stars is
\begin{equation}
    R = \frac{a\left(1 - e^2\right)}{1 + e\ cos(\nu)},
    \label{eq:separation}
\end{equation}
with $a$ the semi-major axis, $e$ the eccentricity, and $\nu$ the true anomaly indicating in what phase of the orbit we are. Plugging this into Kepler's second law in integral form, we obtain
\begin{equation}
    \frac{2\pi(t_b-t_a)}{P} = \int_{\nu_a}^{\nu_b}\frac{R^2 d\nu}{a^2 \sqrt{1 - e^2}} = 
    \left(1-e^2\right)^{\frac{3}{2}} \int_{\nu_a}^{\nu_b}\frac{d\nu}{(1 + e\ cos(\nu))^2},
    \label{eq:kepler2}
\end{equation}
where $P$ is the orbital period and $(t_b-t_a)$ is a time interval corresponding to the integral boundary values of $\nu$. As can be read in more detail in \citet{Kopal1959}, this integral has an analytic solution:

\begin{equation}
    \frac{2\pi(t_b-t_a)}{P} = \left[2\cdot arctan\left(\frac{\sqrt{1 - e}\ sin\left(\frac{\nu}{2}\right)}{\sqrt{1 + e}\ cos\left(\frac{\nu}{2}\right)}\right) - \frac{e\sqrt{1 - e^2}\ sin(\nu)}{1 + e\ cos(\nu)}\right]_{\nu_a}^{\nu_b},
    \label{eq:kepler3}
\end{equation}
written in a slightly different form than in the aforementioned book. It follows naturally that the time interval is a function of eccentricity and argument of periastron (see Equation \ref{eq:true_anom}). What is not immediately apparent from this is the dependence on other properties like inclination and sum of the scaled radii. As described in the main text, these properties appear in Equations \ref{eq:ecl_minima} and \ref{eq:ecl_edges}, which are used to calculate the true anomaly of our special time points that we need to plug into the integral bounds.

An auxiliary angle $\phi_0$ is defined corresponding to the sum of the scaled radii as in Equation \ref{eq:phi0_radius}, which does not a priori have any physical meaning.\footnote{Note that this equation differs by a factor $\sqrt{1-e^2}$ from the one given in \citet{Kopal1959}.} This equation replaces the identical factor in Equation \ref{eq:ecl_edges} to make
\begin{equation}
    \sqrt{1 - sin^2(i)\ cos^2(\phi)} = \sqrt{1 - sin^2(i)\ cos^2(\phi_0)}\ (1 \pm e\ sin(\omega \pm \phi)),
    \label{eq:ecl_edges_phi0}
\end{equation}
which is in fact four equations for four different $\phi$ angles corresponding to the first and last contact of the primary and secondary eclipse. Thus, these angles correspond to four additional special time points where we know the orbital configuration. These equations do not have analytical solutions.

In every formula above we assume spherical stars, or rather, circular projections of the stars, which is a decent approximation in the context of eclipses where the stars are always nearly aligned with the line of sight\footnote{And stars do not rotate too quickly.}. It is in the phases in between the eclipses that we see the stars `side-on' and their deformation as a result of each other's gravity can be significant.

\newpage

\section{Schematic method overview}
\label{apx:overview}

\begin{figure*}
\resizebox{0.95\hsize}{!}
    {\includegraphics[width=\hsize,clip]{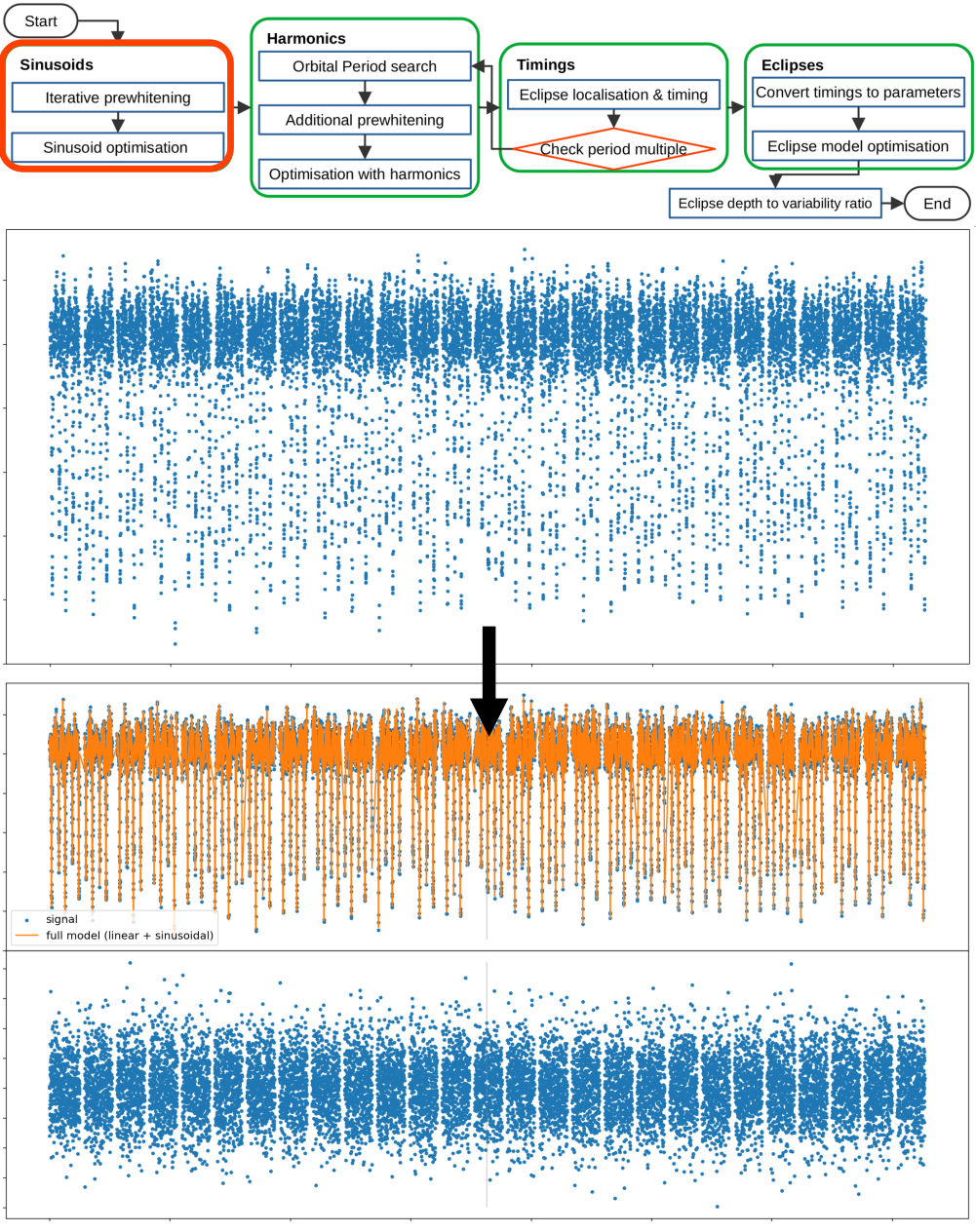}}
    \caption{`Sinusoid' part of the analysis. The input light curve is iteratively pre-whitened, and the resulting parameters are non-linearly optimised. This step results in a mathematical model of the light curve in terms of a sum of sine waves and a linear trend for each sector (separately) and is applicable to any time series. In the second to last panel is the sum of sinusoids light curve model in orange and the bottom panel shows the residuals.}
    \label{fig:schematic_1}
\end{figure*}

\begin{figure*}
\resizebox{0.95\hsize}{!}
    {\includegraphics[width=\hsize,clip]{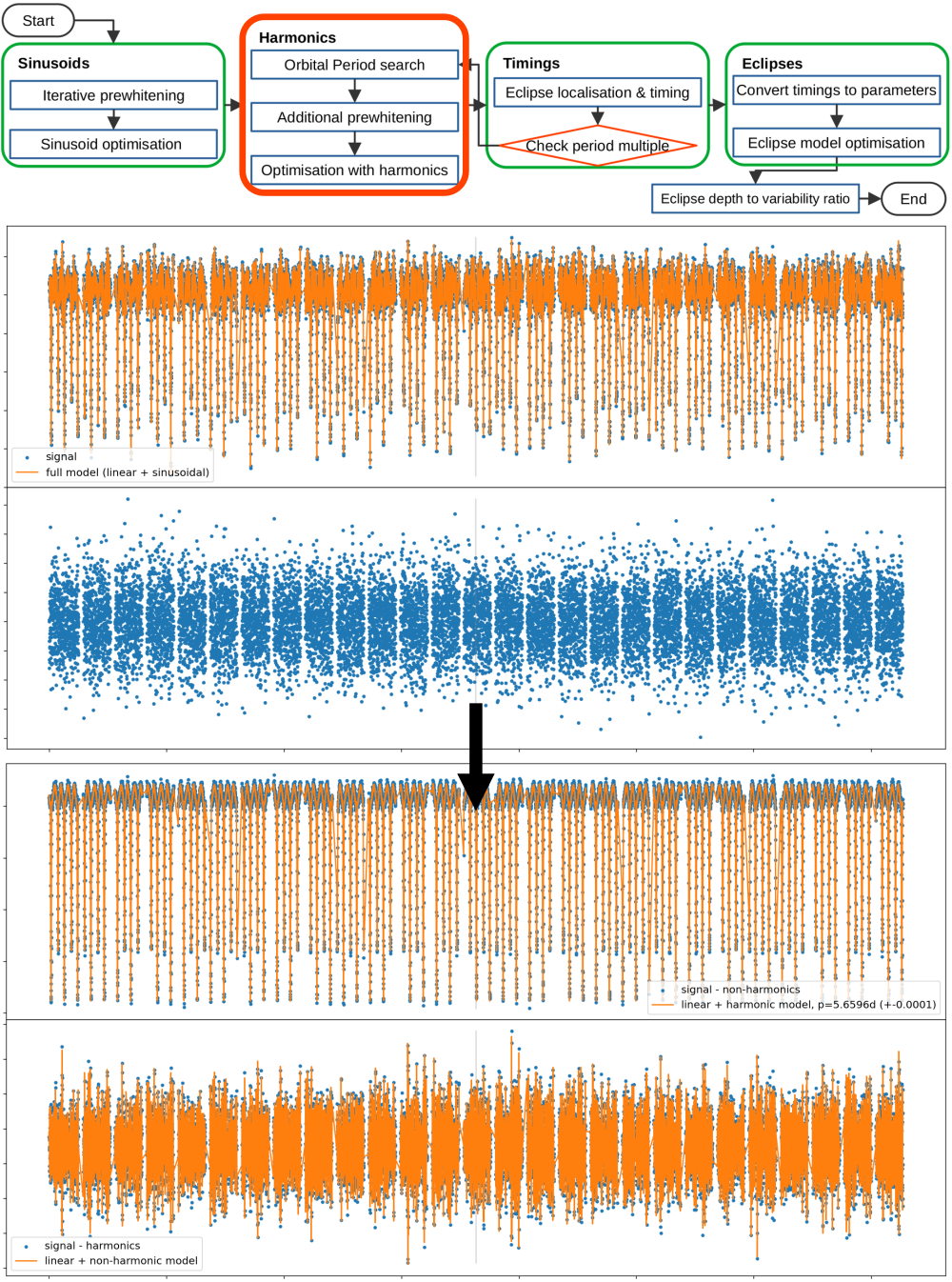}}
    \caption{`Harmonic' part of the analysis.\ The orbital period is determined (if not given), and the orbital harmonics found in the extracted list of frequencies are coupled to the orbital frequency, after which another non-linear optimisation is done. This step is broadly applicable to EBs and other light curves of periodic, non-sinusoidal signal that results in a series of harmonics in the periodogram. The second to last panel shows the orbital harmonic sinusoid model in orange and the bottom panel shows all non-harmonic sinusoids.}
    \label{fig:schematic_2}
\end{figure*}

\begin{figure*}
\resizebox{0.95\hsize}{!}
    {\includegraphics[width=\hsize,clip]{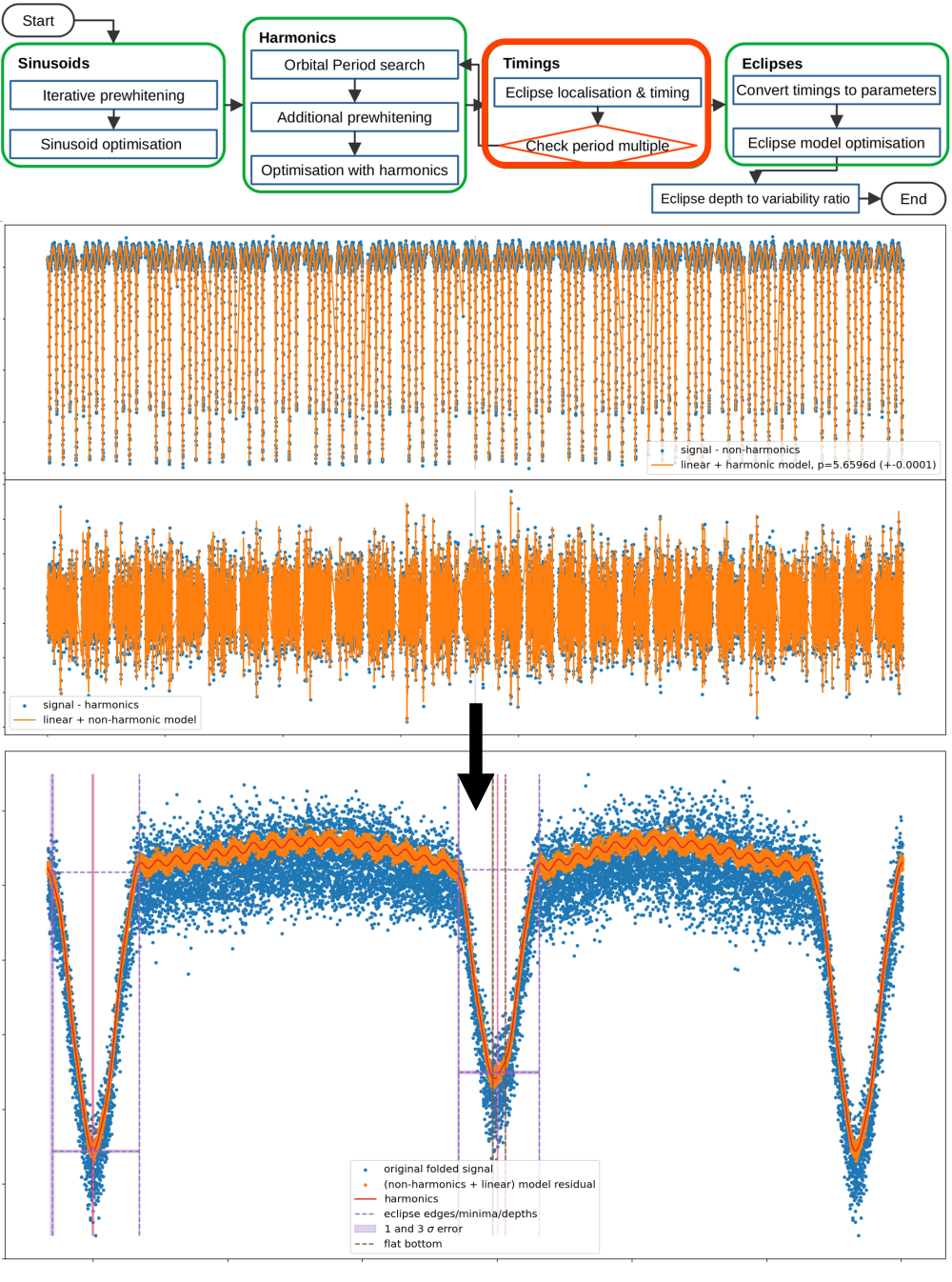}}
    \caption{`Timings' part of the analysis. The eclipses are detected and their positions, durations, and depths are measured through the use of the model of orbital harmonic sinusoids and time derivatives thereof. Also included is a check for the multiplicity of the period that checks for the number of identical eclipses found in one cycle. The bottom panel shows the light curve folded over the orbital period in blue, the harmonic model in red and the non-harmonic model subtracted light curve in orange. The horizontal and vertical dashed lines indicate the eclipse measurements.}
    \label{fig:schematic_3}
\end{figure*}

\begin{figure*}
\resizebox{0.95\hsize}{!}
    {\includegraphics[width=\hsize,clip]{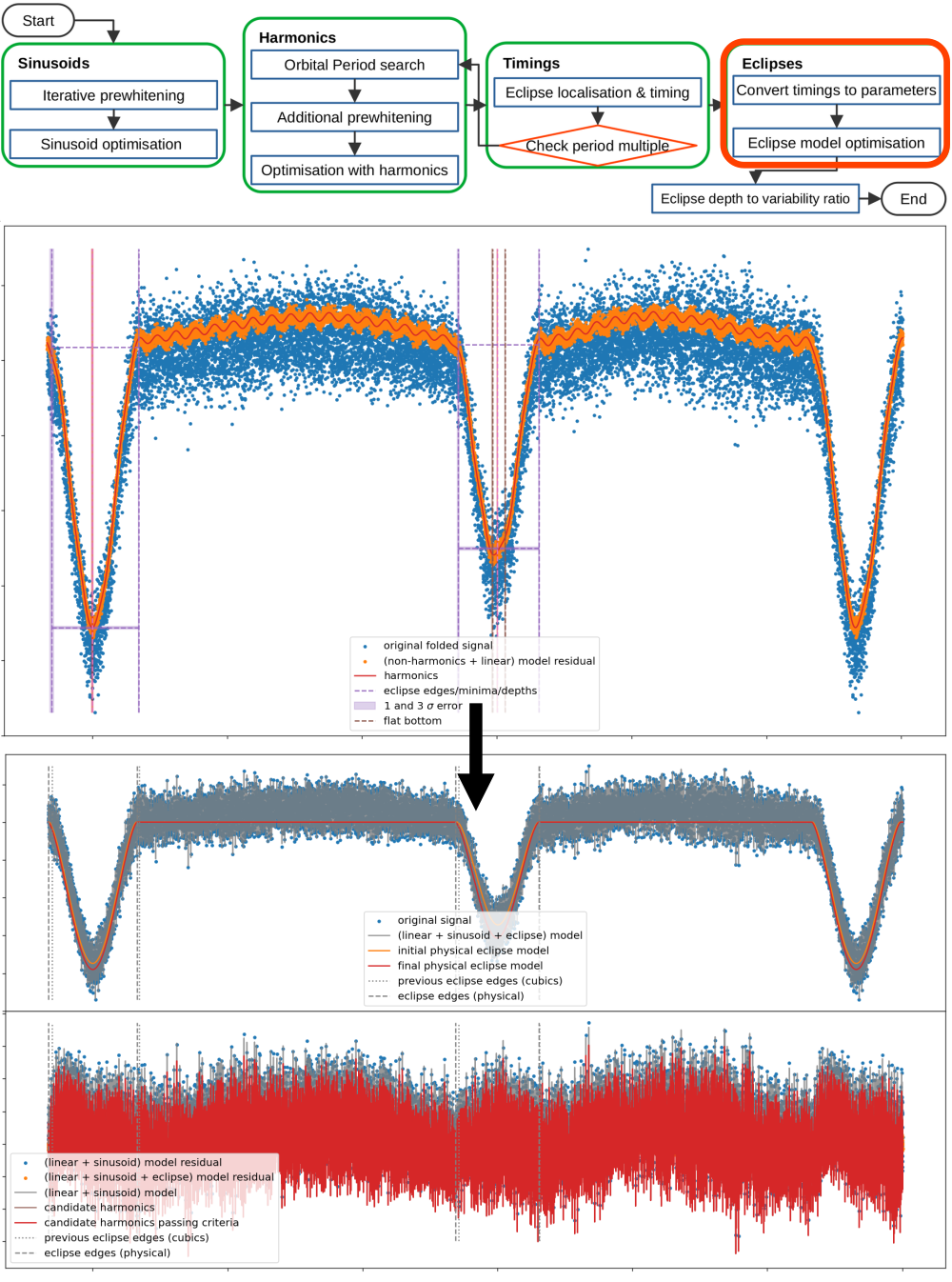}}
    \caption{`Eclipses' part of the analysis. The measurements of eclipse timings and depths are translated to orbital and physical parameters of the system using formulae, and the resulting parameters are used as a starting point for fitting an eclipse model of spherical and uniformly bright stars to the data. The eclipse model is supplemented with a sum of sinusoids, obtained by subtracting the initial eclipse model from the data and pre-whitening the residuals. The second to last panel shows the eclipse model in red and the bottom panel shows the supplemental sinusoid model.}
    \label{fig:schematic_4}
\end{figure*}

\clearpage

\section{Details of the method implementation}
\label{apx:details}

\subsubsection{Iterative pre-whitening}
As already eluded to in the main text, the next frequency, picked by its highest amplitude, is over-sampled in frequency space by a factor of 100 over the default sampling of the full periodogram. To spend as little computation time as possible on calculating periodograms, we use the `fast' implementation of the Astropy Lomb-Scargle periodogram \citep{astropy_ls_fast}. This method extrapolates from a fast Fourier transform and thus is not ideal for high precision. Additionally, it cannot be used on sub-domains in frequency space as it will produce unphysical results. However, since we are using a two-step approach, this poses no problems. The oversampling step makes use of the traditional Lomb-Scargle periodogram, albeit an efficient implementation of it by Cuypers, translated by the authors to Python+Numba from the Fortran implementation in the IvS Python Repository (\href{https://github.com/IvS-KULeuven/IvSPythonRepository/tree/master/timeseries}{GitHub})\footnote{Our translation is tested to be identical in speed.}. The 100 times over-sampled periodogram is only computed for the domain between the point to the left and the point to the right of the topmost point in the original periodogram (the one at the highest amplitude), saving on unnecessary computations. The phase is computed only at the final determined frequency (using the formula from \citealt{Hocke1998}), adding very little extra computational cost. 

Another optimisation in all of the algorithms where sinusoids are iteratively added to or subtracted from a light curve model is to keep a copy of the model as well instead of just keeping a copy of each sinusoid parameter. Calculating sine waves is very time-consuming computationally speaking, so by replacing calculations of sine waves with simple addition and subtraction to the model in memory, we save a lot of computation time. Another benefit is that this step of the iterative process no longer scales in time with the number of sinusoids in the model. The additional step of iterating over closely spaced frequencies does still add a level of dependence on the number of sinusoids, making the overall scaling of iterative pre-whitening with the number of sinusoids slightly more than linear (with the dependence being a function of frequency density in the periodogram). The overall increase in efficiency is large enough that for the data used in testing for the speed (a fairly short light curve), the process even becomes memory-bound instead of CPU-bound, meaning that the speed of the computer memory is now the bottleneck for this part of the process in that scenario. 

\subsubsection{Minimisation with Scipy}

Nelder-Mead \citep[NM;][]{NelderMead1965} is a robust gradient-free method that was tested extensively in this framework, and it performs well in terms of the final objective function value reached. However, it requires many function evaluations and thus is computationally expensive. The optimisation steps in the algorithm are (were) the computational bottleneck, so we looked for an effective way of decreasing this computational cost. Scipy offers a wide range of minimisation methods: in search of a faster but still robust method in our application, we tested all of them. Some methods require a Jacobian or both a Jacobian and a Hessian to be provided, and several cannot handle boundaries. For our application of fitting a sum of sine waves to the light curve, we can compute the partial derivatives constituting the Jacobian and Hessian, and our testing described below refers to this application. Table \ref{tab:minimize} summarises the features of each minimisation method. 

\begin{table}
        \centering
        \caption{Scipy \texttt{minimize} method features.}
        \label{tab:minimize}
        \begin{tabular}{p{2.2cm} p{0.4cm} p{0.4cm} p{0.4cm} p{0.4cm} p{0.4cm} l}
        \hline
        \rotatebox{45}{Method name} & \rotatebox{45}{Bounds} & \rotatebox{45}{Constraints} & \rotatebox{45}{Requires Jacobian} & \rotatebox{45}{Requires Hessian} & \rotatebox{45}{Can use Jacobian} & \rotatebox{45}{Can use Hessian} \\
        \hline
        \texttt{Nelder-Mead} & $\blacksquare$ &  &  &  &  &  \\
        \texttt{Powell} & $\blacksquare$ &  &  &  &  &  \\
        \texttt{CG} &  &  &  &  & $\blacksquare$ &  \\
        \texttt{BFGS} &  &  &  &  & $\blacksquare$ &  \\
        \texttt{Newton-CG} &  &  & $\blacksquare$ &  & $\blacksquare$ & $\blacksquare$ \\
        \texttt{L-BFGS-B} & $\blacksquare$ &  &  &  & $\blacksquare$ &  \\
        \texttt{TNC} & $\blacksquare$ &  &  &  & $\blacksquare$ &  \\
        \texttt{COBYLA} &  & $\blacksquare$ &  &  &  &  \\
        \texttt{SLSQP} & $\blacksquare$ & $\blacksquare$ &  &  & $\blacksquare$ &  \\
        \texttt{trust-constr} & $\blacksquare$ & $\blacksquare$ &  &  & $\blacksquare$ & $\blacksquare$ \\
        \texttt{dogleg} &  &  & $\blacksquare$ & $\blacksquare$ & $\blacksquare$ & $\blacksquare$ \\
        \texttt{trust-ncg} &  &  & $\blacksquare$ & $\blacksquare$ & $\blacksquare$ & $\blacksquare$ \\
        \texttt{trust-exact} & $\blacksquare$ &  & $\blacksquare$ & $\blacksquare$ & $\blacksquare$ & $\blacksquare$ \\
    \texttt{trust-krylov} & $\blacksquare$ &  & $\blacksquare$ & $\blacksquare$ & $\blacksquare$ & $\blacksquare$ \\
        \hline
        \end{tabular}
        \tablefoot{The methods of the function \texttt{scipy.optimize.minimize} have different features and requirements in terms of using the Jacobian or Hessian matrix and in being able to handle bounds or constraints.}
\end{table}

Although it is not strictly necessary for every minimisation step in the algorithm, we do want to impose boundaries on our parameters. We tested methods that have the option to numerically compute gradients with and without the analytical Jacobian to see what the performance gain would be. Both Powell and Sequential Least SQuares Programming (SLSQP) (without Jacobian) are more than an order of magnitude faster than NM, but both diverge from the true solution instead of converging on it. SLSQP becomes slightly faster when the Jacobian is provided and diverges slightly less dramatically. Method trust-constr (without Jacobian) takes a factor of a few longer than NM, while reaching worse cost-function values. However, when provided with the Jacobian it redeems itself by becoming faster than NM by a factor of 2 and obtaining the same cost-function value. Unfortunately providing the Hessian as well makes it much slower again; this could be due to the high computational cost of the Hessian function. Methods Conjugate Gradient (CG), Broyden–Fletcher–Goldfarb–Shanno (BFGS), L-BFGS-B, and Truncated Newton Constrained (TNC) perform similar to NM in terms of cost function, but only BFGS and L-BFGS-B are faster by a noticeable factor of a few without the Jacobian. CG does not benefit time-wise from the Jacobian, but the three others do: they reach similar speeds of more than an order of magnitude faster than NM. Also, Newton-CG, which required the Jacobian to function, falls into the same performance category as the three others in terms of both cost function and speed. It does not benefit meaningfully from the Hessian. All four of the methods that require the Hessian take a longer time than NM, by a factor of more than 2. Since CG, BFGS, and Newton-CG cannot handle bounds or constraints, L-BFGS-B and TNC go on to a second round of testing. Constrained Optimization by Linear Approximation (COBYLA) also makes it to the second round by being fast and only marginally worse in terms of the cost function, and its lack of support for bounds can be mitigated by its support for constraints.

In a more demanding test with a longer light curve, the COBYLA method (which does not use a Jacobian) shows comparable speed to L-BFGS-B and TNC (with Jacobian), but diverges and does not return a sensible result. L-BFGS-B and TNC (both compared with and without Jacobian) perform very similarly: the final choice for L-BFGS-B is made based on its faster speed, and even though the cost-function value it reaches is not quite as good, the difference in the actual light curve model is imperceptibly small.

\subsubsection{Analytical gradients}

As mentioned above, the optimisation makes good use of analytical gradient information. We include their mathematical form here. Our objective function is the negative log-likelihood:

\begin{equation}
    f = -ln(l) = \frac{n}{2}\left[ln\left(\frac{2\pi}{n}\sum_i\left(y_i - \hat{y}_i\right)^2\right) + 1\right],
\end{equation}with symbol definitions as in Appendix \ref{apx:formulae}. The Jacobian is defined as 

\begin{equation}
    J = \left[\frac{\partial f}{\partial \theta_1}, \frac{\partial f}{\partial \theta_2}, ..., \frac{\partial f}{\partial \theta_n}\right], 
\end{equation}where each $\theta_j$ stands for one of the model parameters. Using the generalised product rule the derivatives can be written as

\begin{equation}
    \frac{\partial f}{\partial \theta_j} = \sum_i \frac{\partial f}{\partial \hat{y}_i} \frac{\partial \hat{y}_i}{\partial \theta_j},
\end{equation}

\begin{equation}
    \frac{\partial f}{\partial \hat{y}_i} = \frac{-n \left(y_i - \hat{y}_i\right)}{\sum_i\left(y_i - \hat{y}_i\right)^2} ,
\end{equation}

\begin{equation}
    \frac{\partial \hat{y}_i}{\partial \theta_j} = \frac{\partial}{\partial \theta_j} a_j sin\left(2\pi f_j (t_i - t_{mean}) + \phi_j\right).
\end{equation}Making the derivatives explicit, we get
\begin{equation}
    \frac{\partial \hat{y}_i}{\partial f_j} = 2\pi (t_i - t_{mean}) a_j cos\left(2\pi f_j (t_i - t_{mean}) + \phi_j\right),
\end{equation}

\begin{equation}
    \frac{\partial \hat{y}_i}{\partial a_j} = sin\left(2\pi f_j (t_i - t_{mean}) + \phi_j\right),
\end{equation}

\begin{equation}
    \frac{\partial \hat{y}_i}{\partial \phi_j} = a_j cos\left(2\pi f_j (t_i - t_{mean}) + \phi_j\right), 
\end{equation}And for the orbital period,

\begin{equation}
    \frac{\partial \hat{y}_i}{\partial p_{orb}} = \sum_h 2\pi \frac{-n_h}{p_{orb}^2} (t_i - t_{mean}) a_h cos\left(2\pi \frac{n_h}{p_{orb}} (t_i - t_{mean}) + \phi_h\right),
\end{equation}in which we are now summing over all harmonic sinusoids. 
We note that for the eclipse model parameters we cannot compute gradients analytically. However, this does not prevent us from using the gradients that we do know during the optimisation of the eclipse model plus sinusoids. We use the \texttt{scipy.optimize.approx\_fprime} to obtain numerical gradients for the 6 parameters of the eclipse model and fill those in at their respective locations in the Jacobian. This provides a similar, although smaller, speed increase as for the pure sinusoid optimisations.

\subsubsection{Parameter space}
\label{apx:param_space}
The parameter space we work in for the six free parameters in our eclipse model is made up of $e cos(\omega)$, $e sin(\omega)$, $cos(i)$, $\phi_0$, $log_{10}\left(\frac{r_2}{r_1}\right)$, and $log_{10}\left(\frac{sb_2}{sb_1}\right)$. This choice of parameters ensures two things: minimal correlation between the free parameters and a flatter space across the relevant domain. In the case of $e cos(\omega)$, $e sin(\omega)$, and $\phi_0$ both of these reasons apply, while the reduction of correlation is the driving factor compared to their counterparts $e$, $\omega$ and $r_{sum}$. Reducing correlation has a strong effect on the ability of optimisation algorithms to accurately and quickly converge. Another benefit is that our correlation structures, and thus error estimates, in the least correlated parameter space are most reliable and simple to interpret.

Parameters $\frac{r_2}{r_1}$ and $\frac{sb_2}{sb_1}$ have an extremely curved parameter space in the sense that a step of 0.1 has a completely different meaning depending on where in the parameter domain we are situated. Using logarithms, $log_{10}\left(\frac{r_2}{r_1}\right)$ and $log_{10}\left(\frac{sb_2}{sb_1}\right)$, has two main advantages that are similar to the ones mentioned for the reduction in correlation. It makes it simpler for many optimisation algorithms to traverse the parameter space for a more accurate and sometimes speedier result, as well as improving the reliability and interpretability of the error estimates (in that parameter space).

\subsubsection{Translating eclipse timings to eccentricity}
\label{apx:translation}
Here, we describe the steps for obtaining physical parameters from time measurements in more detail. 

Although a multitude of approaches was tested, we settled on one that provides the best combination of speed and accuracy. The initial step is to set $cos(i) = 0$ and compute the approximate $\phi_0$ with Equation \ref{eq:phi0_approx}. Sufficient accuracy is retained when solving Equation \ref{eq:psi} and plugging in the approximate Equations \ref{eq:ecosw} and \ref{eq:esinw}. Iteratively updating $\phi_0$ using exact formulae for the eclipse durations does, however, provide a noticeable improvement in the measurement of the sum of the scaled radii through Equation \ref{eq:phi0_radius}. 

To find inclination, ratio of radii and ratio of surface brightness we construct a cost function based on iteratively solving Equations \ref{eq:ecl_minima} and \ref{eq:ecl_edges_phi0} and plugging the obtained angles into \ref{eq:true_anom} and then \ref{eq:kepler3} to obtain accurate theoretical eclipse times, and computing theoretical depths with Equation \ref{eq:simple_model}. Minimisation is done with a global optimiser from Scipy, called simplicial\footnote{Indeed, simplicial, not simplistic.} homology global optimisation (SHGO), with reasonable boundaries on the parameters for feasible systems. 

Minimisation is performed in the parameter space: $cos(i), log_{10}\left(\frac{r_2}{r_1}\right), log_{10}\left(\frac{sb_2}{sb_1}\right)$. This has the benefit of homogenising the step size across the parameter space, where especially the two parameter ratios suffer from a large difference in step size on either end of their physical ranges. Since we also compute components of eccentricity $e cos(\omega)$ and $e sin(\omega)$, we also need to convert them to $e$ and $\omega$:
\begin{equation}
\begin{split}
    e &= \sqrt{(e\ cos(\omega))^2 + (e\ sin(\omega))^2}, \\
    \omega &= arctan\left(\frac{e\ sin(\omega)}{e\ cos(\omega)}\right),
    \label{eq:e_approx}
\end{split}
\end{equation}
where code-wise an arctan2 function is used instead of arctan to obtain correct angles across the full range. $\phi_0$ is converted to the sum of radii with
\begin{equation}
    r_{sum} = \sqrt{1 - sin^2(i) cos^2(\phi_0)}\ (1 - e^2).
\end{equation}

Error estimation was done with a combination of analytical error formulae and the Monte Carlo approach of importance sampling. The analytical formulae are used as minimum error estimates, but we do not have approximate analytical formulae for all parameters. Furthermore, for the analytical error formulae, we do not have an a priori value for the inclination error, so we estimate one that is informed by the tests on synthetic light curves. Starting with an error estimate of 0.035 radians in inclination (2 degrees), we use the following formulae as errors on the given subset of parameters:

\begin{equation}
\begin{split}
    \sigma_{ecos(w)} &= \sqrt{\partial_p^2 \sigma_p^2 + \partial_t^2 \left(\sigma_{t,1}^2 + \sigma_{t,2}^2\right) + \partial_i^2 \sigma_i^2}, \\
    \partial_p &= \frac{-\pi(t_2 - t_1) sin^2(i)}{p_{orb}^2 (1 + sin^2(i))}, \\
    \partial_t &= \frac{\pi sin^2(i)}{p_{orb} (1 + sin^2(i))}, \\
    \partial_i &= \frac{2 sin(i) cos(i)}{(1 + sin^2(i))^2},
\end{split}
\end{equation}where we abbreviated the partial derivatives: $\partial_x \equiv \frac{\partial_y}{\partial_x}$.

\begin{equation}
\begin{split}
    \sigma_{esin(w)} &= \sqrt{\partial_p^2 \sigma_p^2 + \sum_{m,n}\partial_\tau^2 \sigma_{\tau,m,n}^2 + \partial_i^2 \sigma_i^2 + \partial_{\phi,0}^2 \sigma_{\phi,0}^2}, \\
    \partial_p &= \frac{-\pi(\tau_{1,2} + \tau_{1,1} - \tau_{2,1} - \tau_{2,1}) sin^2(i) cos^2(\phi_0)}{2 sin(\phi_0) p_{orb}^2 \left(1 + sin^2(i) (1 + cos^2(\phi_0))\right)}, \\
    \partial_\tau &= \frac{\pi sin^2(i) cos^2(\phi_0)}{2 sin(\phi_0) p_{orb} \left(1 - sin^2(i) (1 + cos^2(\phi_0))\right)}, \\
    \partial_i &= \frac{2 sin(i) cos(i) cos^2(\phi_0)}{\left(1 - sin^2(i) (1 + cos^2(\phi_0))\right)^2}, \\
    \partial_{\phi,0} &= \frac{2 sin^2(i) (1 - sin^2(i)) sin(\phi_0) cos(\phi_0)}{\left(1 - sin^2(i) (1 + cos^2(\phi_0))\right)^2},
\end{split}
\end{equation}where the sum goes over primary ($m=1$) and secondary ($m=2$) ingress ($n=1$) and egress ($n=2$). 

\begin{equation}
    \sigma_{e} = \sqrt{cos^2(w) \sigma_{ecos(w)}^2 + sin^2(w) \sigma_{esin(w)}^2},
\end{equation}

\begin{equation}
    \sigma_{w} = \sqrt{\frac{sin^2(w)}{e^2} \sigma_{ecos(w)}^2 + \frac{cos^2(w)}{e^2} \sigma_{esin(w)}^2},
\end{equation}

\begin{equation}
\begin{split}
    \sigma_{\phi,0} &= \sqrt{\partial_p^2 \sigma_p^2 + \sum_{m,n}\partial_\tau^2 \sigma_{\tau,m,n}^2}, \\
    \partial_p &= \frac{\pi (\tau_{1,2} + \tau_{1,1} + \tau_{2,1} + \tau_{2,1})}{2 p_{orb}^2}, \\
    \partial_\tau &= \frac{\pi}{2 p_{orb}},
\end{split}
\end{equation}

\begin{equation}
\begin{split}
    \sigma_{r,sum} &= \sqrt{\partial_e^2 \sigma_e^2 + \partial_i^2 \sigma_i^2 + \partial_{\phi,0}^2 \sigma_{\phi,0}^2}, \\
    \partial_e &= 2 e \sqrt{1 - sin^2(i) cos^2(\phi_0)}, \\
    \partial_i &= \frac{sin(i) cos(i) cos^2(\phi_0) (1 - e^2)}{\sqrt{1 - sin^2(i) cos^2(\phi_0)}}, \\
    \partial_{\phi,0} &= \frac{sin^2(i) sin(\phi_0) cos(\phi_0) (1 - e^2)}{\sqrt{1 - sin^2(i) cos^2(\phi_0)}},
\end{split}
\end{equation}

The error formulae for $e sin(w)$ and $\phi_0$ are found lacking in an earlier version of our synthetic test (Section \ref{sec:testing}), which likely is because they both rely on the measurement of the widths of the eclipses, which can be inaccurate and is more easily disturbed by additional variability. Based on the same test, we constructed simple scaling relations that apply to the minimum error in the concerned parameters. Their functional form is 

\begin{equation}
    minimum\ error = C \cdot exp\left(-\frac{depth_2}{0.3 \sum_h a_h}\right),
\end{equation}where the constant $C$ is 0.2 for $e sin(\omega)$ and 0.05 for $\phi_0$. The specific scaling variable used in the exponent was found to correlate strongest with absolute deviation from the input of these parameters (correlation of -0.45 for $e sin(\omega)$), much more so than secondary eclipse depth alone (correlation of -0.24 for $e sin(\omega)$). The inclusion of the depth of the secondary eclipse is intuitive as shallower secondary eclipses are harder to measure. The other parameter is less intuitive: it is the sum over amplitudes of low and high-frequency harmonics. The sum actually goes over harmonic orders up to and including 5, and harmonic orders above 15. Dependence on the higher frequency harmonics may be understood in terms of the presence of additional variability in the light curve, which reduces the accuracy of the measurements of eclipse edge points. The dependence on the low-frequency harmonics is likely to do with a stronger presence of these harmonics, indicating either very wide eclipses or out-of-eclipse variability, which cross over into the regime where the simplified physics of the applied formulae starts to lose accuracy.

To obtain minimum error values for the two ratios and their logarithms, we compute a scale factor by dividing the computed value for $\sigma_{e}$ by the error estimate obtained for $e$ in the importance sampling (explained below). We take the maximum of the upper and lower error value for this, and multiply the importance sampling errors of the logarithm of radius ratio and the logarithm of surface brightness ratio, again taking the maximum of upper and lower error value. Finally, computing the relative error scaling in log space, and multiplying by the respective parameters in linear space, we obtain the minimal errors for both representations of these two parameters. 

In the importance sampling, we generated input normal distributions of a thousand samples for the measured values of eclipse times (minima, contact, and tangency) within their respective error regions and computed the output of the iterative `translation' scheme outlined above for each of the generated input vectors. This results in a set of output distributions for each computed parameter. Estimates of the errors are then obtained using the highest density interval at 0.683 probability. We note that the resulting error region can exclude the point estimate made before. This is fixed by imposing the minimum errors. The choice not to use the mean of the distribution as a point estimate is made because the distributions are often multi-modal and may not accurately represent the optimal solution.

\newpage

\section{Additional plots}
\label{apx:plots}

\subsubsection{Synthetic test set}
\label{apx:plots_synth}

\begin{figure*}
\resizebox{\hsize}{!}
    {\includegraphics[width=\hsize,clip]{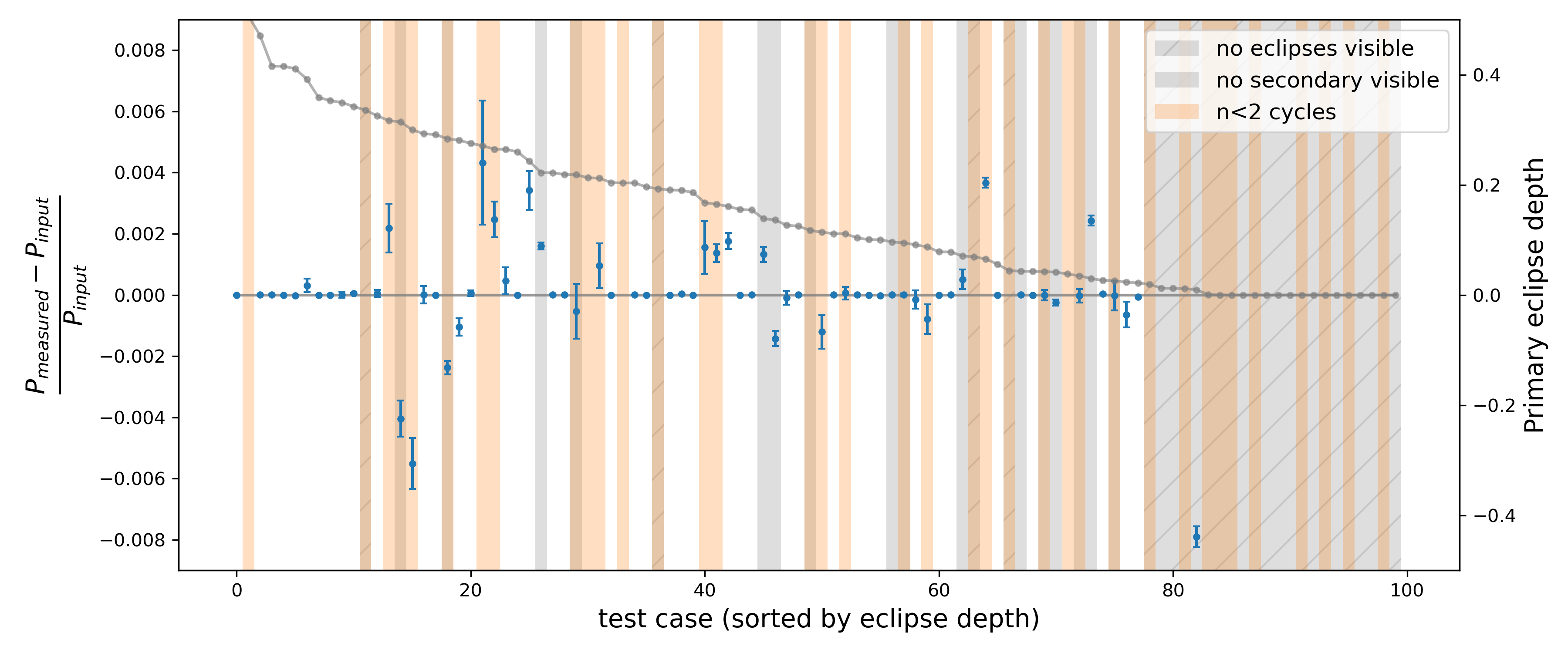}}
    \caption{Performance of the period search algorithm as a function of the primary eclipse depth. Blue points with error bars show the fractional difference between the measured and the input orbital period with SLR error estimates, sorted by descending primary eclipse depth. The grey line with points shows primary eclipse depths, plotted on the right axis. Note that cases shaded grey and hatched do not show eclipses, and cases shaded orange have fewer than three eclipse cycles.}
    \label{fig:period_depths}
\end{figure*}

These additional figures aim to give a more complete view of the output of our analysis of the synthetic light curves discussed in Section \ref{sec:testing}. The first, Figure \ref{fig:period_depths}, shows the same period measurements as Figure \ref{fig:period_cycles}, except now ordered by primary eclipse depth. We note that the eclipse depth does not have any significant influence on the period accuracy except in the most extreme cases where no eclipses can be made out (shaded and hatched dark grey area). 

The majority of the remaining figures show the absolute and relative distributions of the parameter outputs compared to the input values. The relative distributions (denoted with $\chi$) use the estimated error values obtained from the importance sampling and after imposing our error minima (from the formal error calculations and scaling relations). The two colours, blue and orange, in each plot show values obtained at the eclipse timing translation step and the eclipse model fitting step, respectively. 

\begin{figure}
\centering
\includegraphics[width=\hsize]{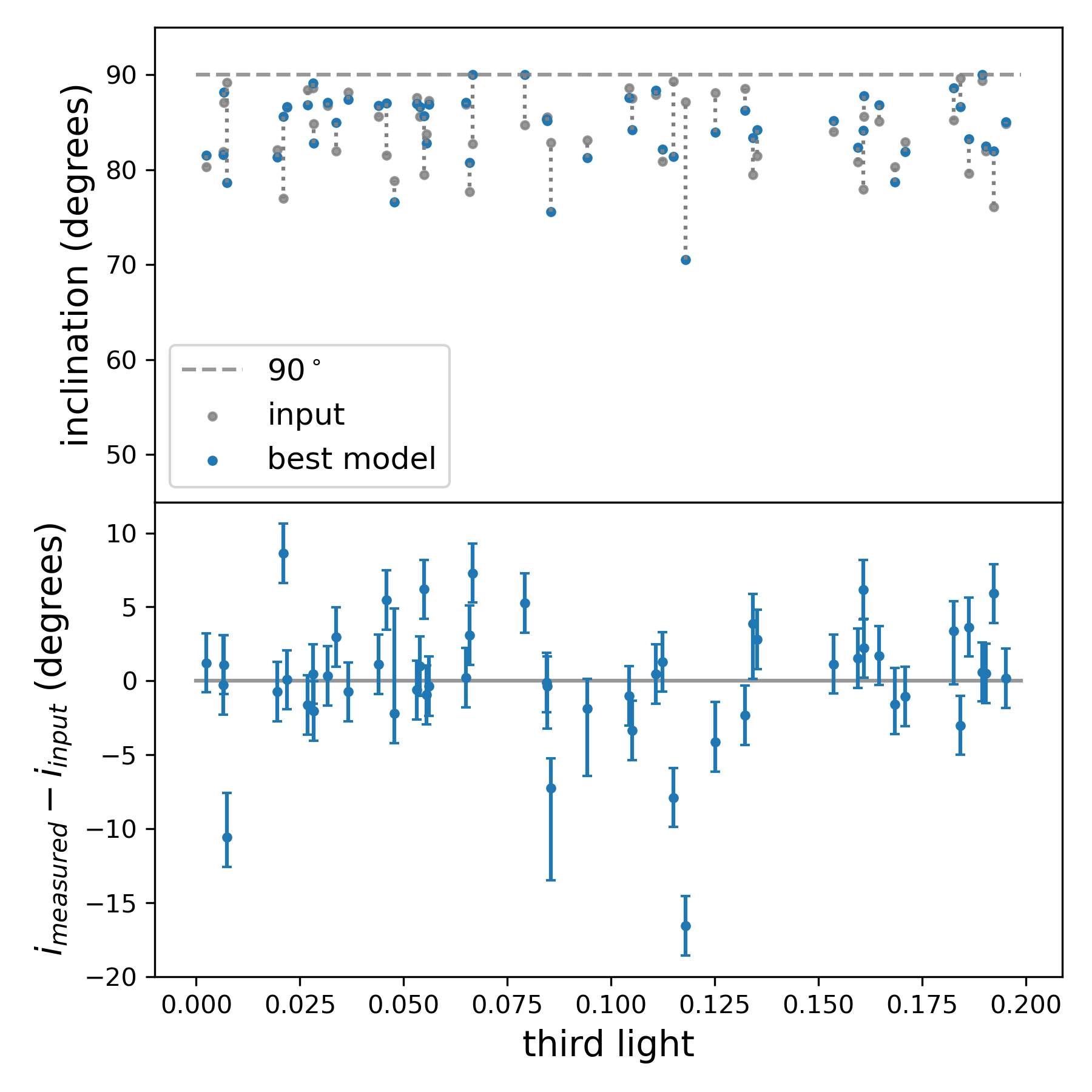}
    \caption{Inclinations (top panel) and deviation from input (bottom panel) versus third light (fraction of total light). Note the absence of an apparent correlation between the accuracy of inclination and third light.}
    \label{fig:incl_tl}
\end{figure}

Finally, we show obtained inclinations versus the third light that was added to the artificial light curves. Third light is not modelled, but Figure \ref{fig:incl_tl} shows no sign of a correlation between third light and inclination. These parameters are mutually degenerate, so such a correlation would be expected: possibly the overall (low) accuracy of our inclination measurements hides this dependence. 

\begin{figure}
\centering
\includegraphics[width=\hsize]{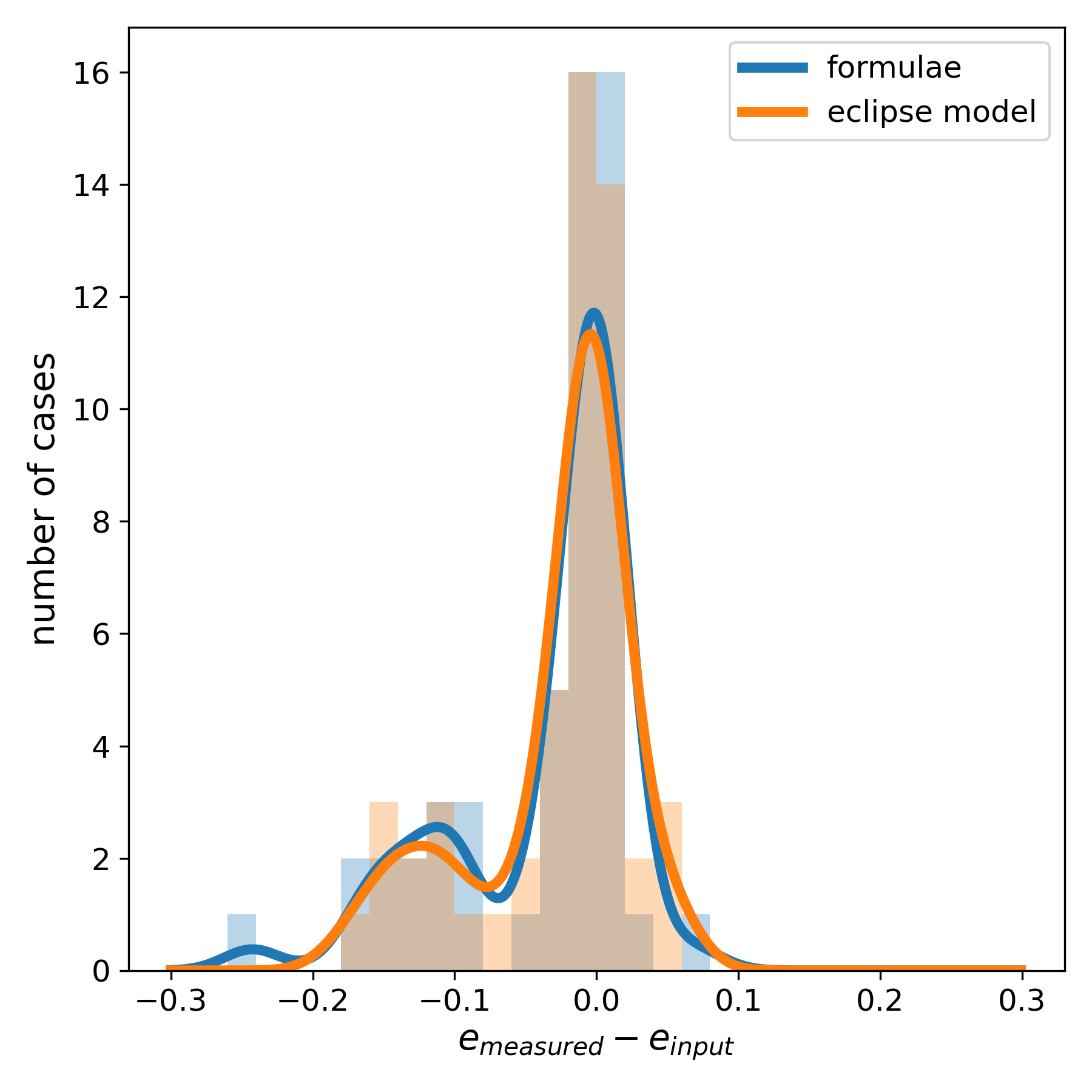}
    \caption{Histogram and KDE of the eccentricity deviations from the input.}
    \label{fig:ecc_dev_abs_b}
\end{figure}

\begin{figure}
\centering
\includegraphics[width=\hsize]{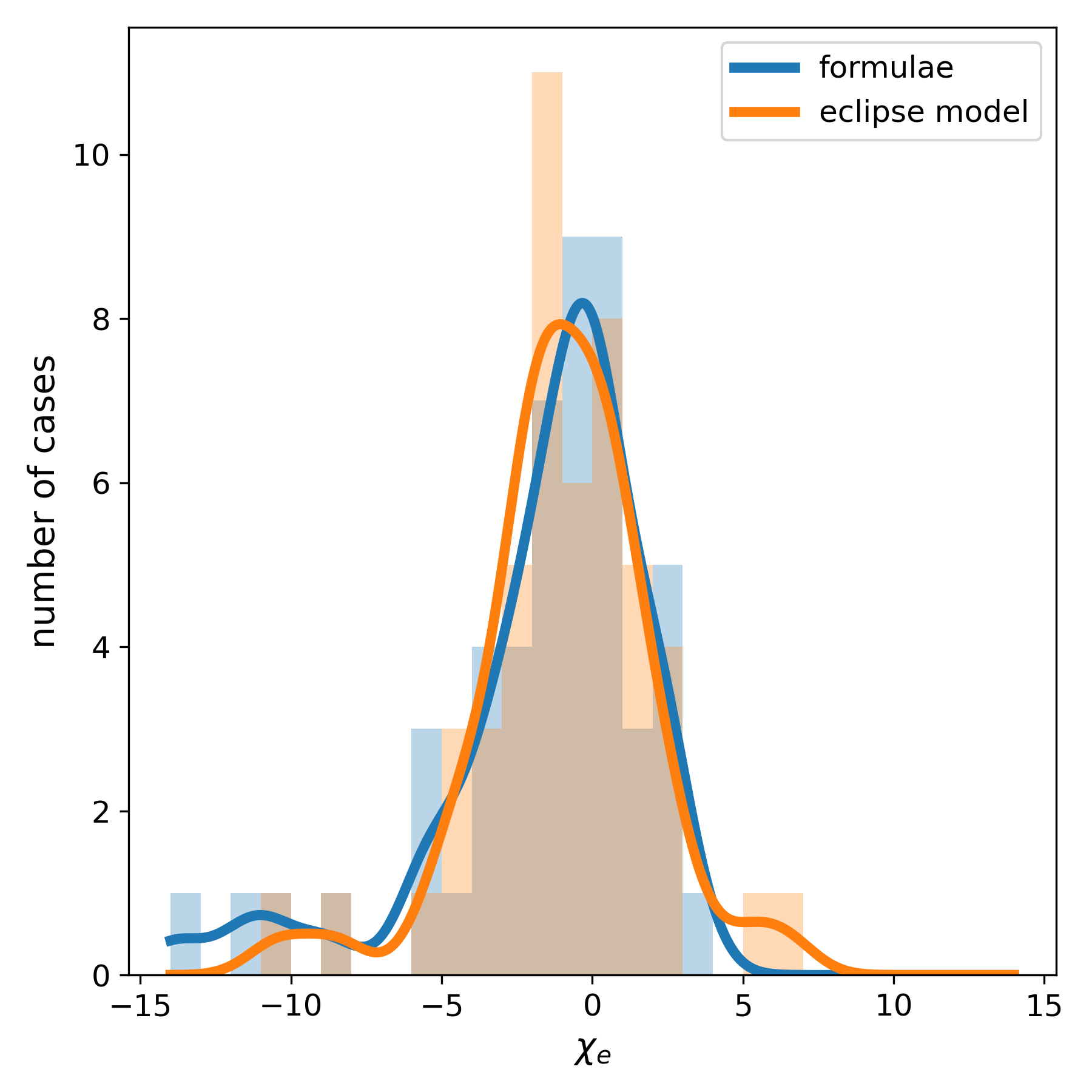}
    \caption{Histogram and KDE of the eccentricity $\chi$ values (deviation from the input divided by the error estimate).}
    \label{fig:ecc_dev_b}
\end{figure}

\begin{figure}
\centering
\includegraphics[width=\hsize]{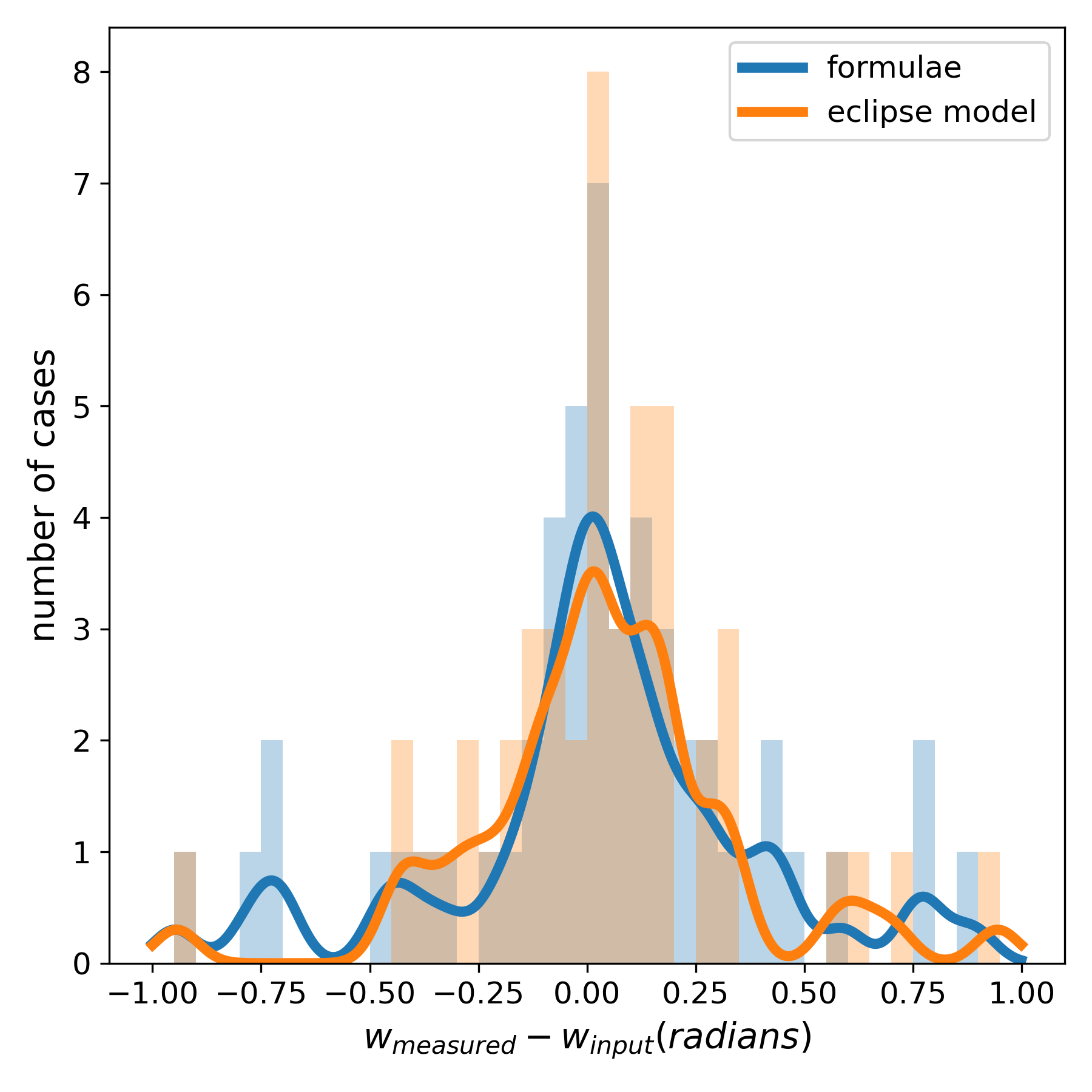}
    \caption{Histogram and KDE of the argument of periastron deviations from the input.}
    \label{fig:omega_dev_abs}
\end{figure}

\begin{figure}
\centering
\includegraphics[width=\hsize]{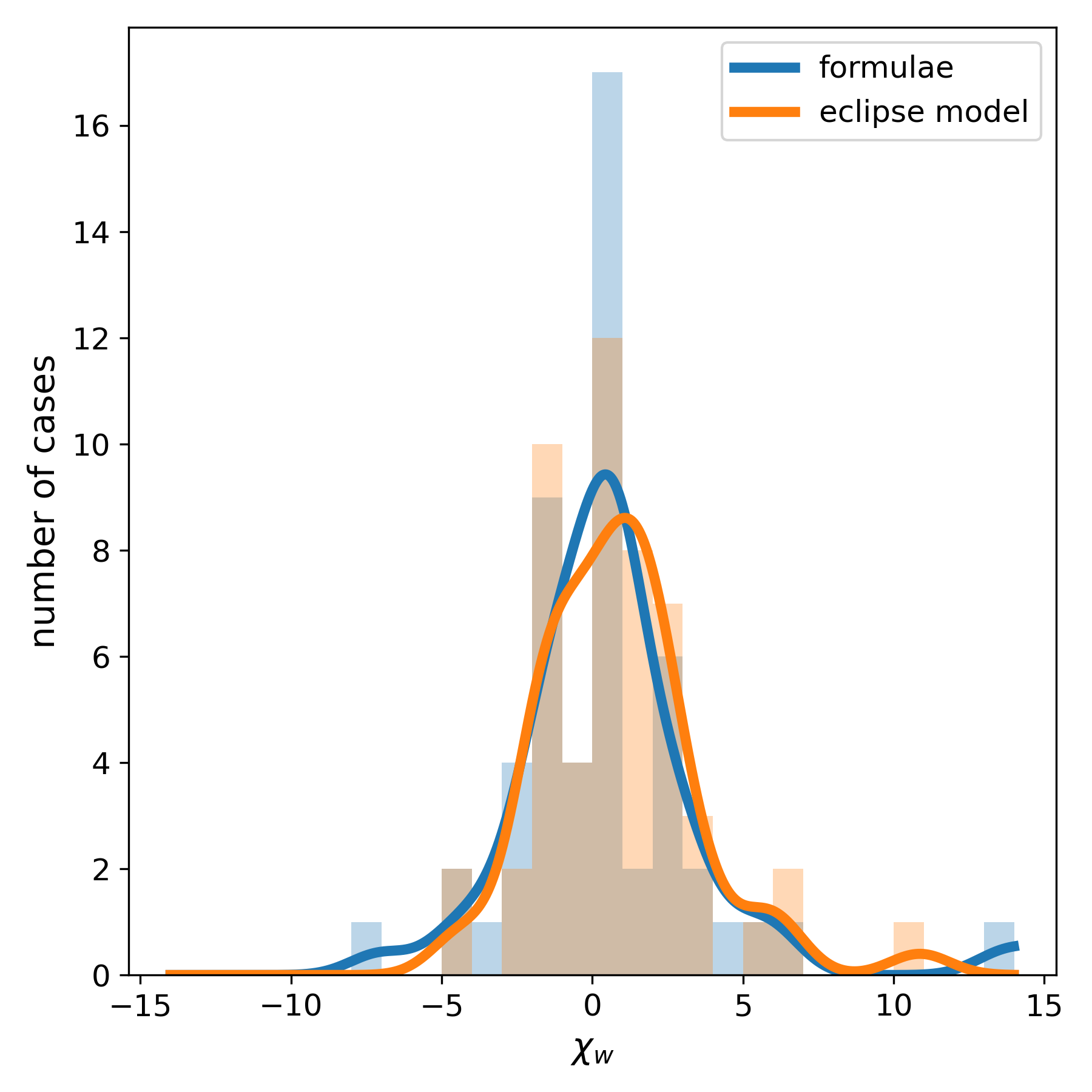}
    \caption{Histogram and KDE of the argument of periastron $\chi$ values (deviation from the input divided by the error estimate).}
    \label{fig:omega_dev}
\end{figure}

\begin{figure}
\centering
\includegraphics[width=\hsize]{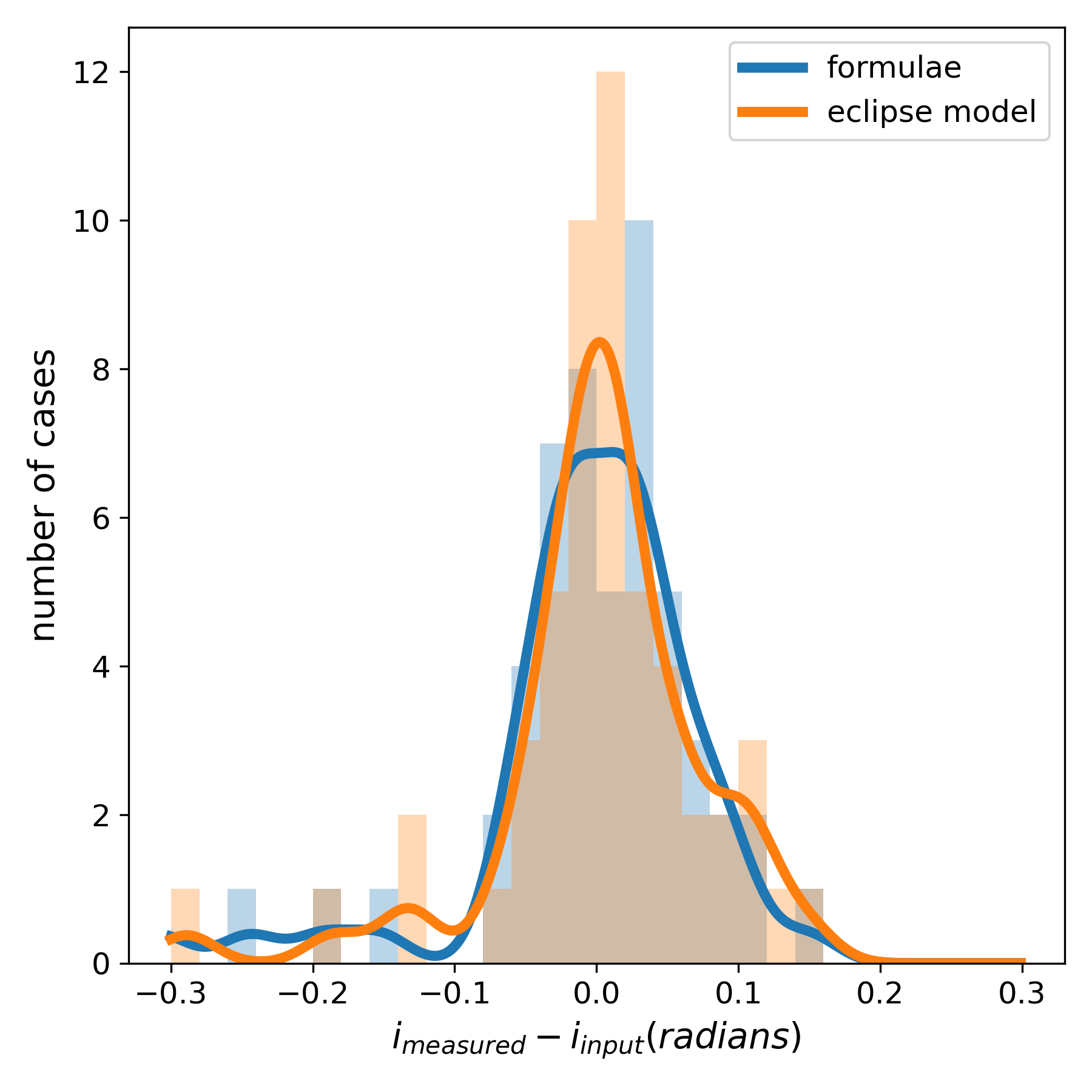}
    \caption{Histogram and KDE of the inclination deviations from the input.}
    \label{fig:incl_dev_abs}
\end{figure}

\begin{figure}
\centering
\includegraphics[width=\hsize]{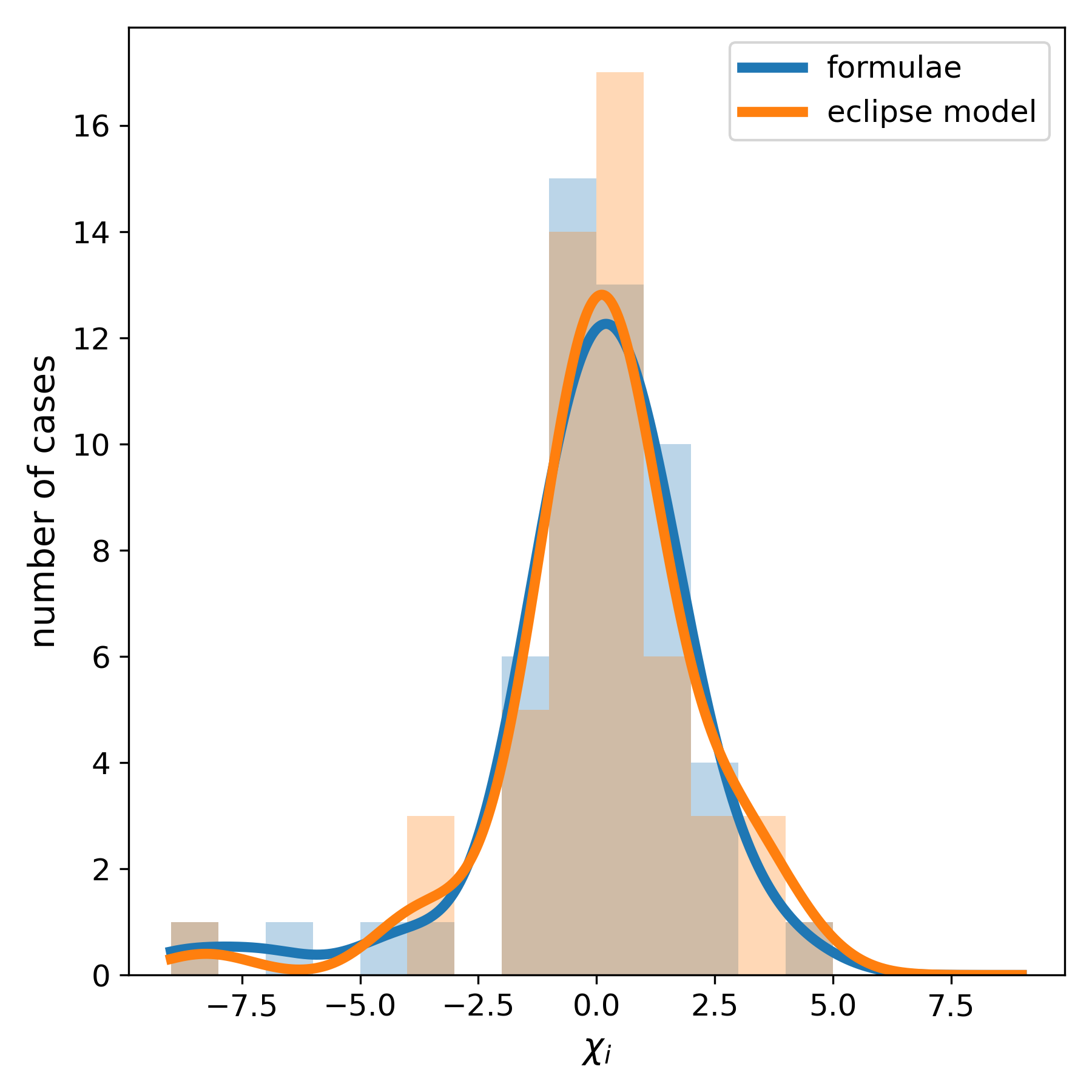}
    \caption{Histogram and KDE of the inclination $\chi$ values (deviation from the input divided by the error estimate).}
    \label{fig:incl_dev}
\end{figure}

\begin{figure}
\centering
\includegraphics[width=\hsize]{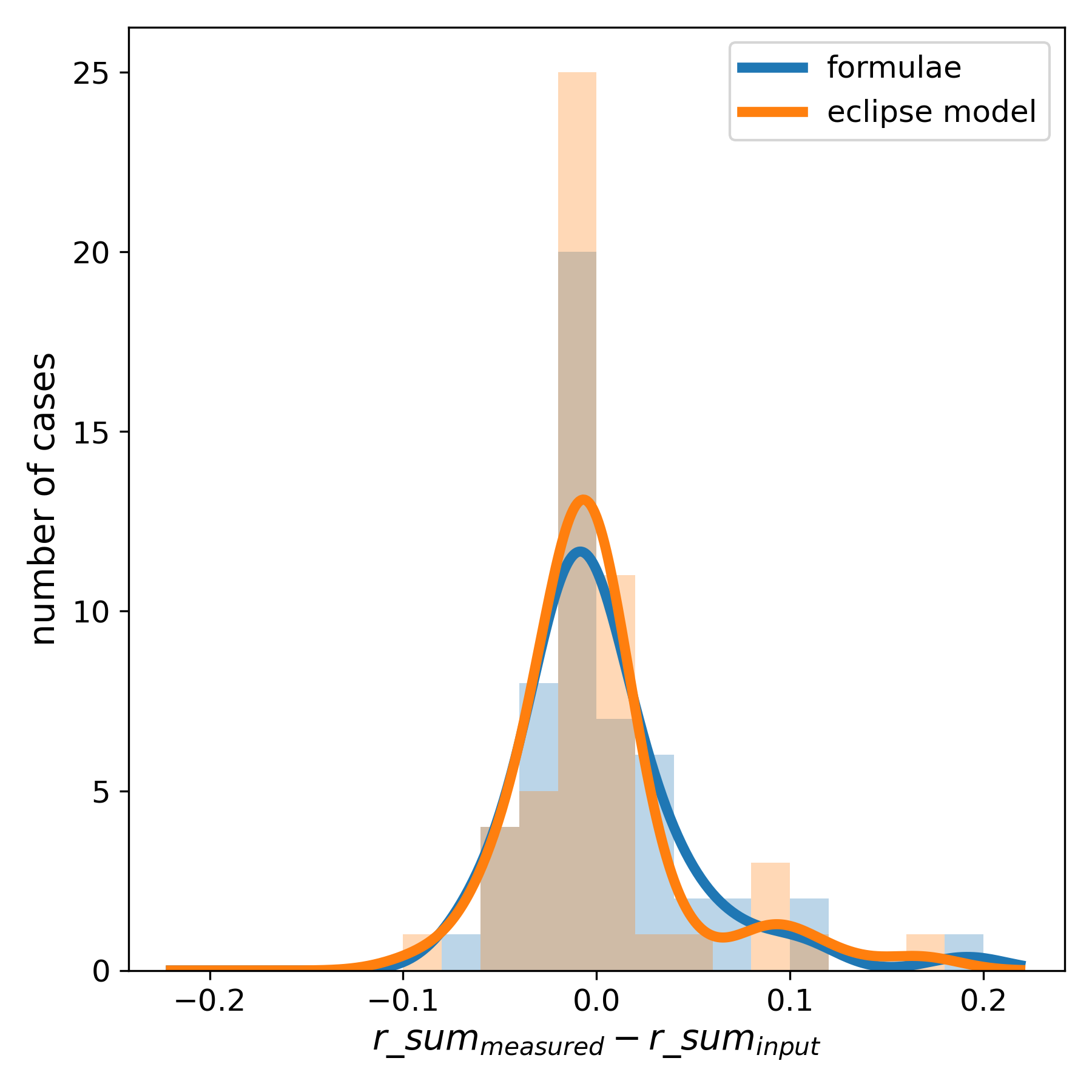}
    \caption{Histogram and KDE of the sum of scaled radius deviations from the input.}
    \label{fig:rsum_dev_abs}
\end{figure}

\begin{figure}
\centering
\includegraphics[width=\hsize]{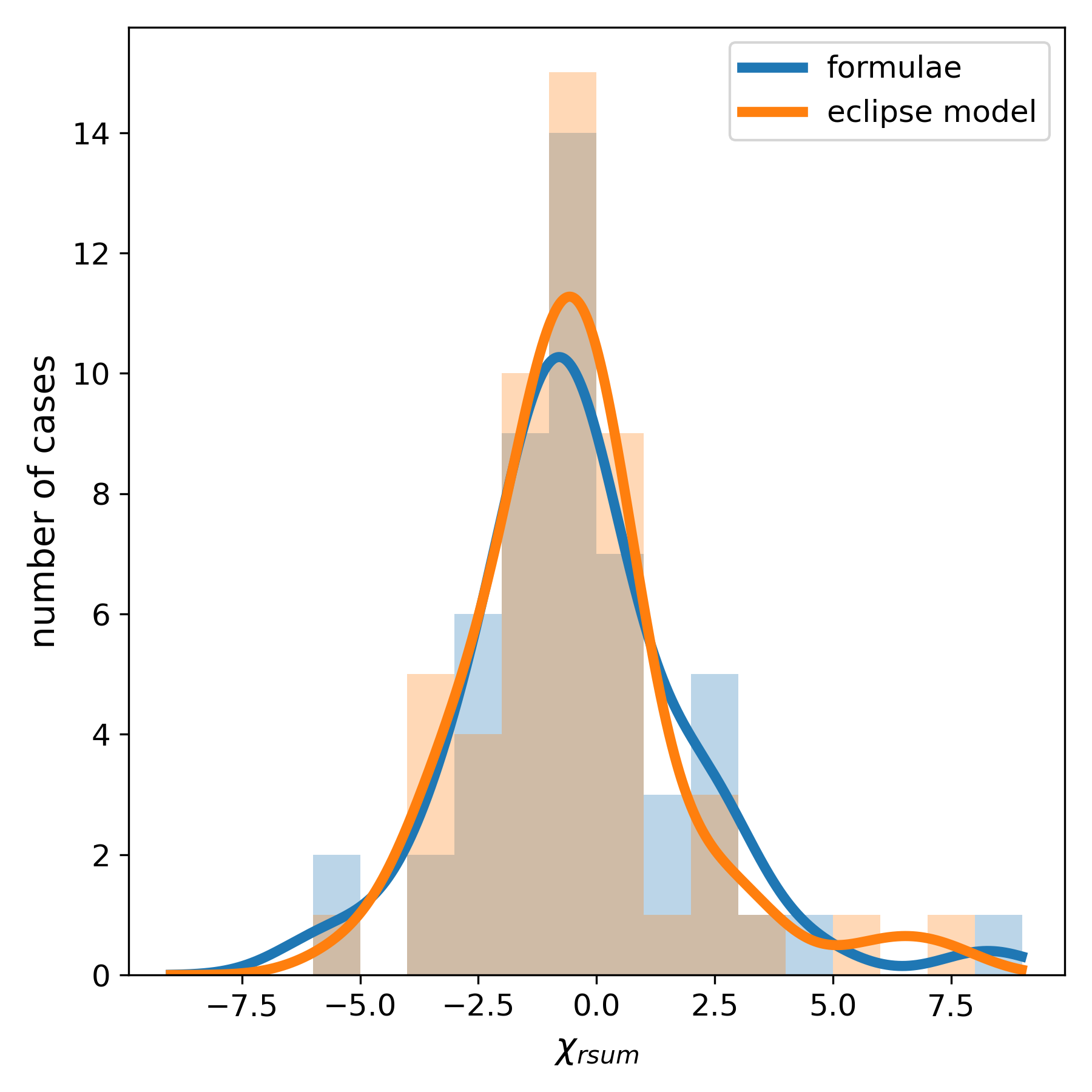}
    \caption{Histogram and KDE of the sum of scaled radius $\chi$ values (deviation from the input divided by the error estimate).}
    \label{fig:rsum_dev}
\end{figure}

\begin{figure}
\centering
\includegraphics[width=\hsize]{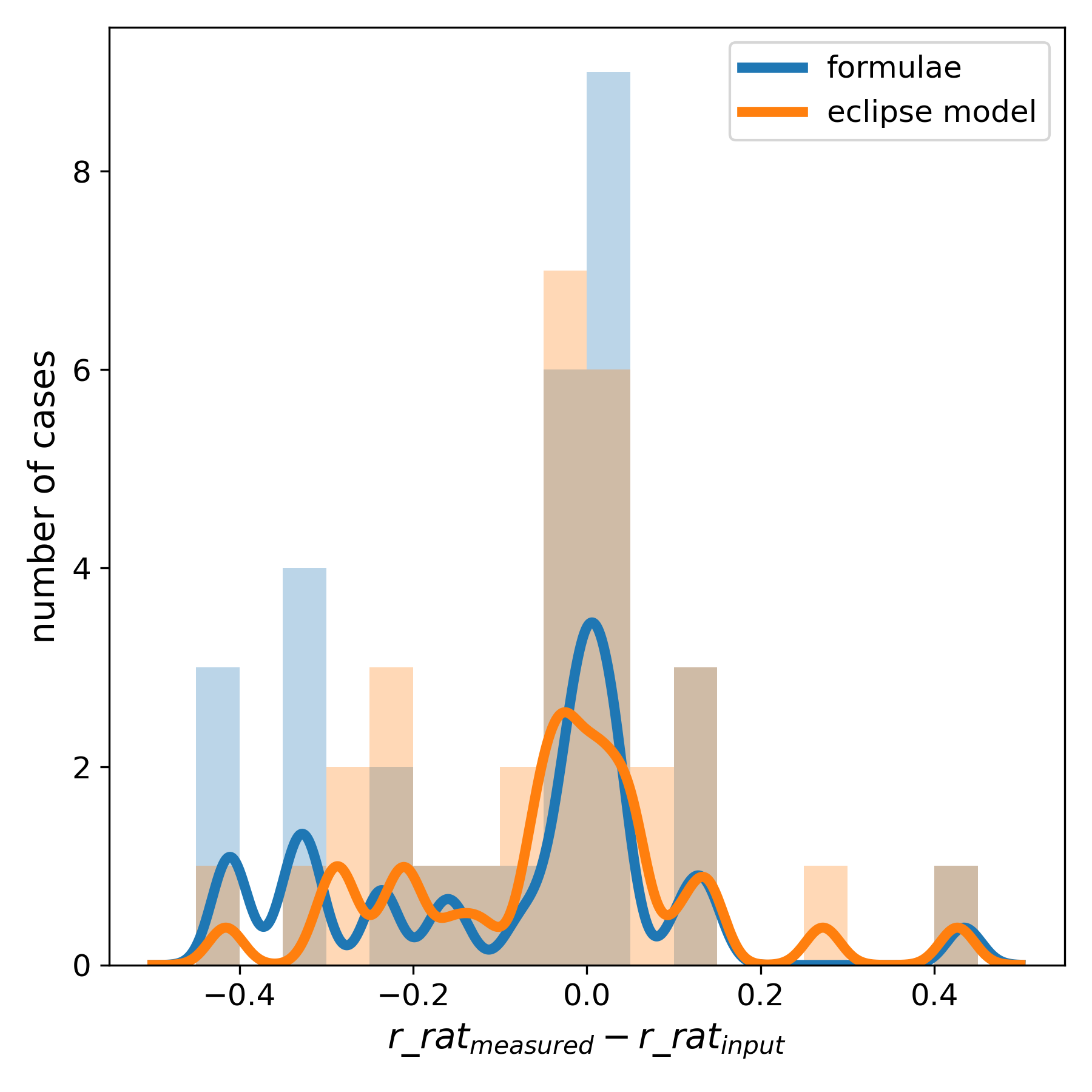}
    \caption{Histogram and KDE of the ratio of radius deviations from the input.}
    \label{fig:rrat_dev_abs}
\end{figure}

\begin{figure}
\centering
\includegraphics[width=\hsize]{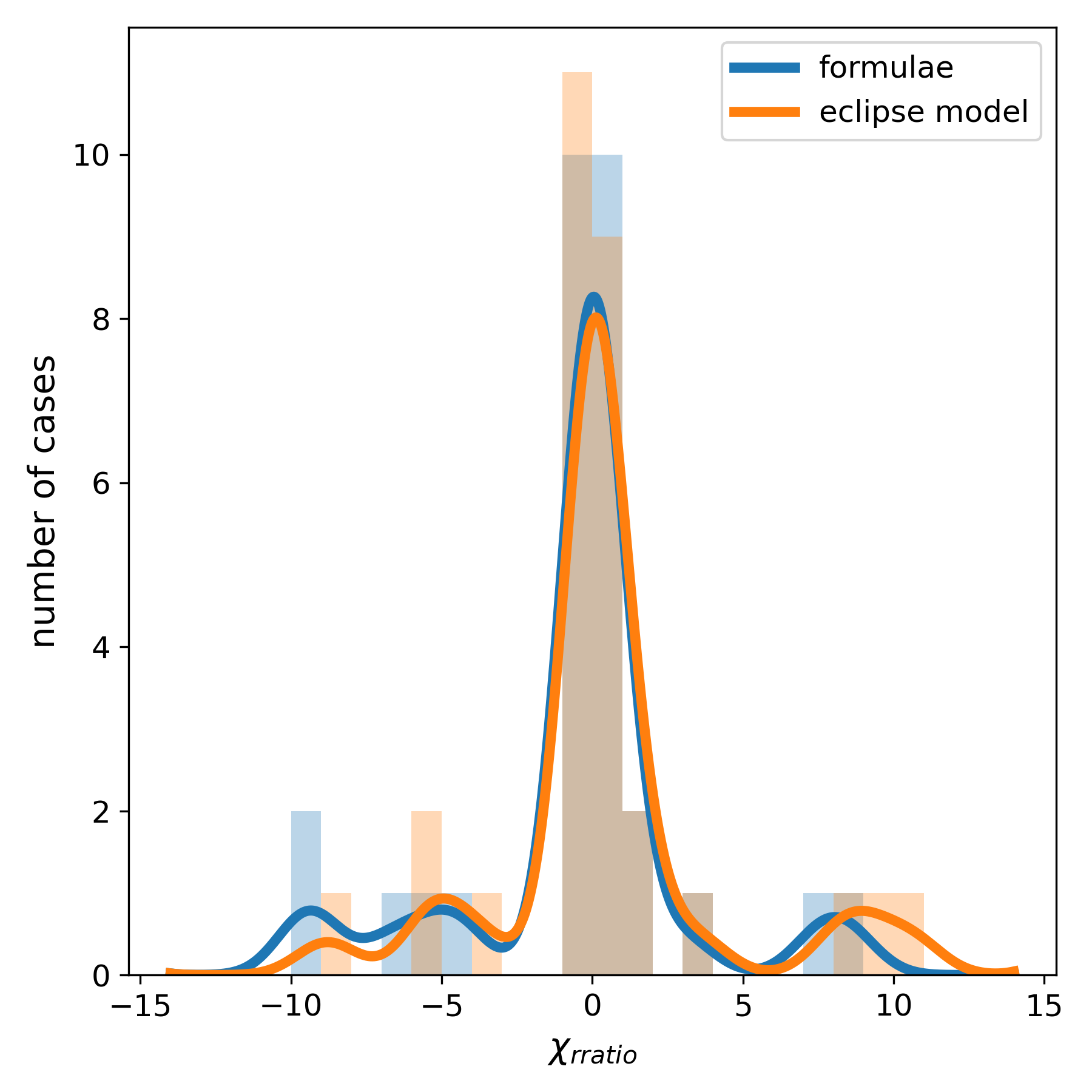}
    \caption{Histogram and KDE of the ratio of radius $\chi$ values (deviation from the input divided by the error estimate).}
    \label{fig:rrat_dev}
\end{figure}

\begin{figure}
\centering
\includegraphics[width=\hsize]{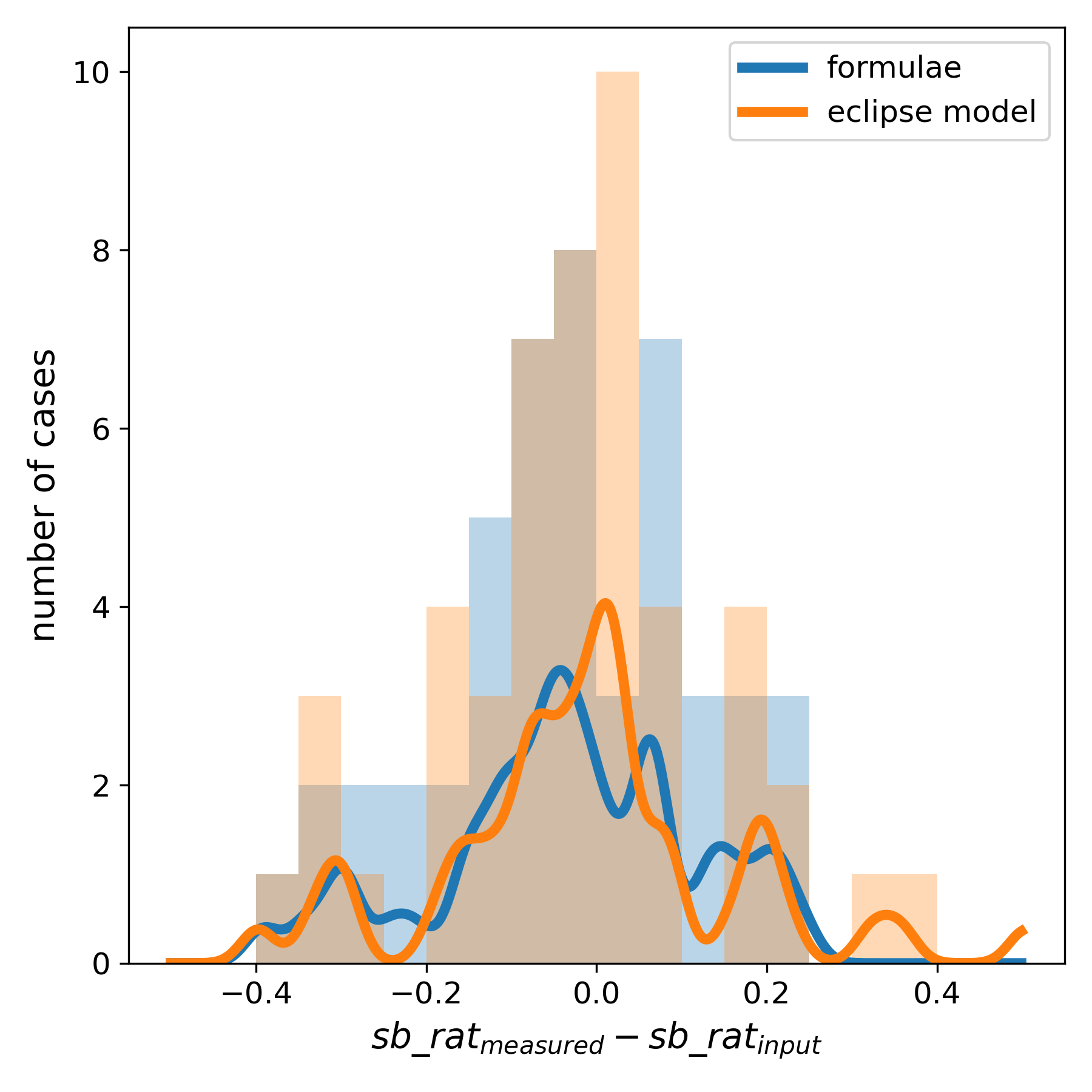}
    \caption{Histogram and KDE of the ratio of surface brightness deviations from the input.}
    \label{fig:sbrat_dev_abs}
\end{figure}

\begin{figure}
\centering
\includegraphics[width=\hsize]{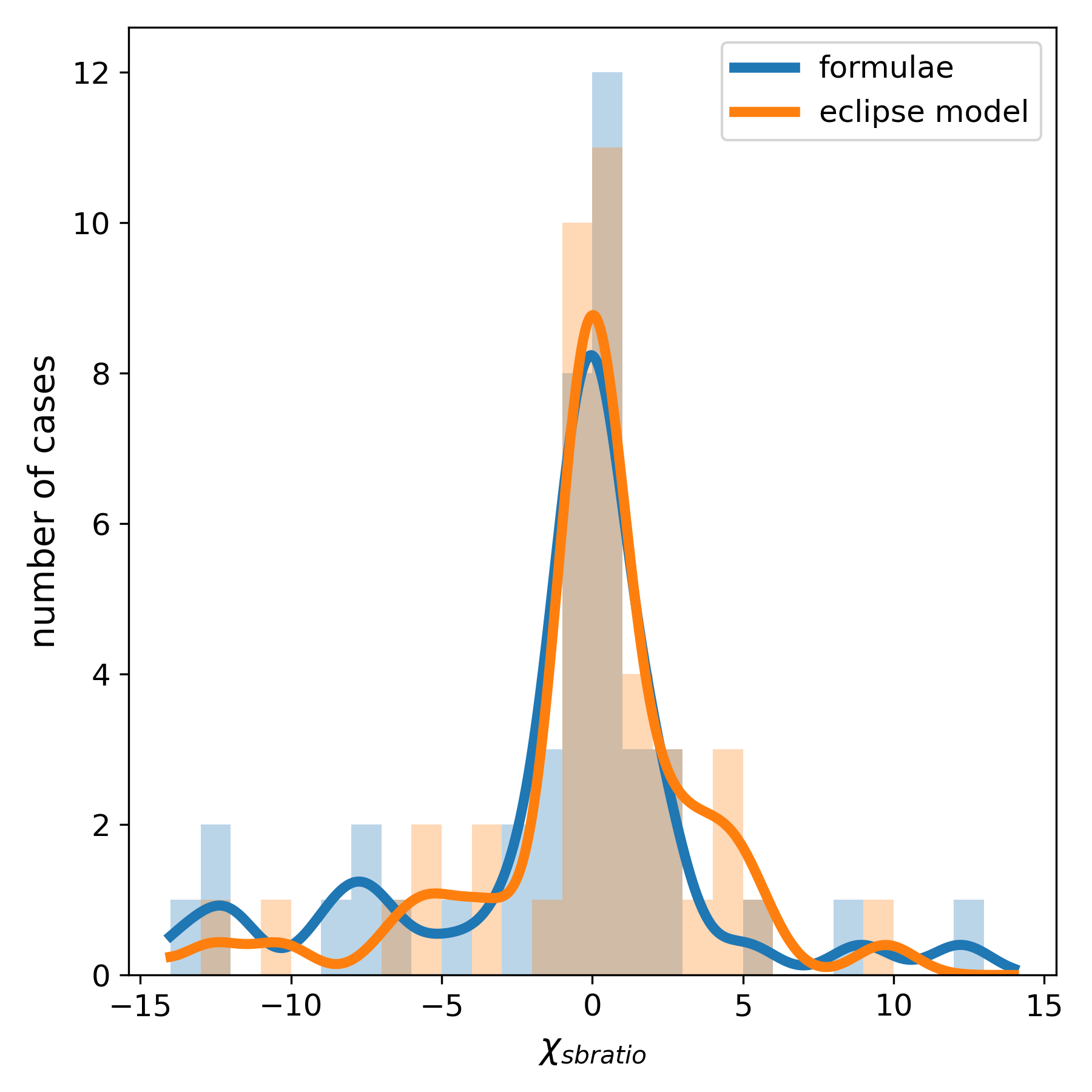}
    \caption{Histogram and KDE of the ratio of surface brightness $\chi$ values (deviation from the input divided by the error estimate).}
    \label{fig:sbrat_dev}
\end{figure}

\begin{figure}
\centering
\includegraphics[width=\hsize]{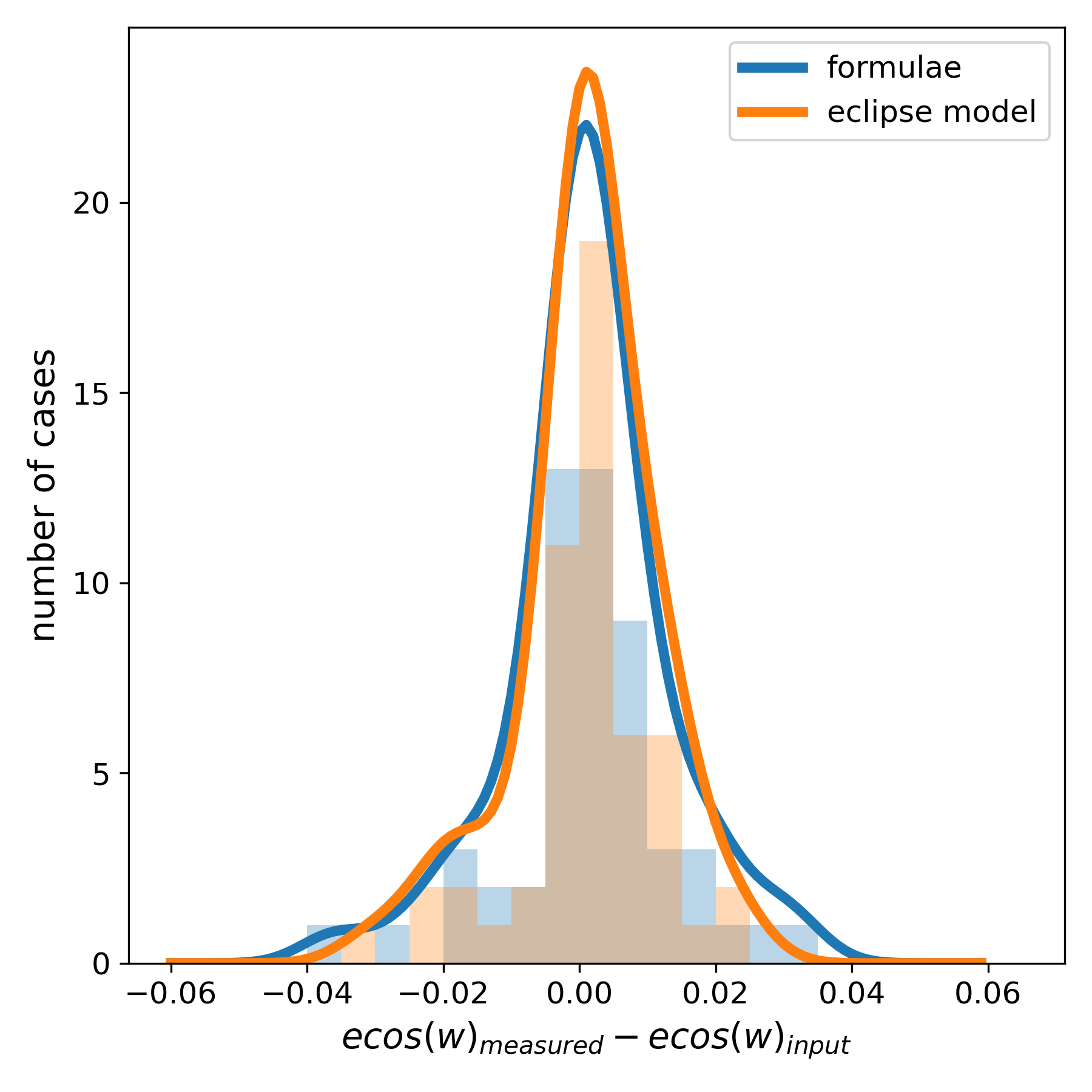}
    \caption{Histogram and KDE of the tangential component of eccentricity deviations from the input.}
    \label{fig:ecosw_dev_abs}
\end{figure}

\begin{figure}
\centering
\includegraphics[width=\hsize]{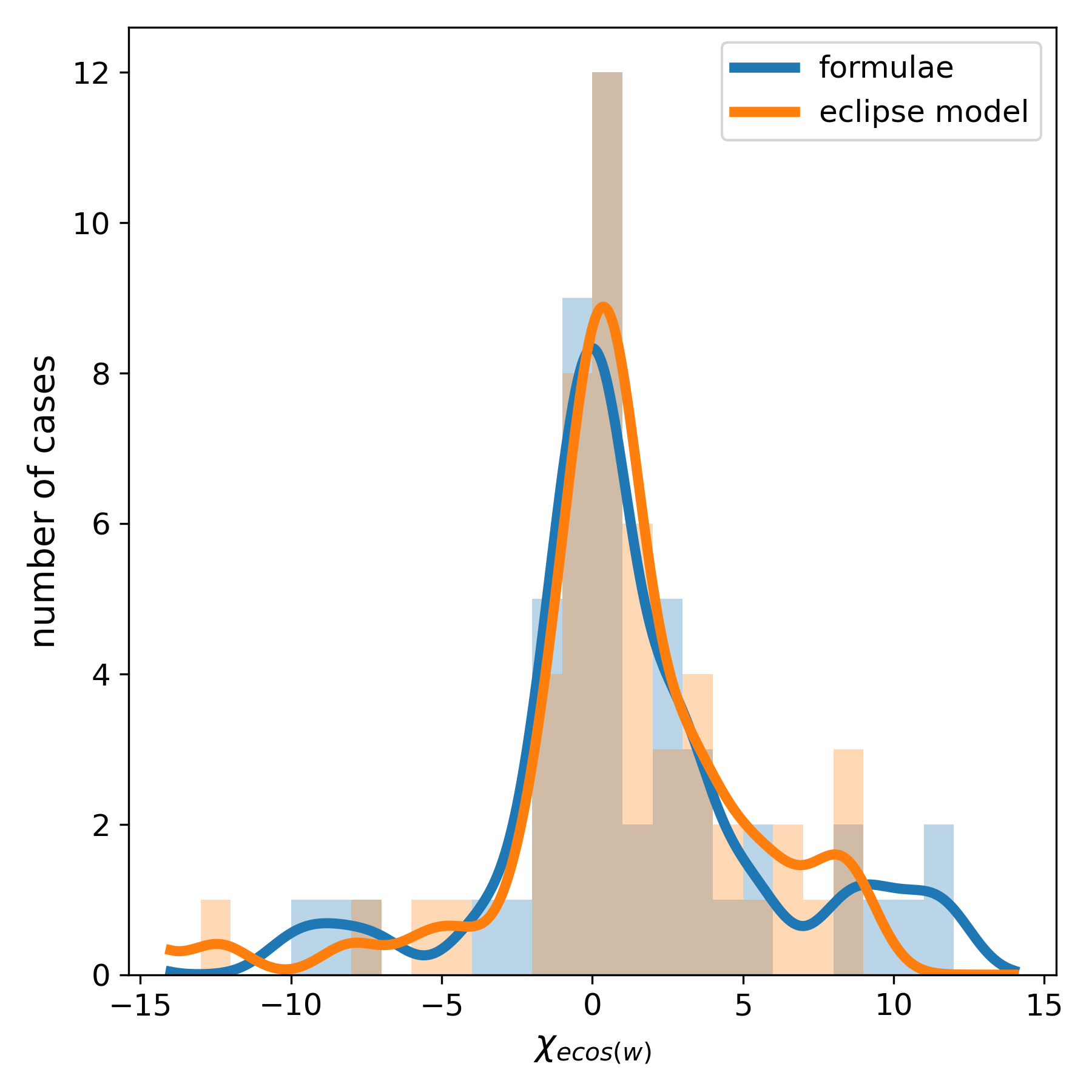}
    \caption{Histogram and KDE of the tangential component of eccentricity $\chi$ values (deviation from the input divided by the error estimate).}
    \label{fig:ecosw_dev}
\end{figure}

\begin{figure}
\centering
\includegraphics[width=\hsize]{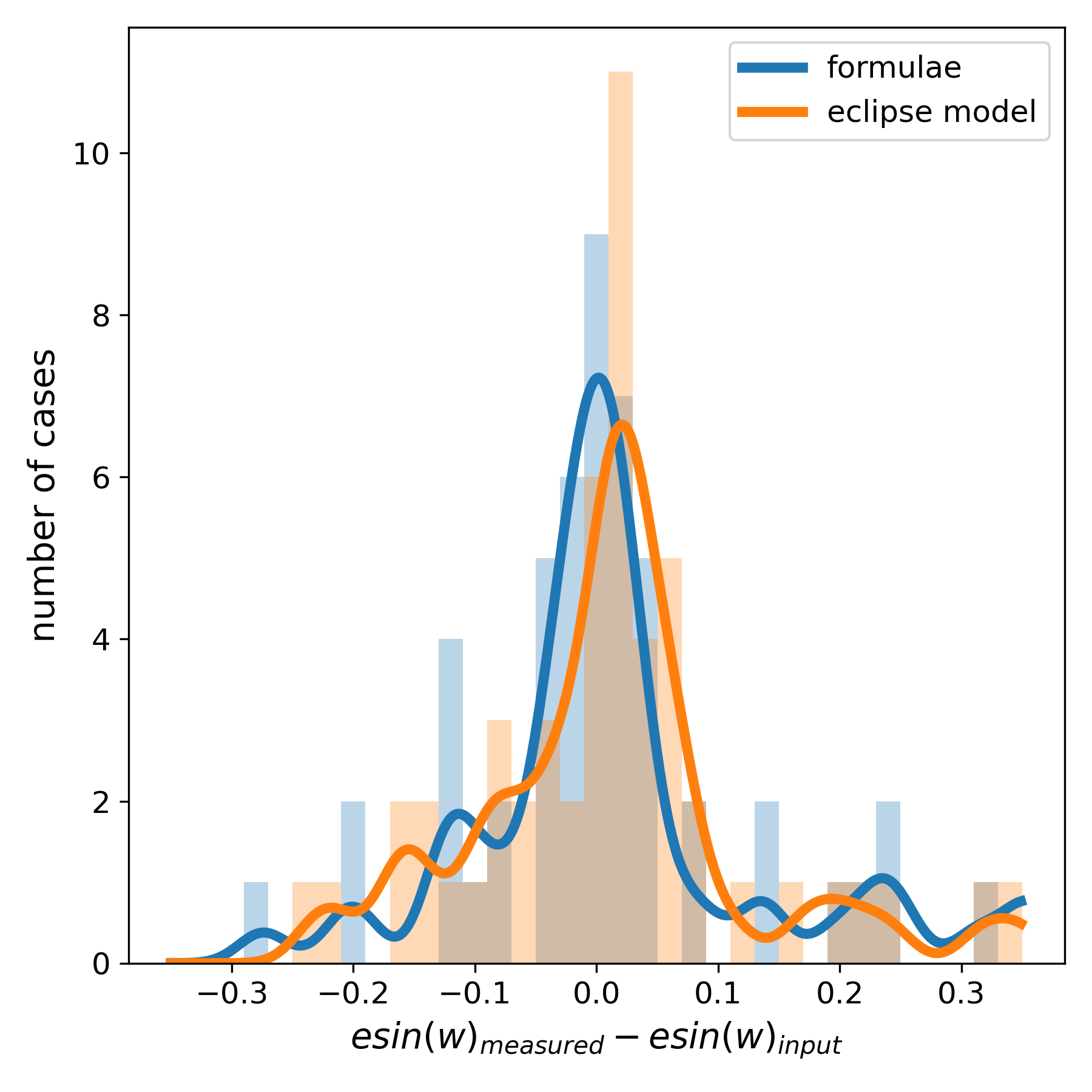}
    \caption{Histogram and KDE of the radial component of eccentricity deviations from the input.}
    \label{fig:esinw_dev_abs}
\end{figure}

\begin{figure}
\centering
\includegraphics[width=\hsize]{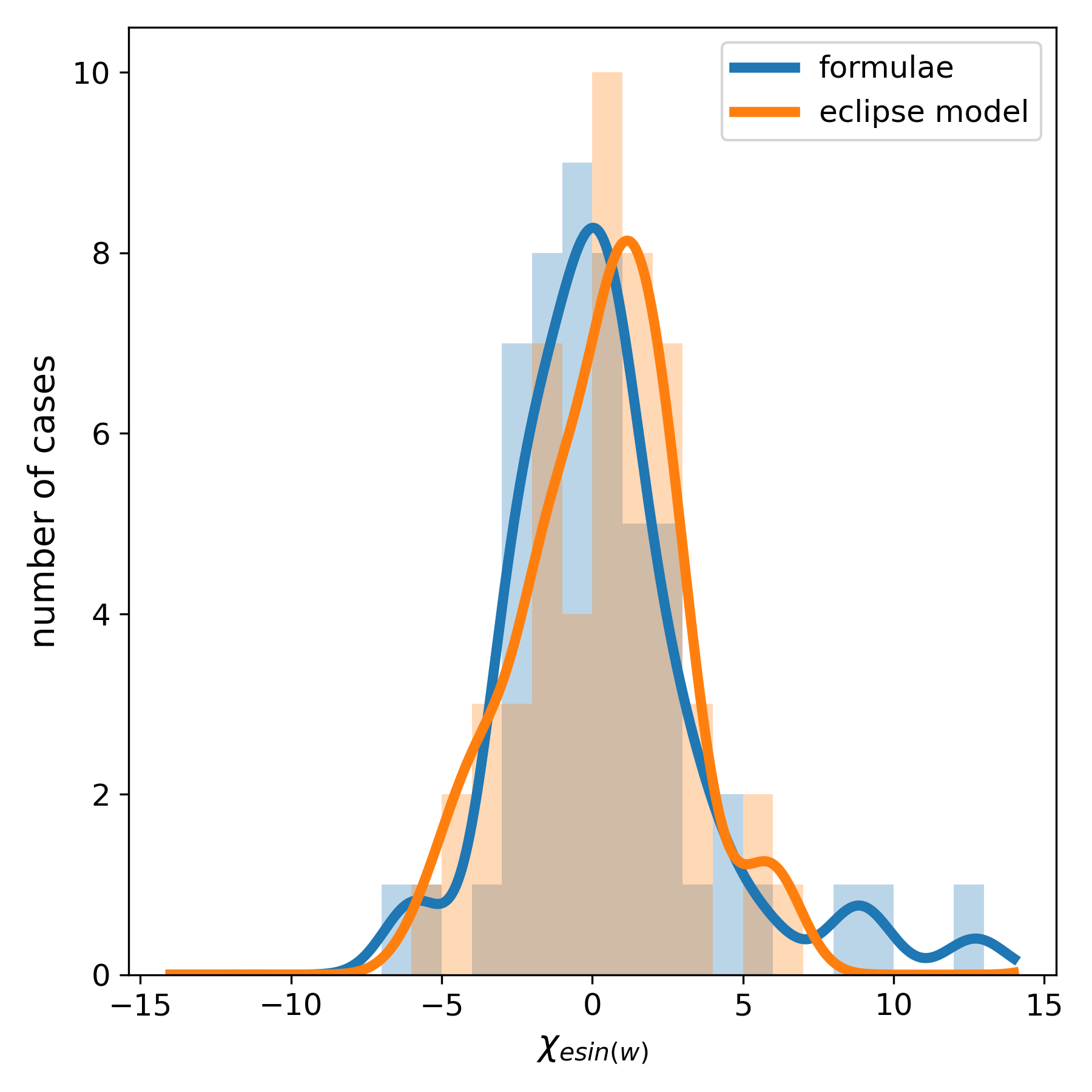}
    \caption{Histogram and KDE of the radial component of eccentricity $\chi$ values (deviation from the input divided by the error estimate).}
    \label{fig:esinw_dev}
\end{figure}

\begin{figure}
\centering
\includegraphics[width=\hsize]{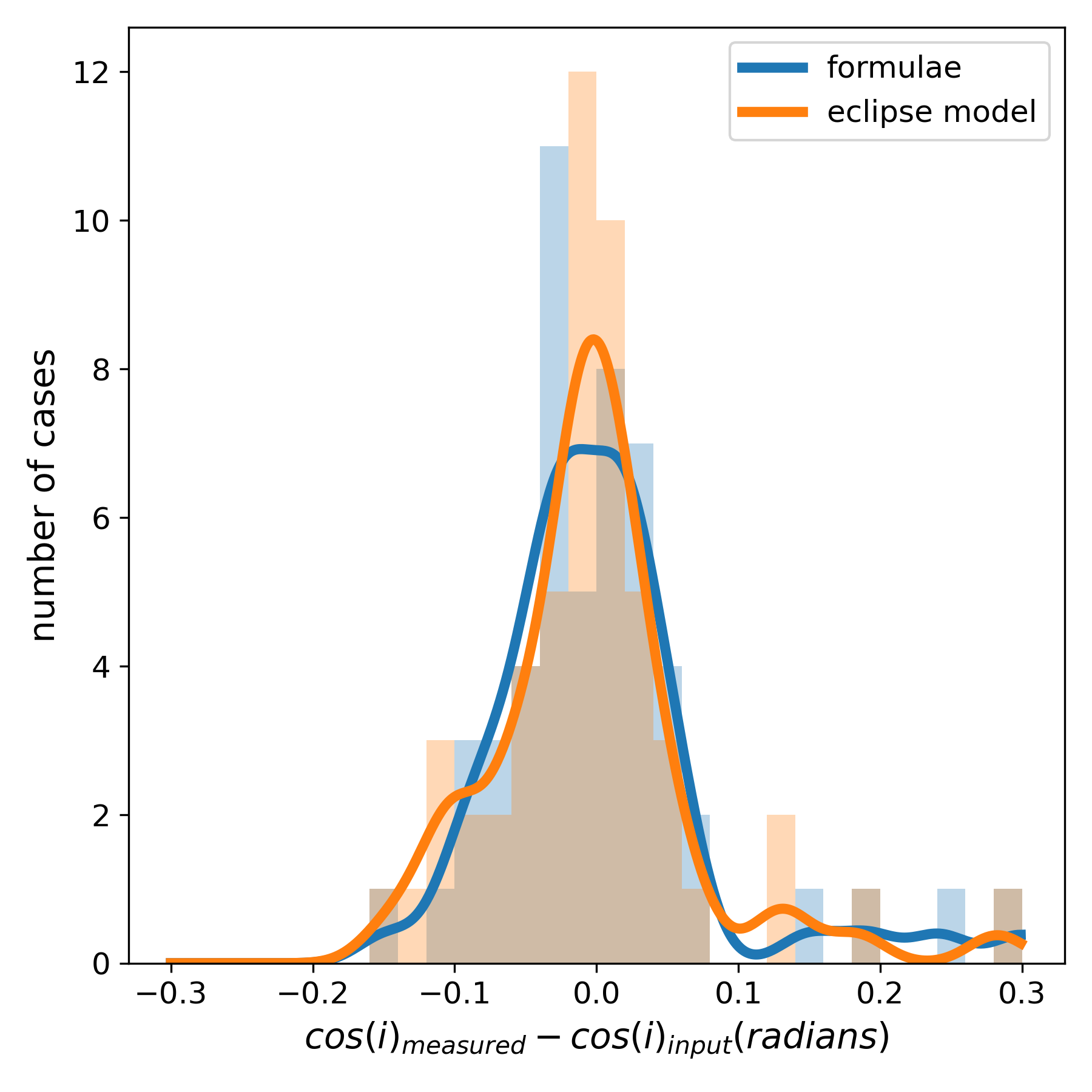}
    \caption{Histogram and KDE of the cosine of inclination deviations from the input.}
    \label{fig:cosi_dev_abs}
\end{figure}

\begin{figure}
\centering
\includegraphics[width=\hsize]{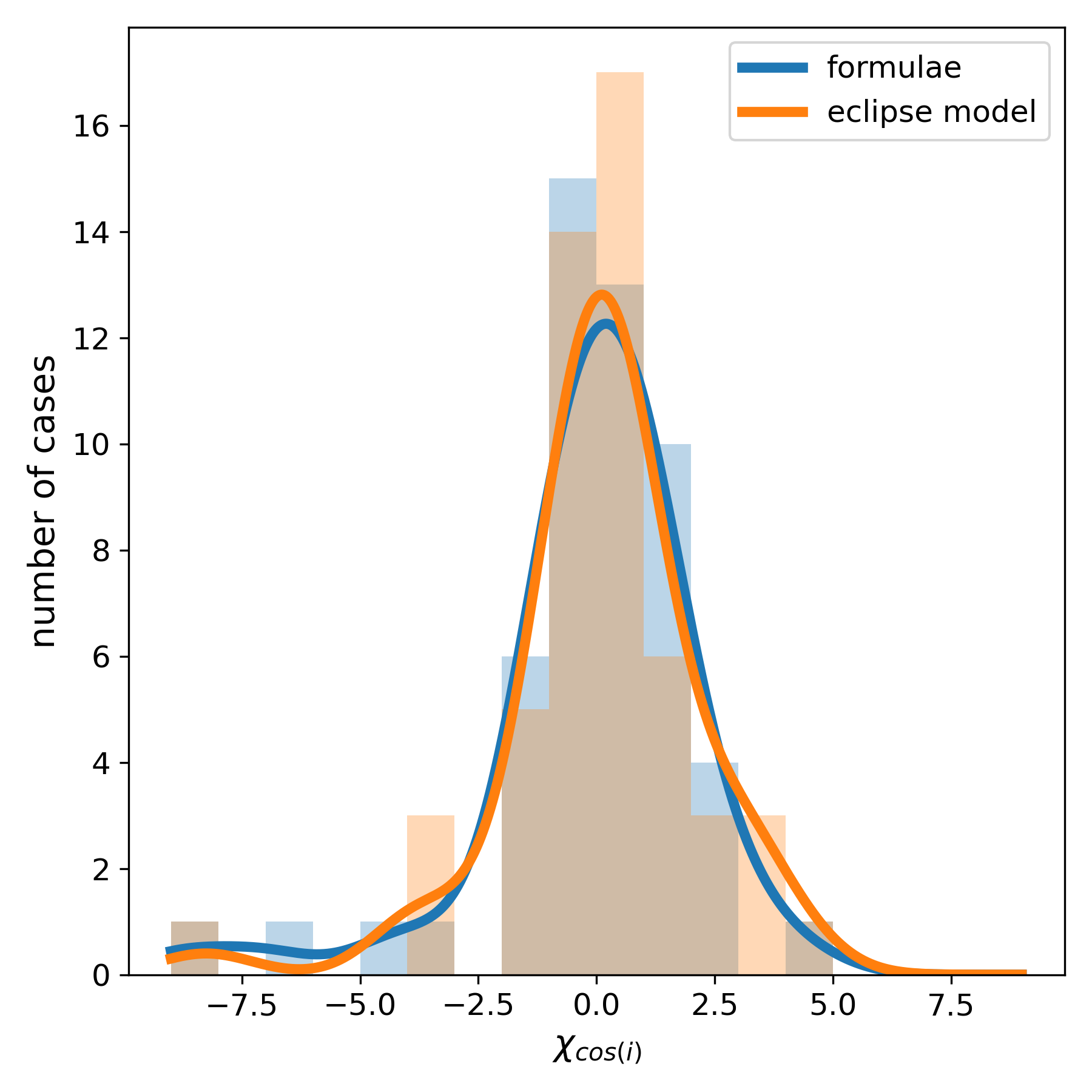}
    \caption{Histogram and KDE of the cosine of inclination $\chi$ values (deviation from the input divided by the error estimate).}
    \label{fig:cosi_dev}
\end{figure}

\begin{figure}
\centering
\includegraphics[width=\hsize]{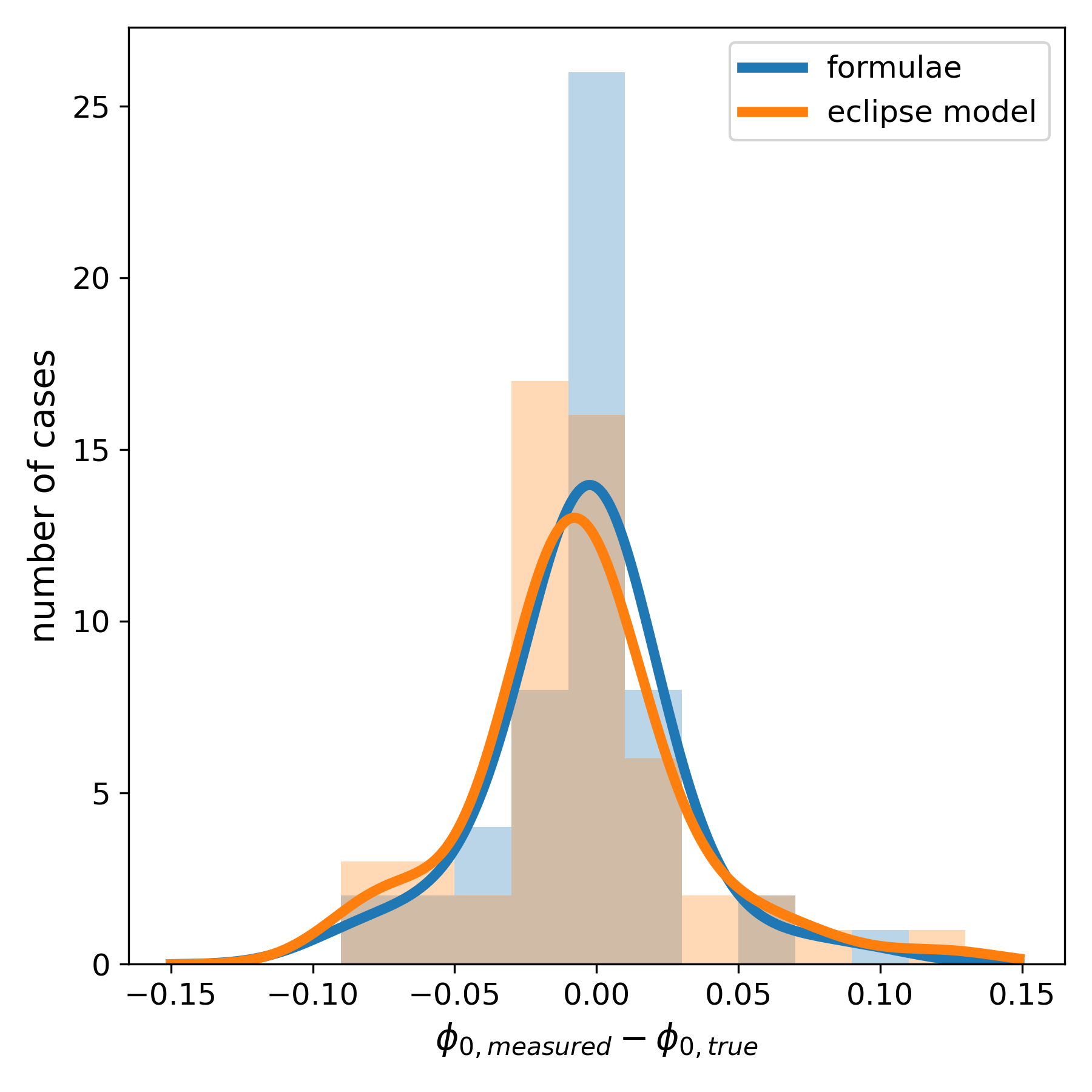}
    \caption{Histogram and KDE of the auxiliary angle, $\phi_0$, deviations from the input.}
    \label{fig:phi_0_dev_abs}
\end{figure}

\begin{figure}
\centering
\includegraphics[width=\hsize]{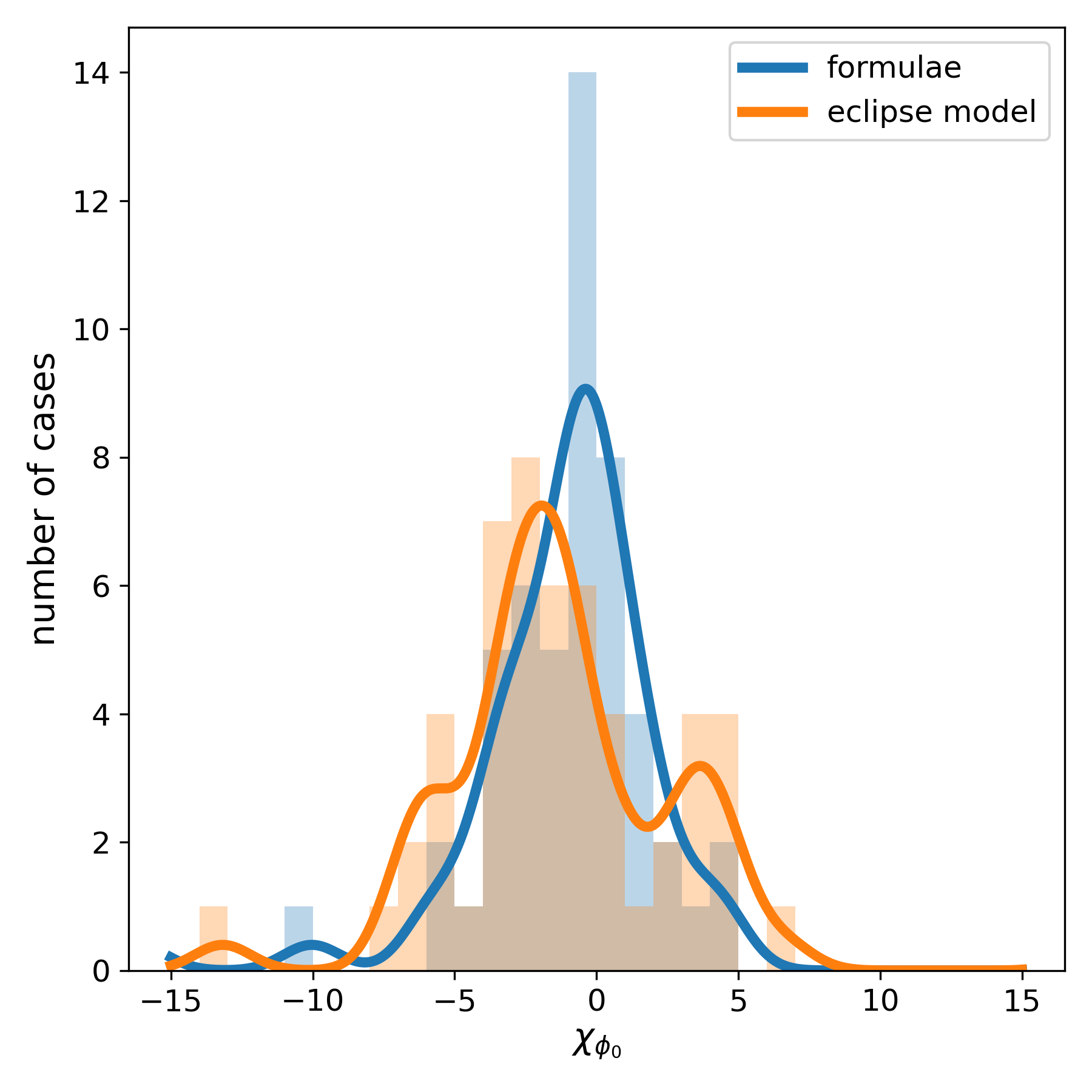}
    \caption{Histogram and KDE of the auxiliary angle, $\phi_0$, $\chi$ values (deviation from the input divided by the error estimate).}
    \label{fig:phi_0_dev}
\end{figure}

\begin{figure}
\centering
\includegraphics[width=\hsize]{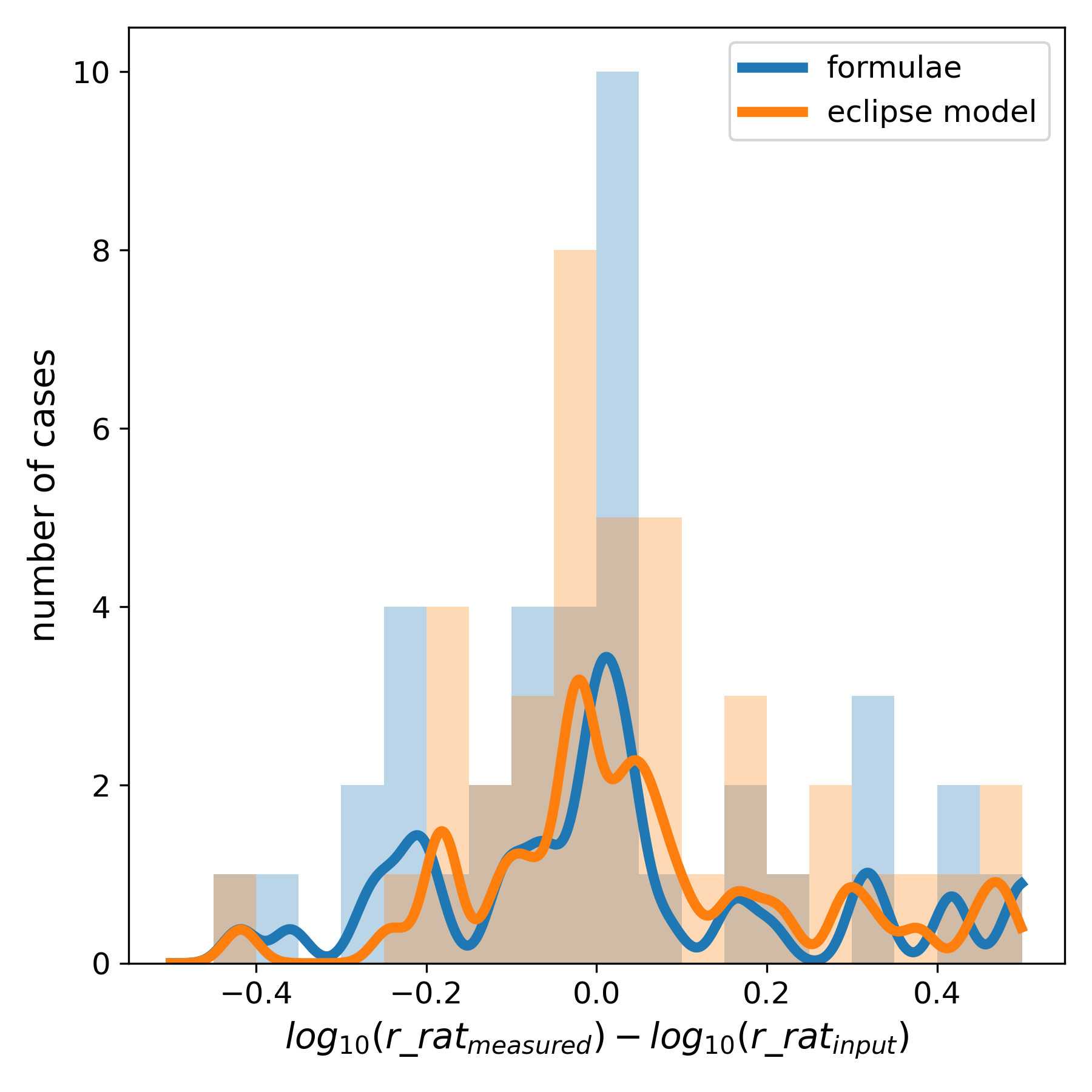}
    \caption{Histogram and KDE of the logarithm of the ratio of radius deviations from the input.}
    \label{fig:logrr_dev_abs}
\end{figure}

\begin{figure}
\centering
\includegraphics[width=\hsize]{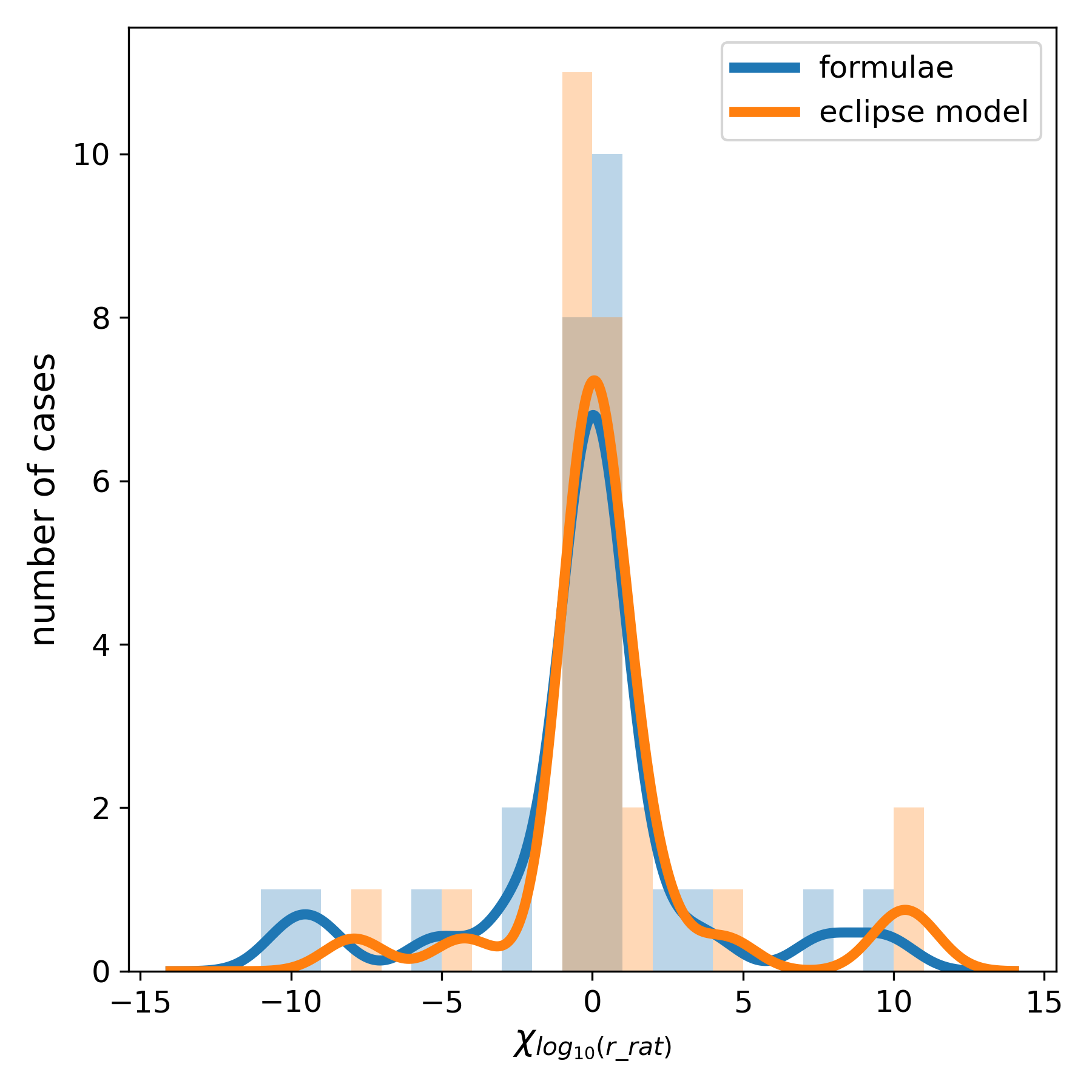}
    \caption{Histogram and KDE of the logarithm of the ratio of radius $\chi$ values (deviation from the input divided by the  error estimate).}
    \label{fig:logrr_dev}
\end{figure}

\begin{figure}
\centering
\includegraphics[width=\hsize]{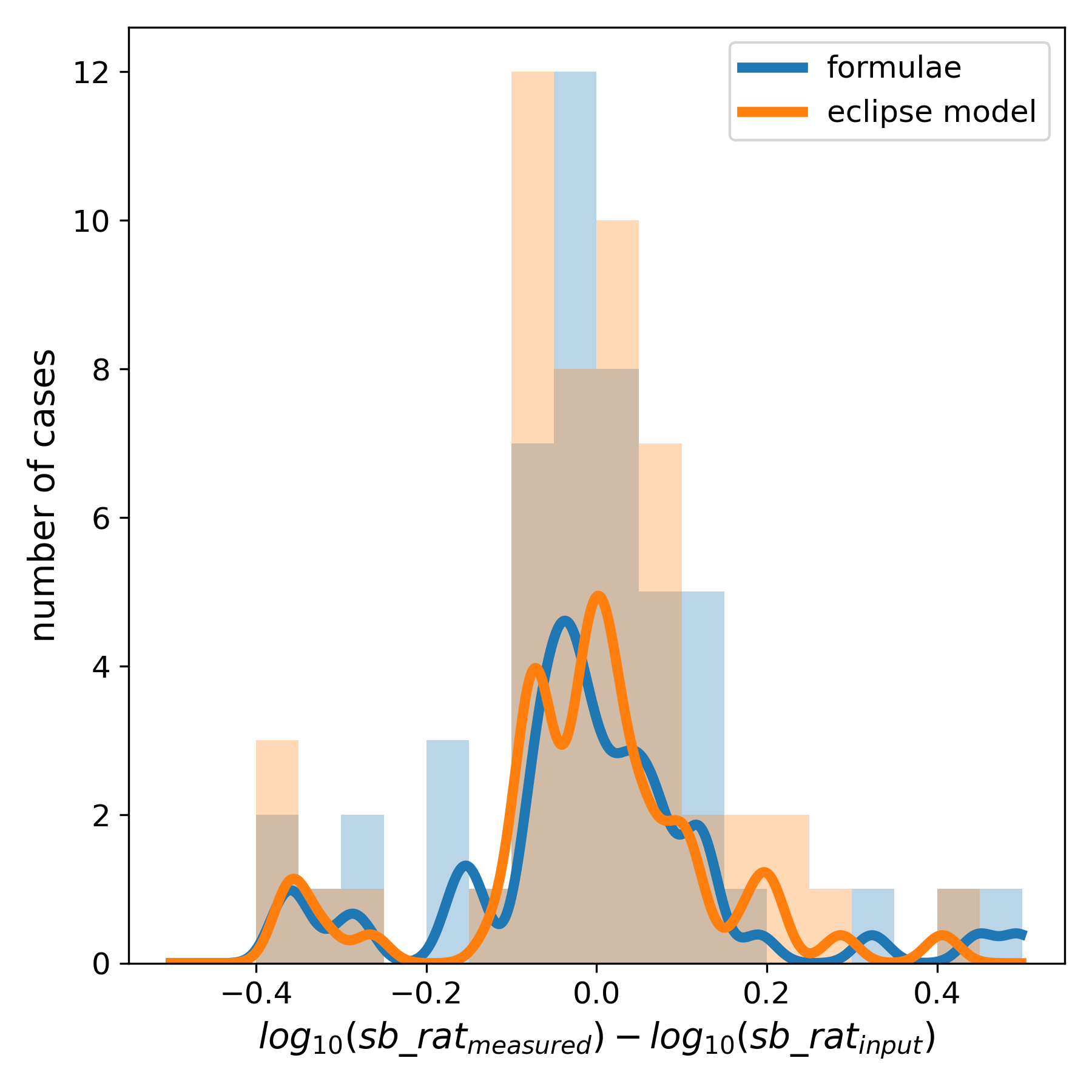}
    \caption{Histogram and KDE of the logarithm of the ratio of surface brightness deviations from the input.}
    \label{fig:logsb_dev_abs}
\end{figure}

\begin{figure}
\centering
\includegraphics[width=\hsize]{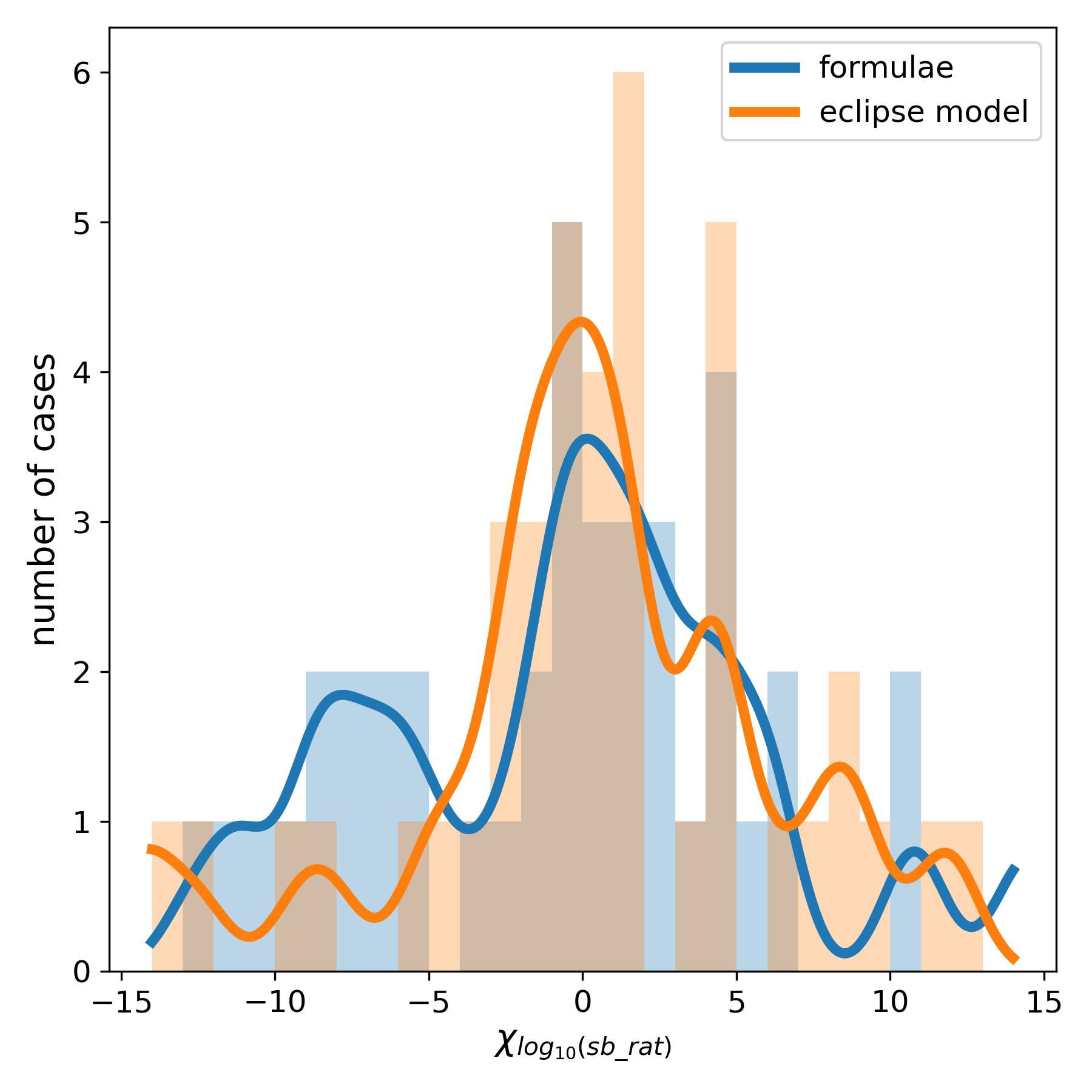}
    \caption{Histogram and KDE of the logarithm of the ratio of surface brightness $\chi$ values (deviation from the input divided by the  error estimate).}
    \label{fig:logsb_dev}
\end{figure}

\begin{figure}
\centering
\includegraphics[width=\hsize]{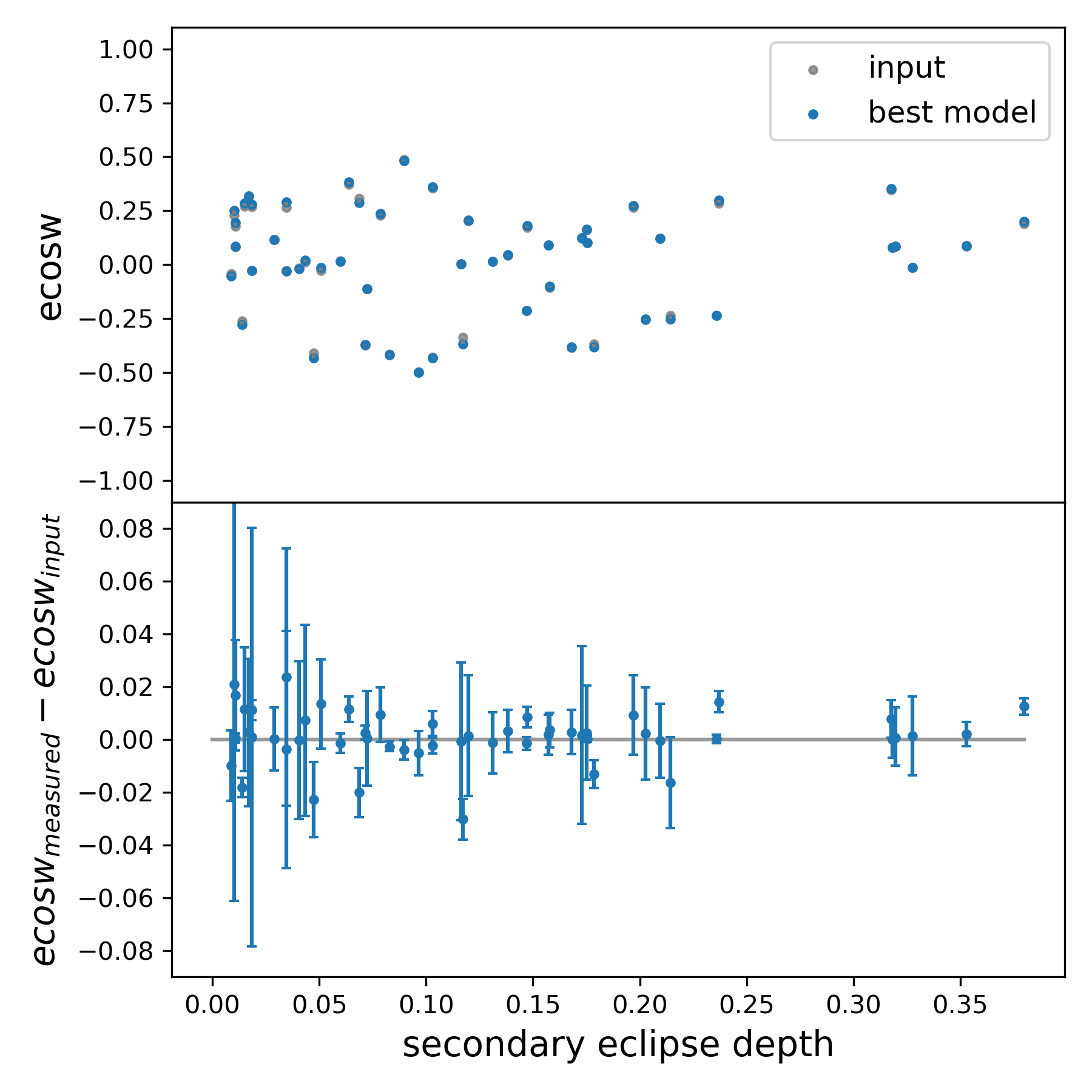}
    \caption{Tangential part of eccentricity (top panel) and deviation from input (bottom panel) versus secondary eclipse depth. }
    \label{fig:ecosw_depth}
\end{figure}

\clearpage

\subsubsection{Example light curves}
\label{apx:plots_examples}

\begin{figure*}
\resizebox{\hsize}{!}
    {\includegraphics[width=\hsize,clip]{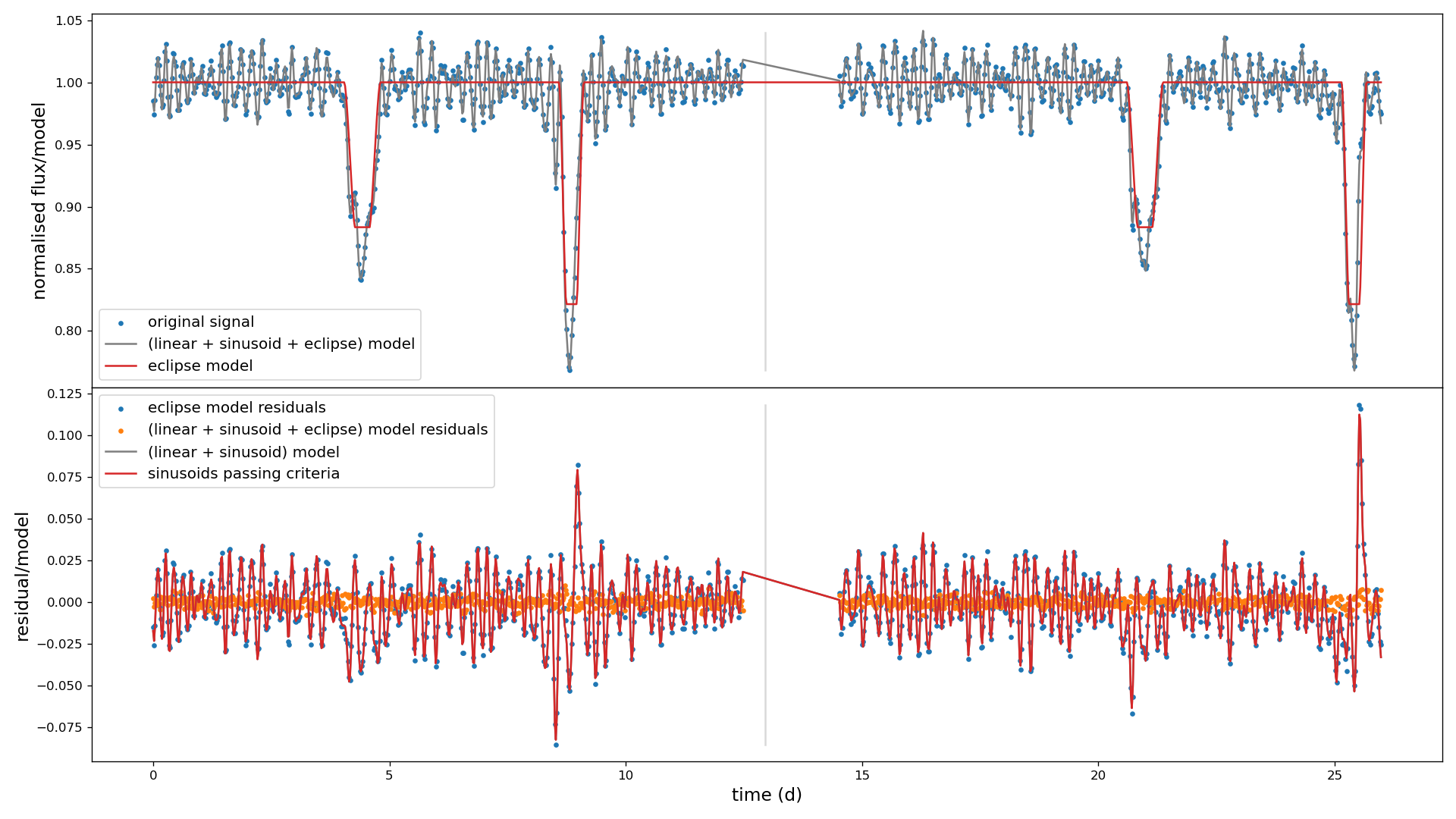}}
    \caption{Example (full) synthetic light curve of case 7. The top panel shows the final eclipse model of the eclipses in red and the full model of the light curve including sinusoids in grey. The bottom panel shows the residuals of subtracting the eclipse model (blue) and those of subtracting the full model (orange). The model of sinusoids is also plotted separately in red here. The slight overestimation of the primary eclipse width can be identified by sharp residual peaks here and leads to an underestimation of the radial part of eccentricity of about 0.2 for this case.}
    \label{fig:sim7}
\end{figure*}

\begin{figure*}
\resizebox{\hsize}{!}
    {\includegraphics[width=\hsize,clip]{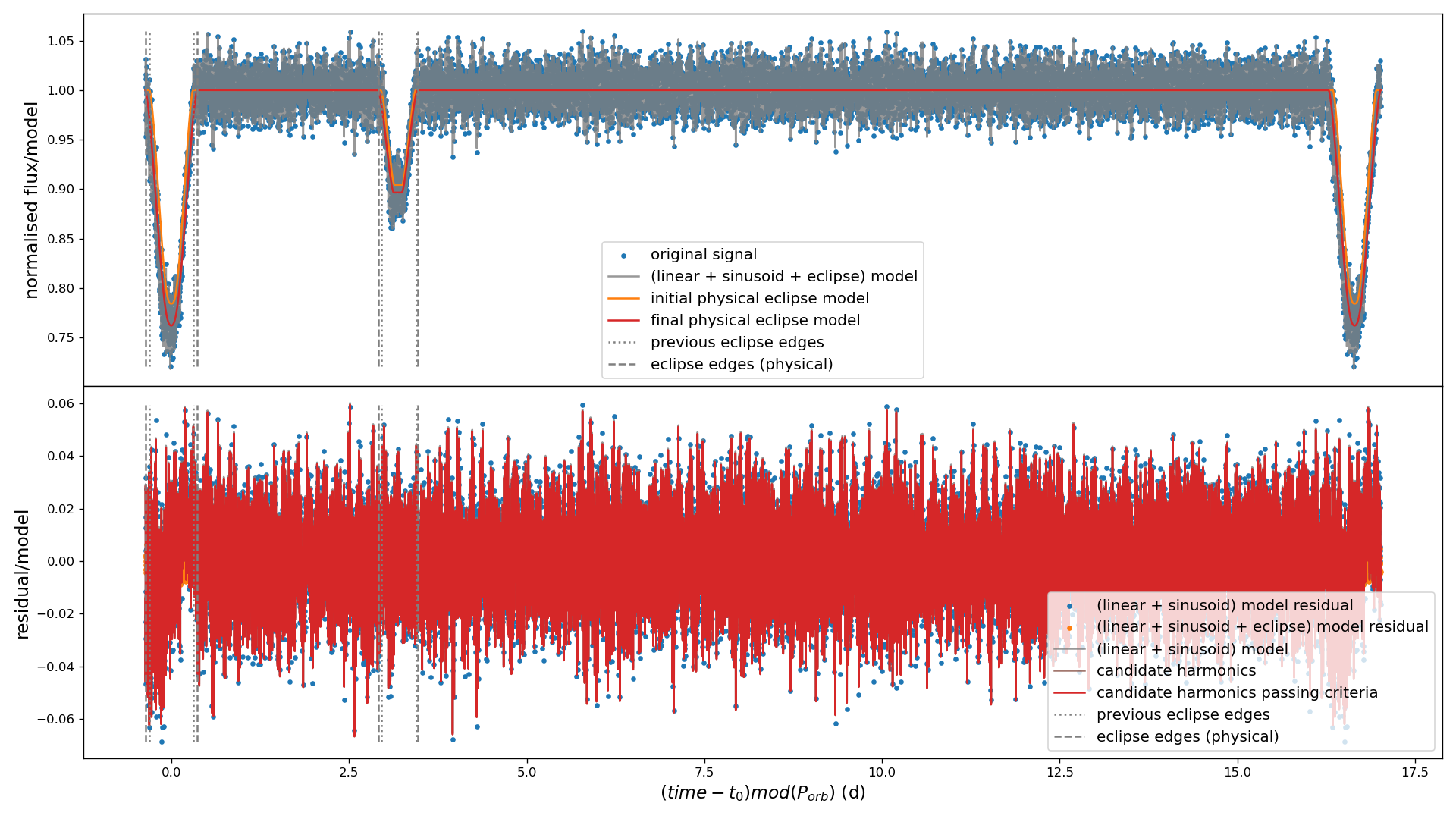}}
    \caption{Synthetic light curve of case 26 folded over the orbital period. The top panel shows the final eclipse model of the eclipses in red and the full model of the light curve including sinusoids in grey. The bottom panel shows the residuals of subtracting the eclipse model (blue) and those of subtracting the full model (orange). The model of sinusoids is also plotted separately in red here and covers up the data points. }
    \label{fig:sim26}
\end{figure*}

\begin{figure*}
\resizebox{\hsize}{!}
    {\includegraphics[width=\hsize,clip]{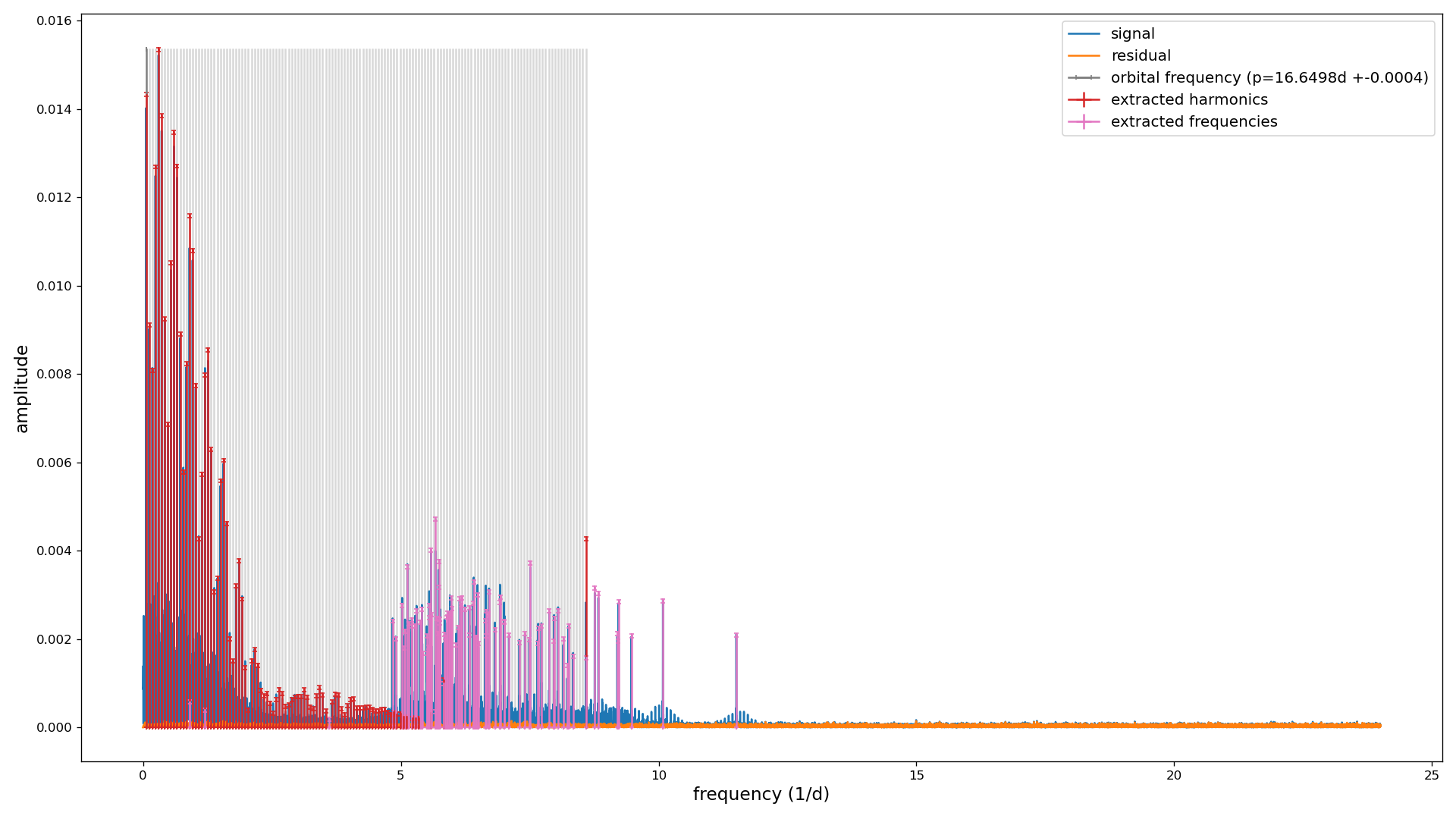}}
    \caption{ Fourier amplitude spectrum of case 26. In red and pink are the extracted sinusoids, of which the red ones qualify as harmonics. The peaks around 5 to 10 cycles per day are from sine waves put into the synthetic light curve and recovered accurately and precisely by our analysis pipeline.}
    \label{fig:sim26b}
\end{figure*}

Here we show a few examples of the analysis results from the synthetic set of test light curves discussed in Section \ref{sec:testing}. The average level of added sinusoid amplitudes is high in this set, challenging the algorithm for eclipse detection and measurement.  In Figure \ref{fig:sim7} the algorithm was successful in detecting and measuring the eclipses, perhaps with the exception of a remaining residual spike on the egress of the primary. Figures \ref{fig:sim26} and \ref{fig:sim26b} show the eclipse model of the final stage of the analysis and the periodogram with extracted sinusoids, respectively. This is the synthetic case used for the analysis of the accuracy and precision of the recovery of input sinusoids (most of them pink in the periodogram).

\begin{figure*}
\resizebox{\hsize}{!}
    {\includegraphics[width=\hsize,clip]{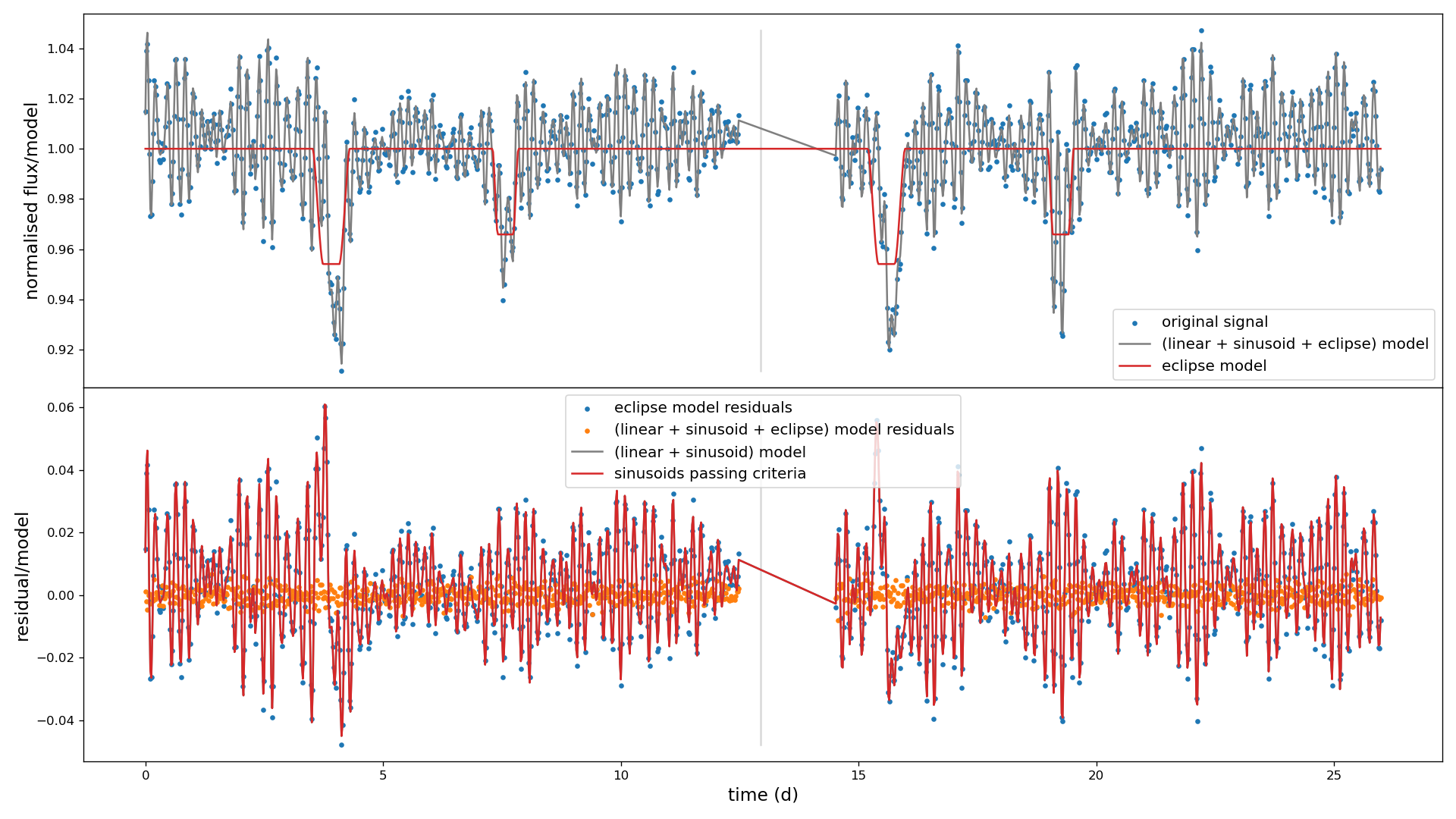}}
    \caption{Example (full) synthetic light curve of case 23. The top panel shows the final eclipse model in red and the full model of the light curve including sinusoids in grey. The bottom panel shows the residuals of subtracting the eclipse model (blue) and those of subtracting the full model (orange). The model of sinusoids is also plotted separately in red here. We can see that the primary eclipse is too wide as the detection algorithm picked up on other variability close to it.}
    \label{fig:sim23}
\end{figure*}

\begin{figure*}
\resizebox{\hsize}{!}
    {\includegraphics[width=\hsize,clip]{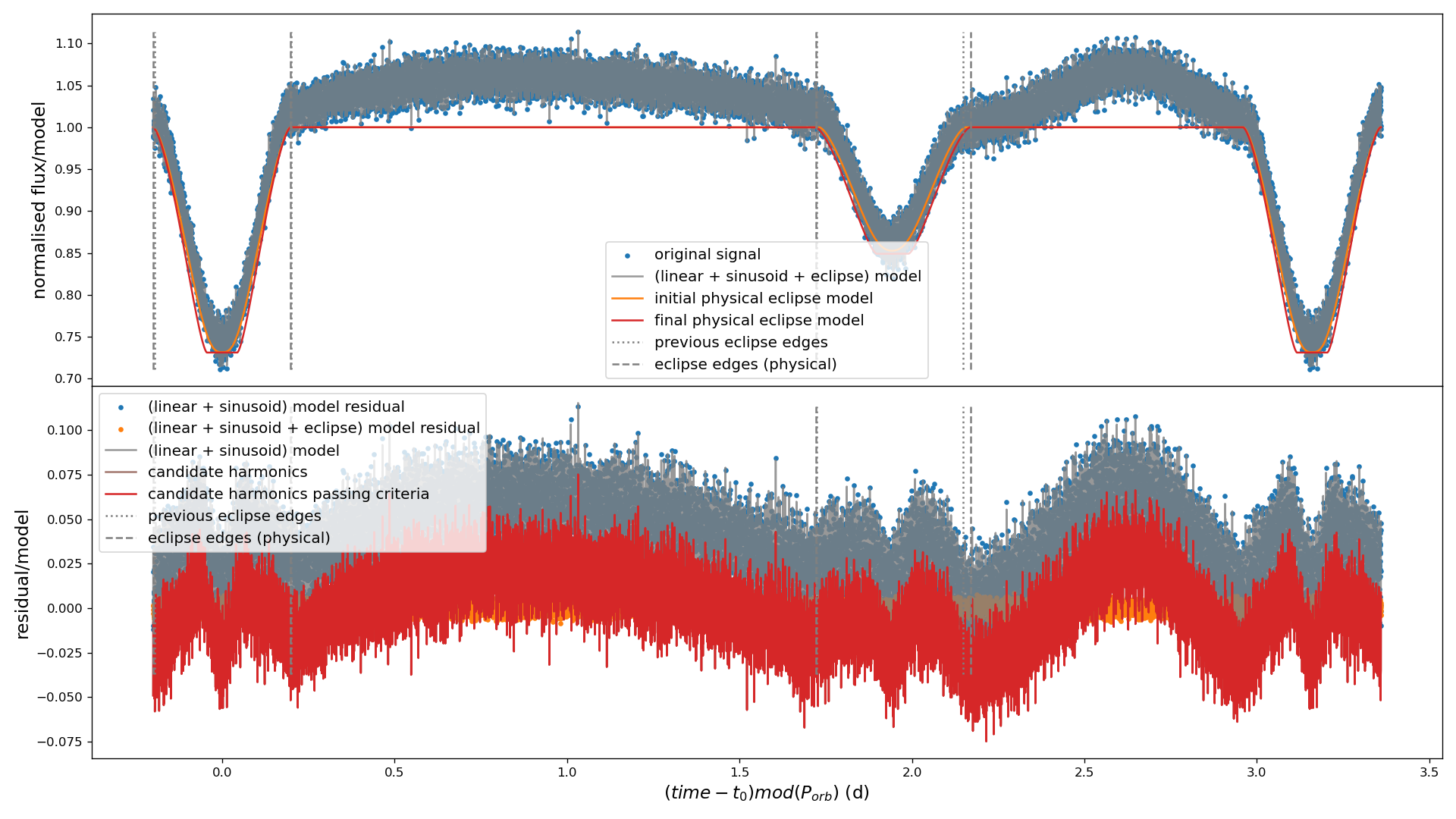}}
    \caption{Example (folded over the period) synthetic light curve of case 24. The top panel shows the final eclipse model in red and the full model of the light curve including sinusoids in grey. The bottom panel shows the residuals of subtracting the eclipse model (blue) and those of subtracting the full model (orange). The model of sinusoids is also plotted separately in red here. The eclipse model has trouble fitting these wide eclipses and does not capture any out-of-eclipse variability.}
    \label{fig:sim24}
\end{figure*}

We see better how the added variability affects the measurement in Figure \ref{fig:sim23}, where the primary is wider to one side than it should be. The resulting large residuals are captured by the sinusoids that are complementary to the eclipse model to make up the full variability of the light curve. This results in harmonic sinusoids that in this case model artificial signal, but in other cases can also help model, for example, the physical shape of the eclipses or the ellipsoidal variability not captured by the eclipse model (see Figure \ref{fig:sim24}). 

\begin{figure*}
\resizebox{\hsize}{!}
    {\includegraphics[width=\hsize,clip]{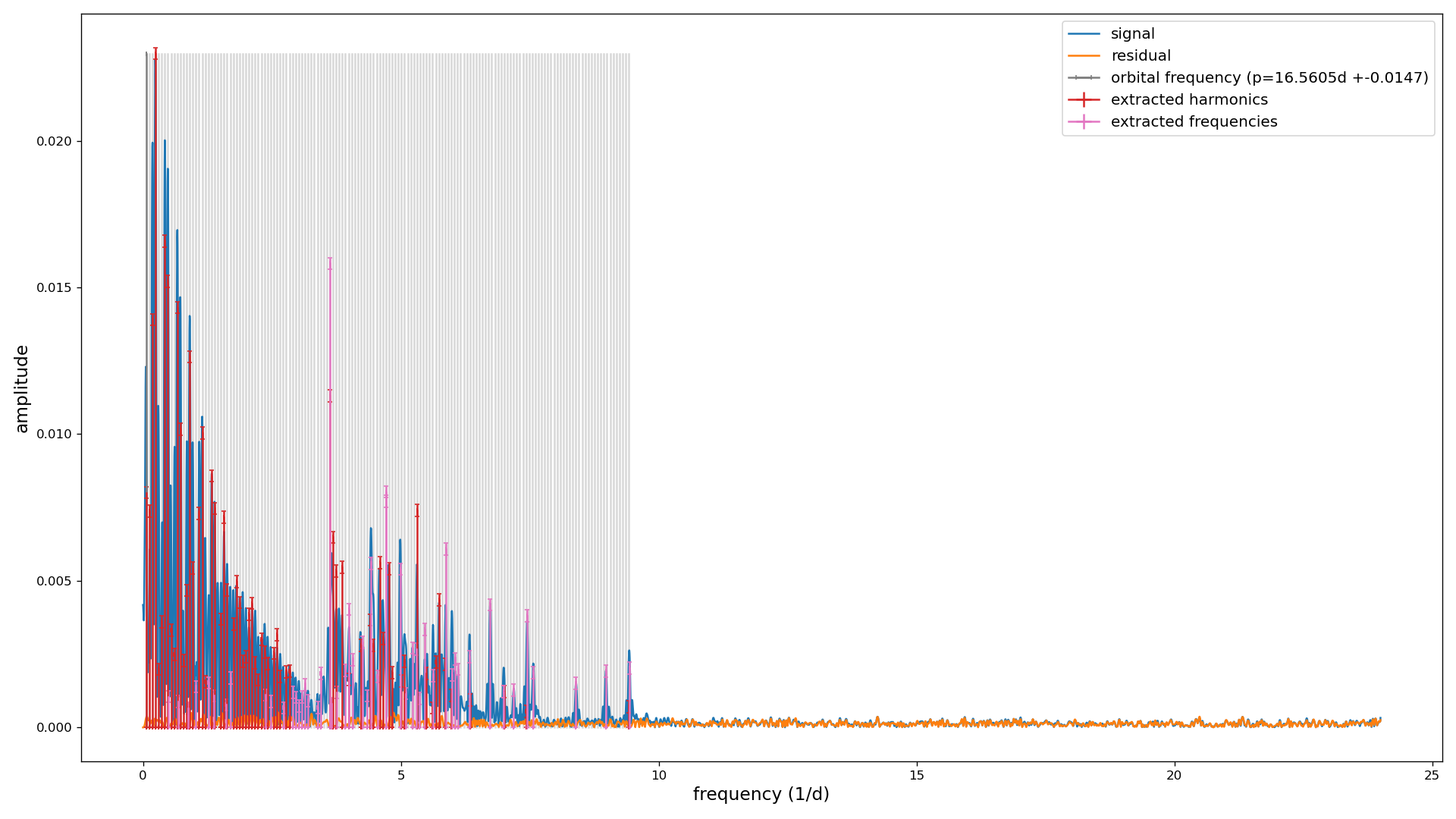}}
    \caption{ Fourier amplitude spectrum of case 7. In red and pink are the extracted sinusoids, of which the red ones qualify as harmonics. At an orbital period of 16.56, the spacing of harmonics is quite dense (grey grid lines).}
    \label{fig:sim7b}
\end{figure*}

\begin{figure*}
\resizebox{\hsize}{!}
    {\includegraphics[width=\hsize,clip]{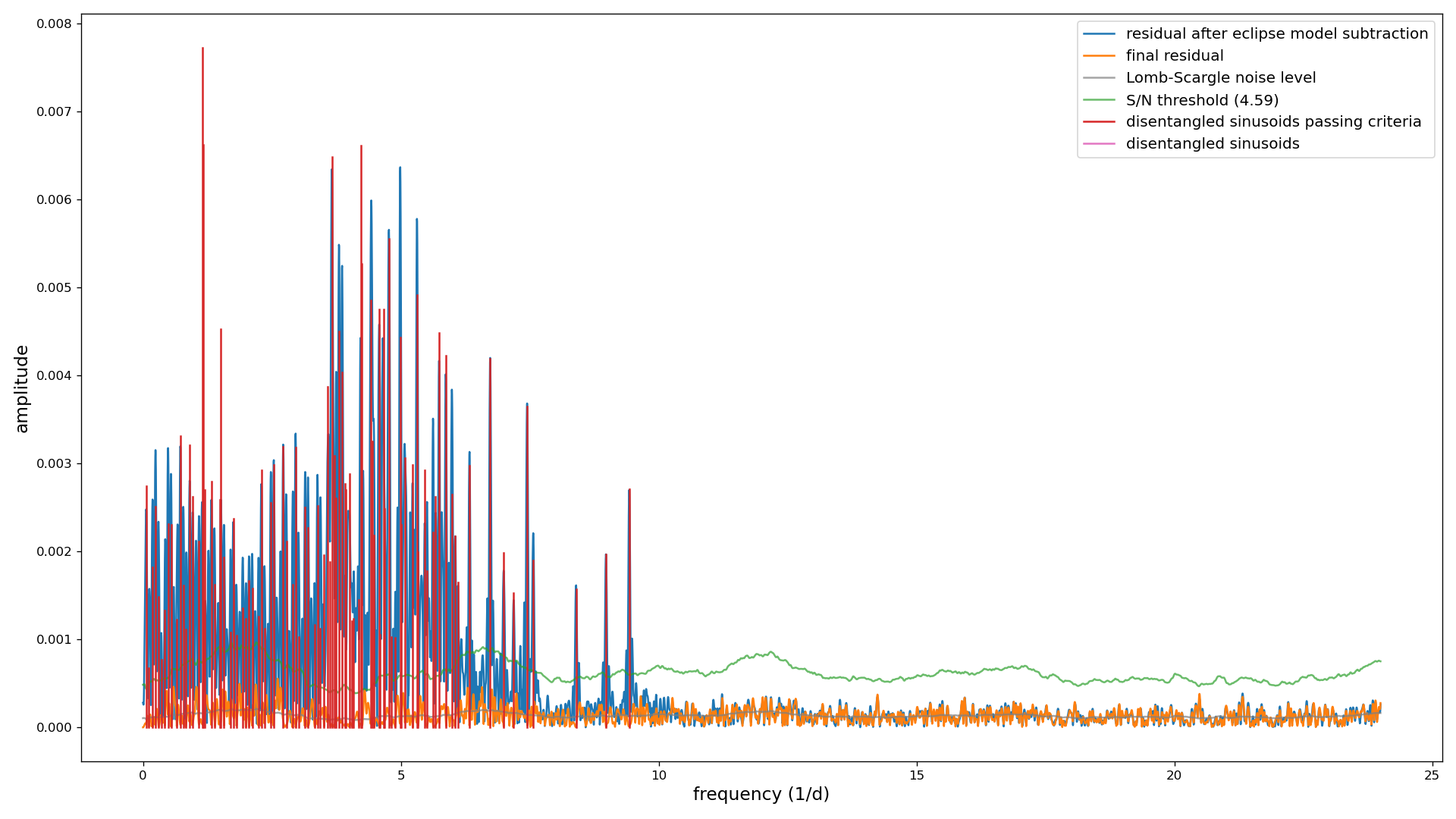}}
    \caption{ Fourier amplitude spectrum of case 7 after subtracting the eclipse model. In red are the extracted sinusoids that pass all criteria, including the signal-to-noise threshold shown in green (calculated in a window of $1 d^{-1}$). }
    \label{fig:sim7c}
\end{figure*}

\begin{figure*}
\resizebox{\hsize}{!}
    {\includegraphics[width=\hsize,clip]{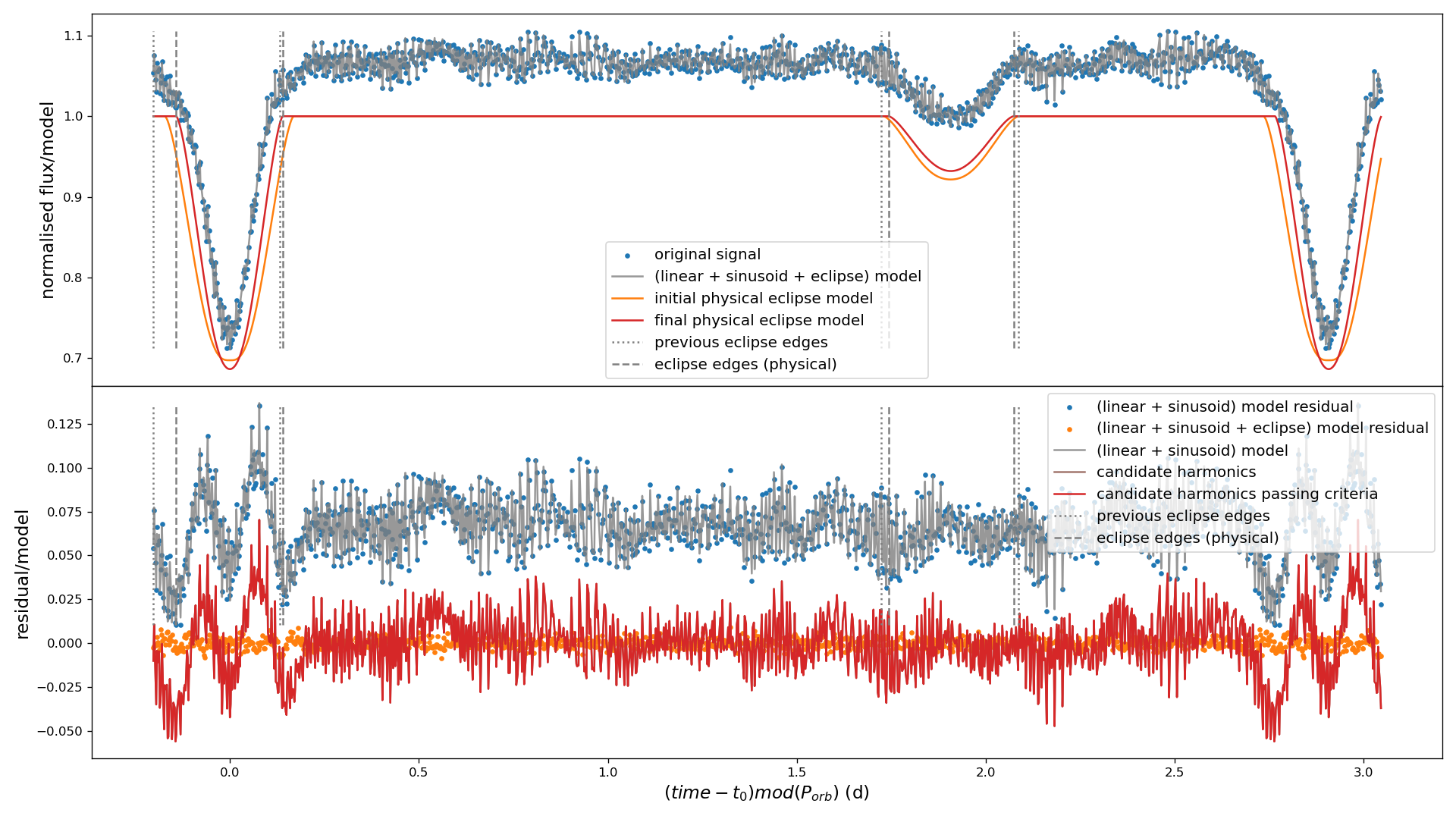}}
    \caption{Example synthetic light curve of case 99 folded over the orbital period. The top panel shows the final eclipse model of the eclipses in red and the full model of the light curve including sinusoids in grey. The orange model is the starting point for the light curve fit. The bottom panel shows the residuals of subtracting the eclipse model (blue) and those of subtracting the full model (orange). The model of sinusoids is also plotted separately in red here. The model starts from an overestimated primary eclipse width, and the fit is unable to fully correct for this as the residuals show. This leads to and underestimated tangential eccentricity by about 0.2 for this case.}
    \label{fig:sim99}
\end{figure*}

\begin{figure*}
\resizebox{\hsize}{!}
    {\includegraphics[width=\hsize,clip]{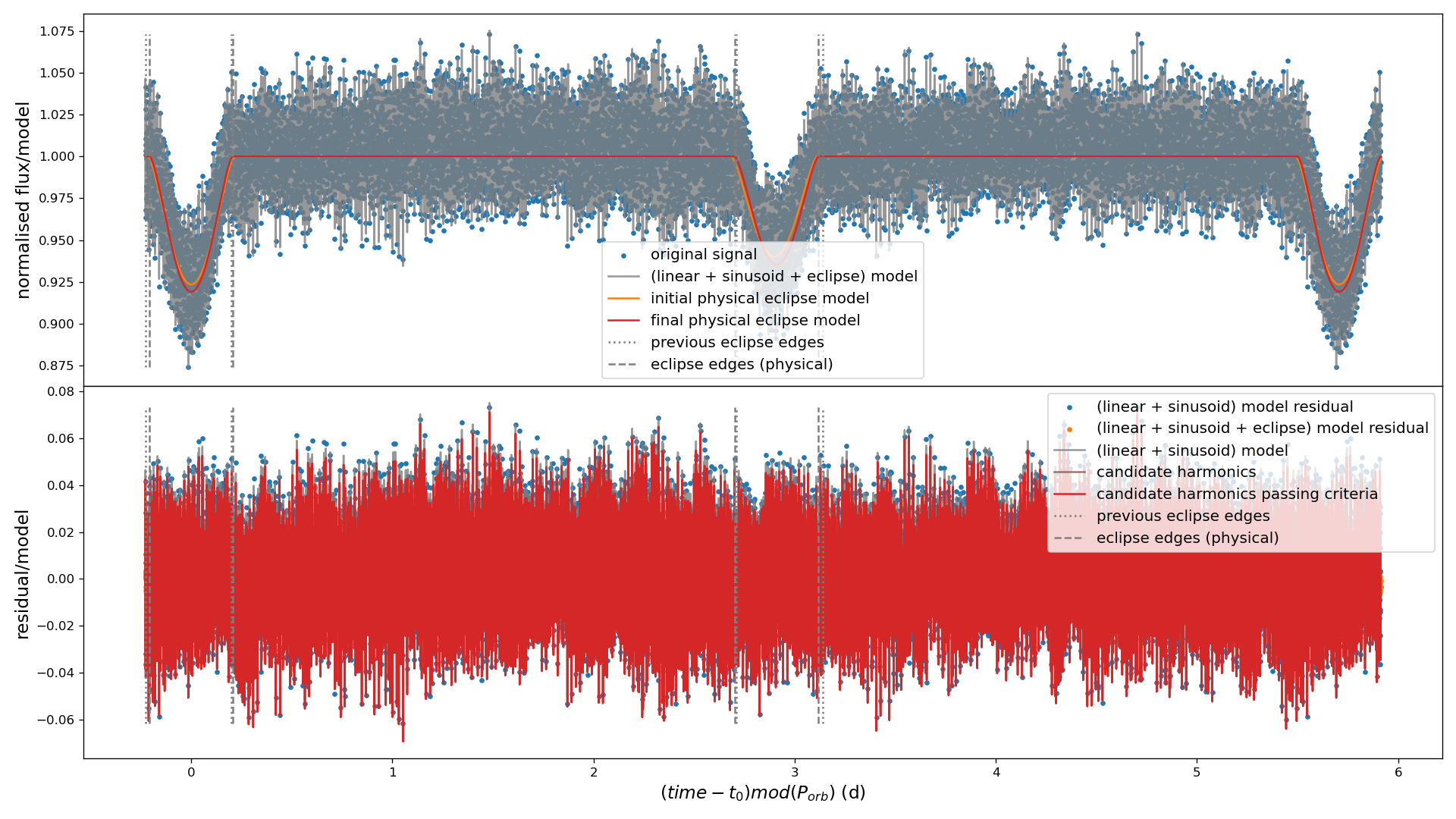}}
    \caption{Example (folded over the period) synthetic light curve of case 88. The top panel shows the final eclipse model in red and the full model of the light curve including sinusoids in grey. The bottom panel shows the residuals of subtracting the eclipse model (blue) and those of subtracting the full model (orange). The model of sinusoids is also plotted separately in red here.}
    \label{fig:sim88}
\end{figure*}

\clearpage

\subsubsection{\textit{Kepler} EB Catalog}
\label{apx:plots_kepler}

\begin{figure}
\centering
\includegraphics[width=\hsize]{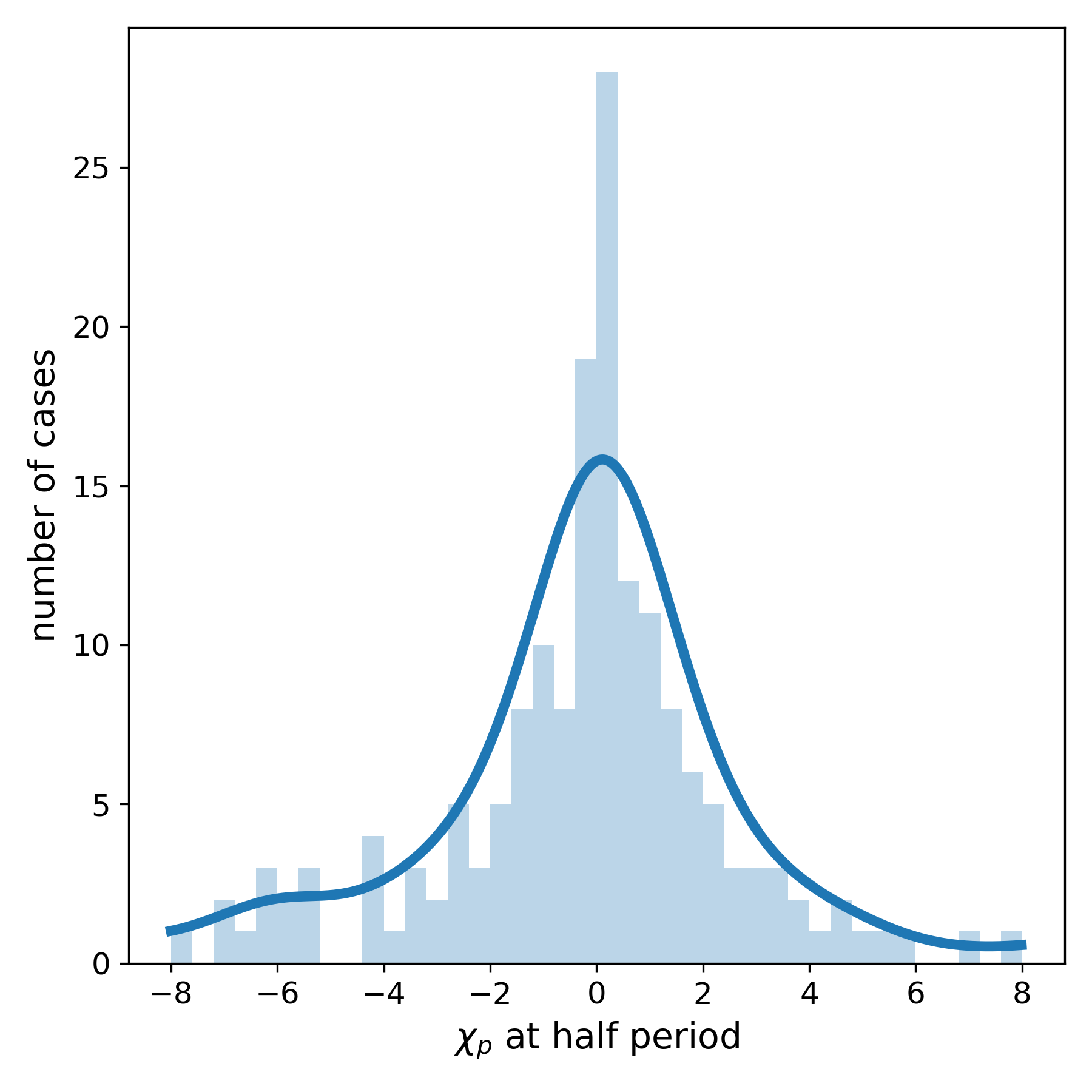}
    \caption{Histogram and KDE of the orbital period $\chi$ values, comparing orbital periods measured in this work to those from the \textit{Kepler} catalogue. Here we select the periods within 1\% of half the catalogue values and the errors are taken to be the SLR uncertainties. The KDE is scaled to the height of the histogram.}
    \label{fig:kepler_period_half_dev}
\end{figure}

\begin{figure}
\centering
\includegraphics[width=\hsize]{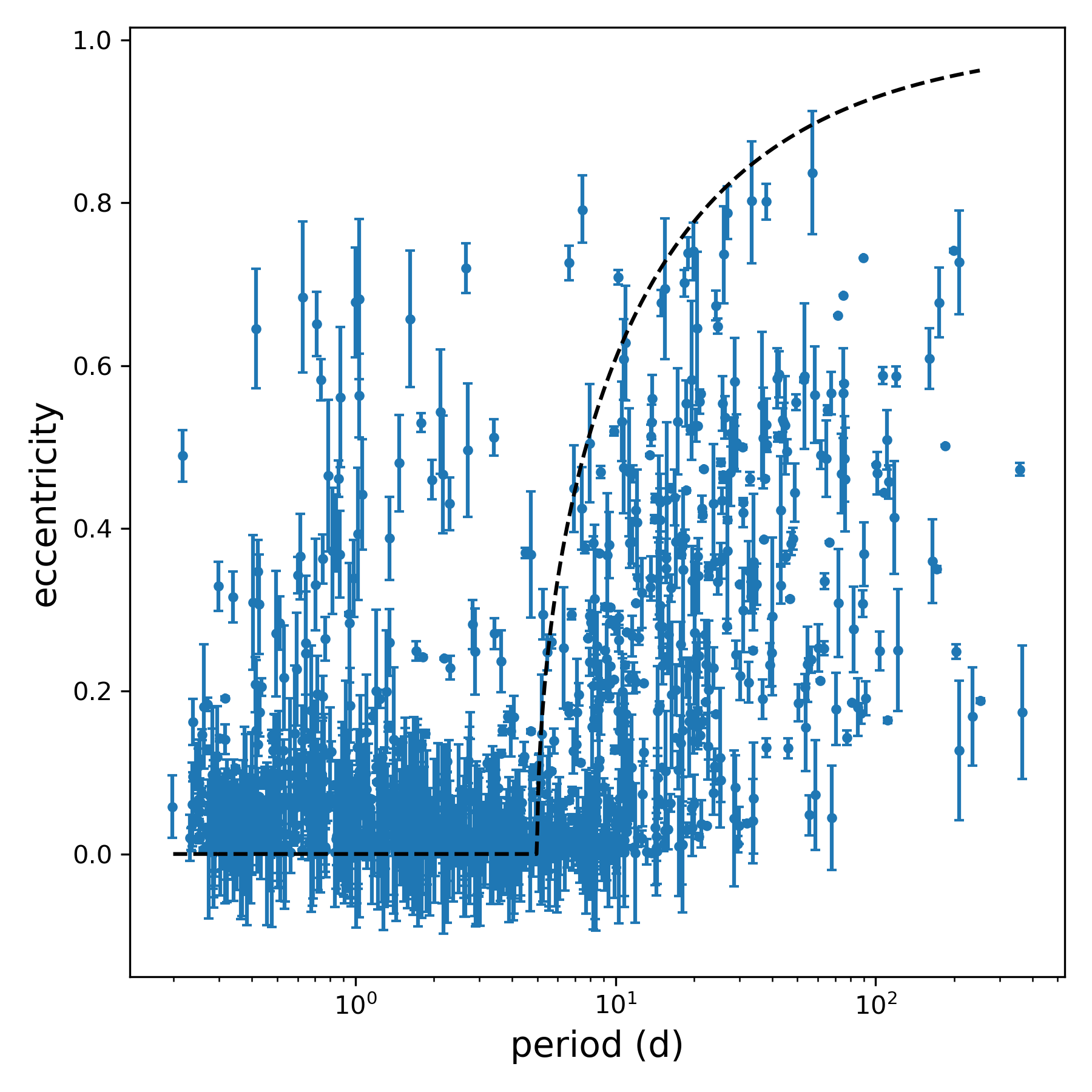}
    \caption{Eccentricity measurements for a subset of targets from the \textit{Kepler} EB Catalog. A cut-off was made in eccentricity error (<0.1) but all morphology values were included. We use the maximum eccentricity as a function of orbital period in \citet[][Equation 3]{Halbwachs2005} with the cut-off period set to 5 days to plot the dashed line.}
    \label{fig:kepler_ecc_morph}
\end{figure}

\begin{figure}
\centering
\includegraphics[width=\hsize]{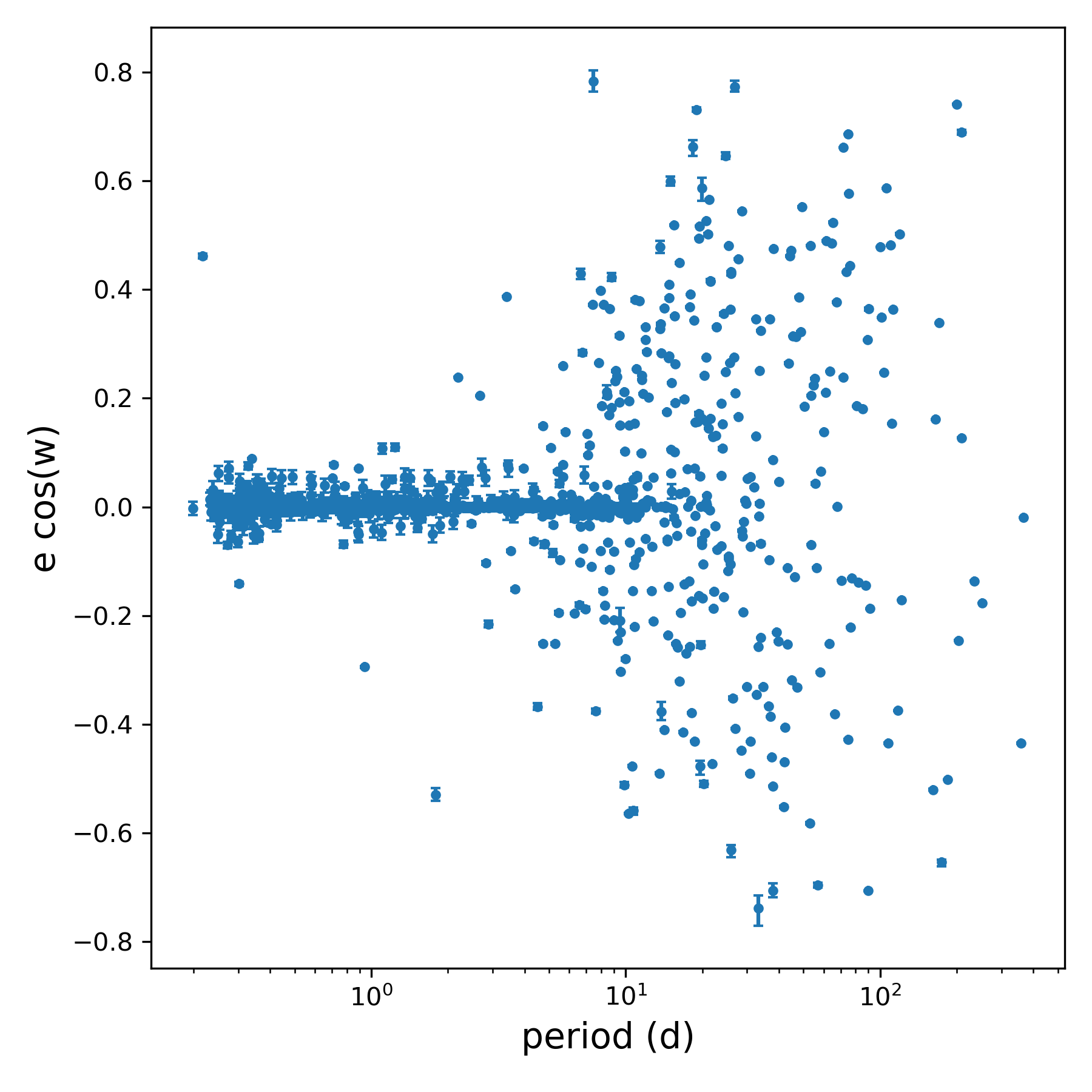}
    \caption{Measurements of the tangential component of eccentricity for a subset of targets from the \textit{Kepler} EB Catalog. The subset results from a cut-off in eccentricity error (<0.1) but all morphology values were included.}
    \label{fig:kepler_ecosw_morph}
\end{figure}

Figure \ref{fig:kepler_period_half_dev} is complementary to Figure \ref{fig:kepler_period_dev}: it shows the distribution of measured periods around half the catalogue values. The shape of the distribution differs little from the one around the catalogue period, with a sharp central peak and heavy tails.

Adding all morphologies from the \textit{Kepler} EB Catalog to the eccentricity-period plot, Figure \ref{fig:kepler_ecc_morph}, extends it substantially to lower periods. We now include 1399 of 2001 targets with an eccentricity measurement, as compared to the 715 we had when limiting to morphologies below 0.5. We see that the majority of additional systems fall at around zero eccentricity, and while the error bars are relatively large, there is more scatter than for the previous selection of targets below ten days orbital period. Especially noticeable are the systems at periods around a day and high eccentricity. These do not indicate physical peculiarities, but rather inaccurate determination of eclipse times. For these low period systems, eclipses are often not sharply defined making it harder for our derivative-based method to accurately determine the edges.

We see from Figure \ref{fig:kepler_ecosw_morph} that most of these outlier systems are not due to large $e cos(\omega)$ values: even though the scatter around the zero line for the tangential component of eccentricity is markedly larger for these high morphology targets, it stays within a value of 0.1 for most of them. That means inaccurate $e sin(\omega)$ measurement is the driving factor for these outliers.

\clearpage

\section{Correlation reduction}
\label{apx:correlation}
Reducing the correlation between optimisation parameters can strongly benefit the ease and accuracy with which the optimisation is done. It may be possible to transform a parameter to this end; one such example is the use of auxiliary angle $\phi_0$ as mentioned in the main text. For the reason that we are trying to solve the general case of an EB light curve, it may not always make sense to use a certain re-parametrisation. For example: when eclipses are grazing, instead of showing the flattened-off phase of either total eclipses or transits, we lose information about the eclipsing stars. In this case, a grazing eclipse provides less information on the radii, but we do not lose all. It may then be appealing to keep the ratio of radii fixed at one, while only fitting for the sum of the scaled radii. We can, however, do better. The only information we actually lose completely is which star is bigger and which is smaller, but we can still recover information about the size of the bigger star and the size of the smaller star. We can represent this loss of knowledge in a new free parameter, which replaces the ratio of radii:
\begin{equation}
    \rho = \frac{r_1^2 + r_2^2}{a^2},
\end{equation}
where $r_i$ is each star's radius and $a$ the semi-major axis, similar to the sum of radii. We keep the sum of radii,
\begin{equation}
    r_{sum} = \frac{r_1 + r_2}{a},
\end{equation}
or, alternatively, angle $\phi_0$ can be chosen. We cannot recover the individual radii $r_1$ and $r_2$ from this combination of parameters, representing the lack of information in the light curve. What we can recover is
\begin{equation}
    r_{s,l} = \frac{1}{2} r_{sum} \pm \frac{1}{2} \sqrt{2 \rho - r_{sum}^2},
\end{equation}
with which we find a small and a large radius (implicitly scaled by $a$). Clearly, for the systems where we have the information available, we would not want to artificially limit ourselves, in addition to actually increasing correlations between the parameters in those cases. Therefore, we chose to accept sub-optimal correlation structures in return for a more generally and homogeneously applicable method. However, for an analysis of a specific binary with grazing eclipses (which is a large fraction of all EBs for geometrical reasons), this approach may aid the specific optimisation procedure utilised, increasing accuracy and potentially speed.

\end{appendix}

\end{document}